\documentclass{article} 
\PassOptionsToPackage{table}{xcolor}
\usepackage{iclr2027_conference,times}

\usepackage{amsmath,amsfonts,bm}

\def\eqref#1{equation~\ref{#1}}

\def\1{\bm{1}}

\DeclareMathAlphabet{\mathsfit}{\encodingdefault}{\sfdefault}{m}{sl}
\SetMathAlphabet{\mathsfit}{bold}{\encodingdefault}{\sfdefault}{bx}{n}

\usepackage{hyperref}
\usepackage{url}

\usepackage{booktabs}
\usepackage{makecell}
\usepackage{xcolor}
\usepackage{tikz}
\usepackage{graphicx}
\usepackage{subcaption}
\usepackage{multirow}
\usepackage{tabularx}
\usepackage{xspace}
\usepackage{listings}
\usepackage{textcomp}
\usepackage{longtable}

\newcommand{\bench}{\textsc{LoLBench}\xspace}

\title{LoLBench: Evaluating Coding Agents with Long-Horizon Proposals on Large Software Systems}

\author{%
\textbf{Yun Peng}$^{1}$, \textbf{Zihan Wu}$^{2}$,
\textbf{Zeyang Zhuang}$^{3}$, \textbf{Xin Zhou}$^{4}$, \textbf{Rui Shu}$^{5}$, \\
\textbf{Xu Han}$^{6}$, \textbf{Chun Yong Chong}$^{7}$,
\textbf{Yuan Wang}$^{5}$, \textbf{Jiakun Liu}$^{8}$ \\
$^{1}$Fudan University, China \quad
$^{2}$City University of Hong Kong, Hong Kong \\
$^{3}$Chinese University of Hong Kong, Hong Kong \\
$^{4}$Singapore Management University, Singapore \\
$^{5}$Independent Researcher, Hong Kong \quad $^{6}$HKUST (GZ), China \\
$^{7}$Monash University Malaysia, Malaysia \quad
$^{8}$Harbin Institute of Technology, China \\
\texttt{yunpeng4@sigsoft.org, zihanwu7-c@my.cityu.edu.hk} \\
\texttt{zyzhuang@link.cuhk.edu.hk, xinzhou.2020@phdcs.smu.edu.sg} \\
\texttt{terryshu512@gmail.com, xhanab@connect.ust.hk} \\
\texttt{chong.chunyong@monash.edu, zhongykau@gmail.com} \\
\texttt{jiakunliu@hit.edu.cn}
}

\iclrfinalcopy 
\begin{document}

\maketitle
\lhead{}
\renewcommand{\headrulewidth}{0pt}

\begin{abstract}
Modern coding agents can deliver increasingly large repository-level changes, and recent benchmarks reflect this by emphasizing long-horizon tasks with large reference implementations. Many benchmarks evaluate coding agents' implementation capability to produce correct code edits from detailed specifications. However, practical modular development tasks also require the perception capability of grounding user intent and high-level design to derive a specification. We introduce \bench to evaluate both capabilities through the entire proposal-to-implementation process on large software systems. It is a multilingual benchmark of 100 tasks across 29 software systems in five domains. Each task provides a human-written enhancement proposal with user intent and high-level design. On average, proposals contain about 5,000 words, software systems contain 2.4 million source lines of code (LoC), and implementation pull requests (PRs) change approximately 5,500 LoC. Across 28 agents we evaluated, the best agent resolves only 14\% of tasks and achieves a 52.7\% Fail-to-Pass (F2P) pass rate. Failure analysis identifies incomplete code localization as a major bottleneck, while providing reference-derived file trees alongside API specifications improves resolved rates by 16--22 percentage points (2.4--17$\times$), reaching at most 34\%. These results show that both perception and implementation remain central challenges for coding agents in practical modular development on large software systems. \bench is available at \url{https://huggingface.co/datasets/lolbench26/LoLBench}.
\end{abstract}

\section{Introduction}\label{sec:intro}

Large language models (LLMs) and coding agents have advanced rapidly in software development. Equipped with tools for planning, exploration, editing, and testing, modern agents can resolve increasingly complex issues and implement large features. Correspondingly, recent benchmarks have expanded evaluation scale through longer specifications and reference solutions~\citep{jimenez2024swe,swebenchpro,deepswe,li2025fea,zhou2026featurebench}. These benchmarks reveal important limitations in agents' ability to carry out large implementations.

However, software development in practice does not begin with detailed specifications. Instead, users and maintainers often begin by expressing the intent and high-level design. Developers must then ground them in existing software systems and decide implementation plans. Many coding benchmarks reduce this uncertainty by providing detailed specifications that contain target interfaces and implementation guidance, or bug reports with local problem scopes. They therefore emphasize a critical but comparatively later stage of software development: producing correct code edits based on clarified specifications or bug reports. This raises a complementary question: \textit{Can coding agents deliver correct implementations based directly on user intent and high-level design?}

This intent-to-implementation process is particularly visible in modular development on large software systems. Projects such as CPython, OpenJDK, Apache Kafka, and Kubernetes use enhancement proposals (EPs) to discuss and govern substantial changes. These proposals express the intent of users and maintainers, including desired functionalities, constraints, and design rationale, together with high-level design decisions. Developers must determine how the proposed design fits into the existing architecture, which modules and interfaces must change, and how to integrate and validate the changes. For example, Python Enhancement Proposal 617~\citep{pep617} proposes replacing CPython's LL(1)-based parser with a PEG-based parser. Realizing this proposal nevertheless requires developers to determine how the new parser should integrate with CPython’s existing components without breaking other functionalities.

Modular development involves two sources of complexity: 1) \textbf{perception complexity}, which captures the difficulty of grounding user intent and high-level design in a large software system to derive a detailed implementation specification. This requires interpreting the proposal, understanding cross-module dependencies, and determining the affected code. 2) \textbf{implementation complexity}, which captures the difficulty of translating that specification into functionally correct and regression-free code changes on an existing large software system instead of from scratch.

\begin{figure*}[t]
    \centering
    \includegraphics[width=0.9\linewidth]{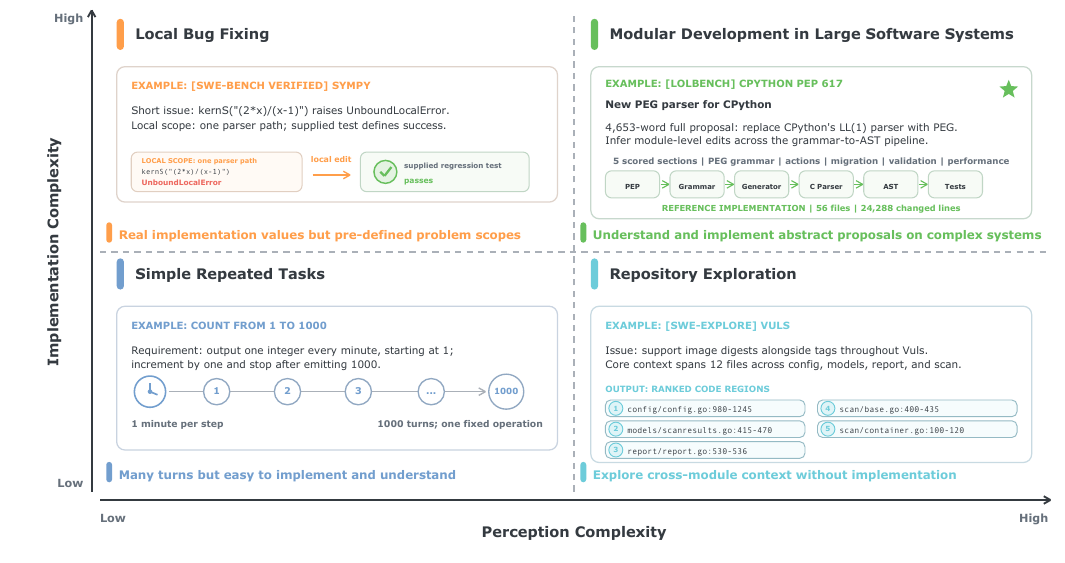}
    \caption{Illustration of tasks with different levels of perception and implementation complexity. Perception complexity captures the difficulty of grounding user intent and high-level design in an existing software system, whereas implementation complexity captures the difficulty of producing correct and regression-free code changes from that understanding.}
    \label{fig:intro}
\end{figure*}

Fig.~\ref{fig:intro} illustrates how these two complexities distinguish different tasks. Simple repeated tasks have low perception and implementation complexity. Although they may require many steps, each step is easy to understand and execute. Repository-exploration tasks have substantial perception complexity with low implementation complexity. For example, SWE-Explore~\citep{zhang2026swe} provides an agent with an issue and a codebase and asks it to identify and rank the relevant code regions under a fixed context budget. Conversely, a local bug-fixing task may require a technically difficult repair but have a clear problem scope, giving it high implementation complexity but low perception complexity. Modular development in large software systems occupies the most challenging region. An agent must first understand an abstract intent in a large software system, then carry that understanding through a large implementation while maintaining compatibility with the rest of the system. These tasks combine high perception complexity with high implementation complexity.

To evaluate current agents in these tasks, we introduce \bench, a multilingual benchmark containing 100 modular development tasks across 29 large software systems in five domains. Each task in \bench includes a human-written enhancement proposal from open-source communities. On average, a proposal contains approximately 5,000 words, the corresponding software system contains 2.4 million source LoC, and its implementation PR changes approximately 5,500 LoC. Agents must interpret the proposal, determine how its intended functionality fits into the existing system, and produce regression-free changes in an offline Docker environment. We evaluate implementations using system-level and end-to-end Fail-to-Pass (F2P) tests through public interfaces, allowing structural divergence from the reference implementation, along with Pass-to-Pass (P2P) regression tests. We further augment the F2P suites through proposal mutation and coverage-guided test generation, and map F2P tests to implementable sections of EPs to enable fine-grained evaluation of partial progress.

We evaluate six powerful LLMs on five scaffolds on \bench. The highest resolved rate is only 14\%, achieved by Opus 5 on Claude Code and mini-SWE-agent. Claude Code with Opus 5 achieves the highest completed rate of 26.3\% and an F2P pass rate of 52.7\%, showing that even the strongest agent implements only a fraction of the proposal. Based on reference solutions from human developers, our failure analysis attributes the largest failures to incomplete code localization. When we additionally give three representative agents reference-derived file trees and API specifications, their resolved rates increase by 16--22 percentage points (2.4--17$\times$), reaching at most 34\%. These results highlight the importance of perception and implementation capability in modular development tasks on large software systems.

We summarize our key contributions as follows:
\begin{itemize}
    \item We introduce modular development tasks on large software systems with two dimensions of complexity: perception complexity, which concerns grounding user intent and high-level design in an existing software system, and implementation complexity, which concerns turning that understanding into verified code changes.
    \item We construct \bench, a multilingual benchmark that consists of 100 modular development tasks from enhancement proposals containing user intent and high-level design, across 29 large software systems in five domains. \bench also includes augmented system-level tests to support flexible implementation and fine-grained evaluation.
    \item We evaluate six frontier LLMs on five popular scaffolds on \bench and demonstrate substantial challenges in both perception and implementation for current coding agents.
\end{itemize}

\section{Related Work}\label{sec:literature}

\subsection{Coding Benchmarks}

Early benchmarks such as HumanEval~\citep{humaneval}, MBPP~\citep{mbpp}, APPS~\citep{apps}, CodeContests~\citep{codecontests}, xCodeEval~\citep{xcodeeval}, CrossCodeEval~\citep{ding2023crosscodeeval} and DevEval~\citep{li2024deveval} evaluate code generation from self-contained problems. NL2Repo-Bench~\citep{nl2repo} extends evaluation to repository generation, but does not require integration into an existing software system. SWE-bench~\citep{jimenez2024swe} introduces real-world repository-level issue resolution, followed by SWE-bench Verified~\citep{swebenchverified}, Multi-SWE-bench~\citep{zan2026multi}, and the longer-horizon SWE-bench Pro~\citep{swebenchpro}. Recent benchmarks further increase implementation complexity through original engineering tasks~\citep{deepswe}, feature implementation~\citep{li2025fea,zhou2026featurebench}, and terminal environments~\citep{merrill2026terminal}. These tasks may require repository exploration, but primarily evaluate the correctness of the resulting implementation. Conversely, SWE-Explore~\citep{zhang2026swe} isolates repository exploration by asking agents to rank relevant code regions without implementing the requested change. \bench complements these benchmarks by jointly evaluating perception and implementation. Its tasks use human-written enhancement proposals that predate their reference implementations. Agents must ground the expressed intent and high-level design in an existing large software system, identify necessary cross-module changes, and produce a functionally correct and regression-free implementation.

\subsection{Coding Agents}

ReAct~\citep{yao2022react} establishes the paradigm of interleaving reasoning with actions in an external environment. Software-engineering agents such as SWE-agent~\citep{yang2024swe}, AutoCodeRover~\citep{zhang2024autocoderover}, live-swe-agent~\citep{xia2025live}, and mini-SWE-agent~\citep{minisweagent} combine repository exploration, editing, testing, and repair. General-purpose scaffolds include OpenHands~\citep{wang2025openhands}, OpenCode~\citep{opencode}, and Pi~\citep{pi}, alongside commercial systems such as Claude Code~\citep{claudecode} and Codex~\citep{codex}. We evaluate representative models and scaffolds on the substantially broader context required by modular development in large software systems.

\section{Benchmark Construction}\label{sec:meth}

\begin{figure*}[t]
    \centering
    \includegraphics[width=0.9\linewidth]{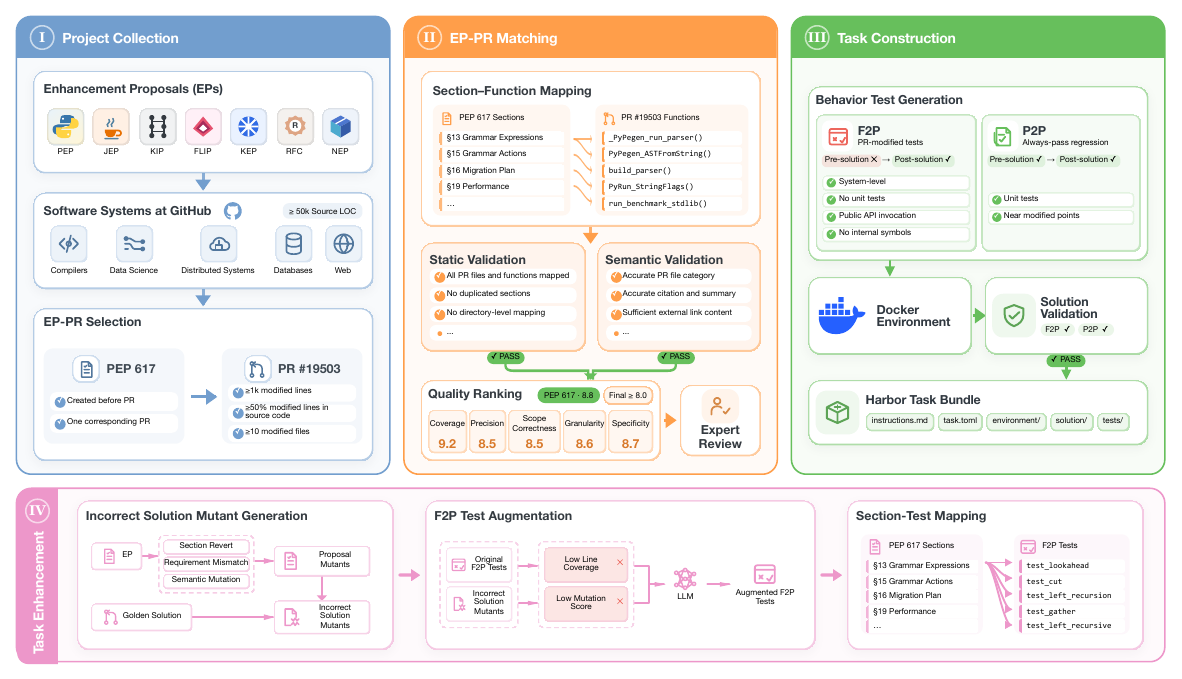}
    \caption{Construction of \bench. We collect enhancement proposals and their corresponding implementations, align proposal sections with implementation changes, construct executable tasks, and augment behavioral tests with incorrect solution mutants and coverage feedback.}
    \label{fig:overview}
\end{figure*}

We implement a four-phase construction process to collect high-quality modular development tasks for \bench, as shown in Fig.~\ref{fig:overview}.

\subsection{Project Collection}

We use \emph{enhancement proposal} (EP) to denote a community-authored artifact that records intended functionality and high-level design before implementation, including formal proposals, RFCs, design documents, and feature requests. For each completed EP, we identify its software system and corresponding implementation PR. We retain an EP--PR pair when: 1) the software has at least 50k source LoC, the EP contains at least one implementable section, and the PR changes at least 1k LoC across at least 10 files, with source code accounting for over half of the changed lines; 2) the EP predates the PR and corresponds to exactly one merged PR, with no competing pending implementation; and 3) the PR contains tests that check the proposed behavior.

\subsection{EP-PR Matching}

We align each EP with its reference implementation to verify that their functional scopes match. We first classify proposal sections as \emph{implementable} when they describe externally observable functionality realized by the PR, or \emph{contextual} when they provide background, rationale or process information. We then map changed files, classes, and functions to the implementable sections they directly implement. We validate each mapping using 28 static rules for structural properties such as completeness, uniqueness, and mapping direction, and eight semantic rules for scope and attribution correctness. An LLM judge subsequently scores each candidate along five dimensions: requirement coverage, mapping precision, scope correctness, granularity, and requirement specificity. We retain candidates with a weighted score of at least 8.0. Appendix~\ref{sec:quality} provides details of this phase.

\subsection{Task Construction}

For each remaining candidate, we build a task for it based on the Harbor framework~\citep{harbor}. Each task provides an agent with the EP and the codebase before implementation. The non-test portion of the corresponding PR forms the reference solution but is hidden from the agent. We evaluate the correctness of solutions based on F2P and P2P tests. An F2P test must fail or error before implementation and pass after applying the reference implementation. To allow semantically equivalent implementations with different structures from the reference, F2P tests check system-level or end-to-end behavior through public interfaces rather than private functions. A P2P test must pass both before and after the reference implementation and serves as a regression guard, and it may range from unit to system level. We package the repository, build dependencies, test runner, and reference solution in an offline Docker environment. We retain a task only when the reference solution applies successfully, all F2P tests pass after the solution, and all P2P tests pass both before and after it. Appendix~\ref{app:task} provides a task example.

\subsection{Task Enhancement}

Because system-level tests are limited in the wild, the original F2P tests may evaluate only part of the functionality in the EP and implementation. We therefore augment each F2P suite along two dimensions: behavioral distinguishability and solution coverage. For behavioral distinguishability, we construct incorrect solution mutants using three proposal-level transformations: \emph{Section Revert}, which removes an implementable capability; \emph{Requirement Mismatch}, which alters a required value, condition, or boundary; and \emph{Semantic Mutation}, which changes the required structure or execution flow. We then generate new F2P tests to kill incorrect solution mutants that the original F2P tests cannot kill. We retain a generated F2P test only if it kills its target mutant and satisfies the same public-behavior restrictions. For solution coverage, we identify tasks whose original F2P suite has line coverage below 50\%, and generate new F2P tests to improve their coverage. Appendix~\ref{app:test-augmentation} details the F2P test augmentation methods. Finally, we map each F2P test to the implementable sections in EPs whose required behavior it asserts, based on the behavior relationships confirmed in the semantic review and the implementation coverage. These mappings enable fine-grained evaluation by supporting section-level assessment of partial progress.

\subsection{Benchmark Statistics}

\begin{table*}[t]
  \centering
  \caption{Comparison with recent coding benchmarks. The last four columns report mean (median) values over tasks. ``Repo Size'' counts source lines excluding comments and blanks. ``Solution Size'' counts inserted and deleted lines across the implementation PR. ``Modified'' counts inserted, deleted, and changed files/classes/functions in solutions. ``Class Mentioned'' is the percentage of target class names that appear in the task description. Lower values indicate less explicit implementation guidance.}
  \label{tab:benchmark-comparison}
  \begingroup
  \small
  \setlength{\tabcolsep}{4pt}
  \renewcommand{\arraystretch}{1.02}
  \resizebox{\linewidth}{!}{%
  \begin{tabular}{@{}lrccrrrr@{}}
    \toprule
    \textbf{Benchmark}
      & \textbf{\#Tasks}
      & \textbf{Lang.}
      & \textbf{Type}
      & \makecell{\textbf{Repo Size}\\\textbf{(k LoC)}}
      & \makecell{\textbf{Solution Size}\\\textbf{(LoC)}}
      & \makecell{\textbf{Modified}\\\textbf{(F/C/Fn)}}
      & \makecell{\textbf{Target Class}\\\textbf{Mentioned (\%)}} \\
    \midrule
    SWE-bench Verified & 500   & Python   & Issue         & 253.6 (277.6) & 38 (25) & 2.6/1.1/6.7 (2/0/5)       & 15.9 (0.0) \\
    SWE-bench Pro      & 731   & Multiple & Issue         & 270.0 (107.6) & 300 (173) & 7.2/2.4/14.5 (5/1/11)     & 13.7 (0.0) \\
    Multi-SWE-bench    & 2,132 & Multiple & Issue         & 205.8 (112.2) & 224 (64) & 6.6/1.8/11.6 (3/0/5)      & 14.0 (0.0) \\
    FEA-Bench          & 1,401 & Python   & Issue         & 174.7 (89.4) & 222 (155) & 5.0/2.8/16.3 (4/2/12)     & 15.5 (0.0) \\
    DeepSWE            & 113   & Multiple & Issue         & 66.7 (29.8) & 1,670 (1,492) & 11.6/11.9/80.7 (8/6/72)   & 31.8 (6.7) \\
    FeatureBench       & 200   & Python   & Specification & 490.6 (460.8) & 2,050 (1,391) & 15.4/11.3/78.3 (10/9/54)  & 38.7 (30.9) \\
    \midrule
    \rowcolor{blue!6}
    \textbf{\textsc{LoLBench}} & \textbf{100} & \textbf{Multiple}
      & \textbf{Proposal} & \textbf{2,393.6 (1,545.1)} & \textbf{5,499 (2,774)}
      & \textbf{64.9/40.1/195.9 (39/21.5/117.5)} & \textbf{7.3 (2.8)} \\
    \bottomrule
  \end{tabular}%
  }
  \endgroup
\end{table*}

We present the comparison between \bench and recent coding benchmarks in Table~\ref{tab:benchmark-comparison}. \bench contains 100 tasks from 29 software systems in five domains and eight programming languages. On average, its proposals contain approximately 5,000 words, its repositories contain 2.4 million source LoC, and its implementation PRs change 5,499 LoC, 65 files, 40 classes, and 196 functions. The repository scale and broad cross-module changes indicate significantly higher implementation complexity compared with other benchmarks. Furthermore, only 7.3\% of target class names appear in the proposals, substantially less than in other benchmarks with large implementations. The limited implementation guidance and large abstract proposals demonstrate the substantial perception complexity of \bench.

\begin{table*}[t]
  \centering
  \caption{Main results of 28 agents on \bench across 100 tasks. ``Resolved'', ``Completed'', ``F2P Pass'', and ``P2P Pass'' report percentages. We average ``turns'', ``time'', ``input tokens'', ``generated tokens'', and ``cost'' over tasks. ``Time'' reports average wall-clock minutes per task. ``Input tokens'' includes cached input and is reported in millions. ``Generated tokens'' sums output and reasoning tokens. All models use their second-highest reasoning effort. Bold and underline denote the best and second-best performance values within each scaffold, respectively. Claude Code with GPT-5.6 Sol and Codex with Opus 5 are unavailable due to provider issues.}
  \label{tab:overall-lolbench}
  \begingroup
  \newcommand{\lbModelLogo}[1]{%
    \raisebox{-0.28ex}{\includegraphics[height=1.22em,keepaspectratio]{figures/#1}}}
  \newcommand{\lbAgentLogo}[1]{%
    \raisebox{-0.32ex}{\includegraphics[height=1.38em,keepaspectratio]{figures/#1}}}
  \newcommand{\lbHarnessCell}[2]{%
    \makecell[c]{\lbAgentLogo{#1}\\[-0.15ex]\textsc{#2}}}
  \newcommand{\lbClaudeOpus}{\lbModelLogo{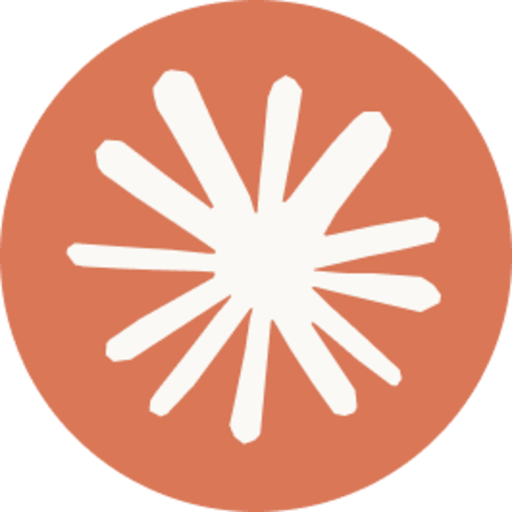}\,Opus 5}
  \newcommand{\lbGPT}{\lbModelLogo{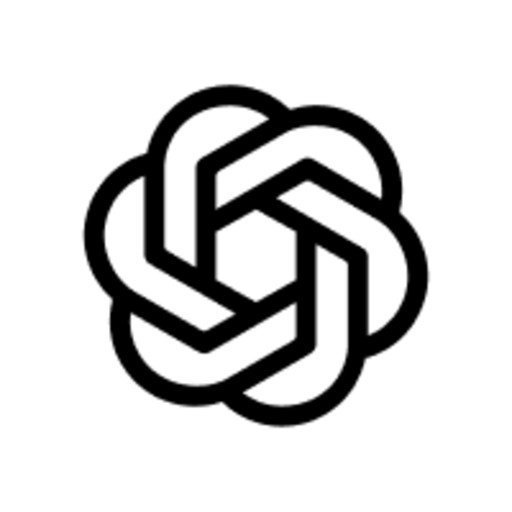}\,GPT-5.6 Sol}
  \newcommand{\lbKimiModel}{\lbModelLogo{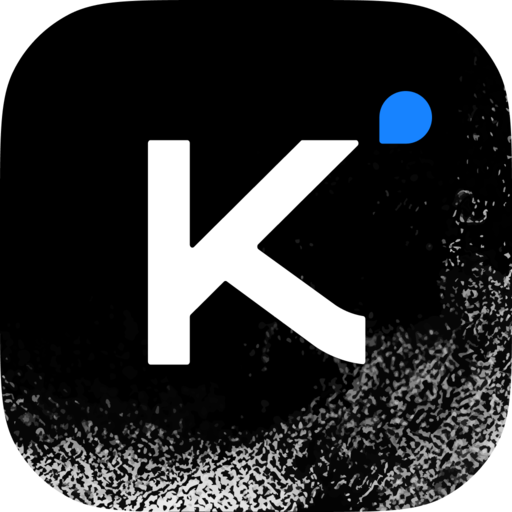}\,Kimi K3}
  \newcommand{\lbGLMModel}{\lbModelLogo{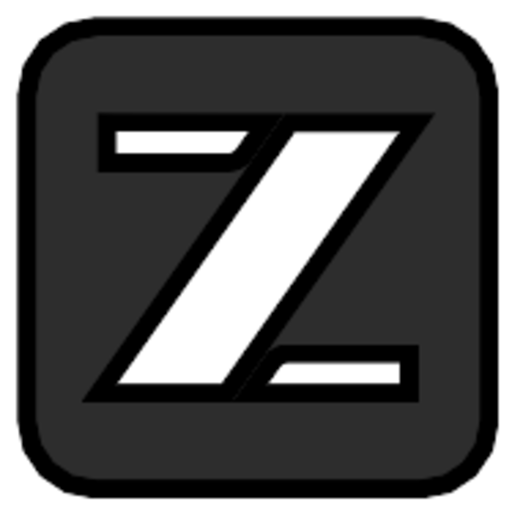}\,GLM-5.2}
  \newcommand{\lbMiniMaxModel}{\lbModelLogo{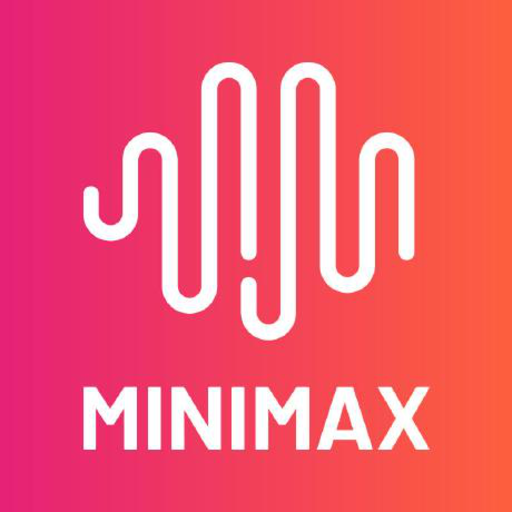}\,MiniMax M3}
  \newcommand{\lbDeepSeekModel}{\lbModelLogo{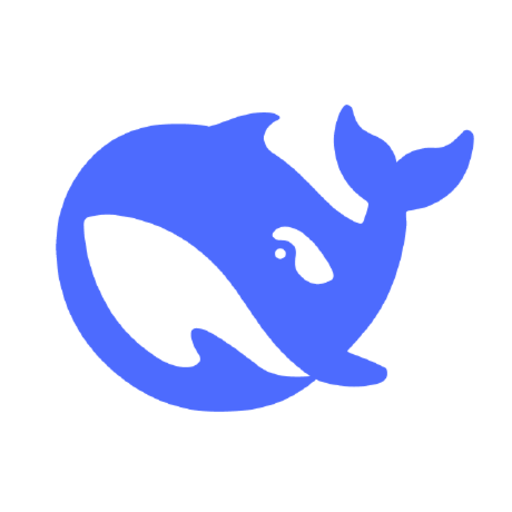}\,DeepSeek V4 Flash}
  \small
  \setlength{\tabcolsep}{3.2pt}
  \renewcommand{\arraystretch}{0.96}
  \scalebox{0.6}{%
  \begin{tabular}{@{}clrrrrrrrrr@{}}
    \toprule
    \textbf{Scaffold}
      & \textbf{Model}
      & \makecell{\textbf{Resolved}\\\textbf{(\%)}}
      & \makecell{\textbf{Completed}\\\textbf{(\%)}}
      & \makecell{\textbf{F2P Pass}\\\textbf{(\%)}}
      & \makecell{\textbf{P2P Pass}\\\textbf{(\%)}}
      & \makecell{\textbf{Avg.}\\\textbf{Turns}}
      & \makecell{\textbf{Avg. Time}\\\textbf{(min)}}
      & \makecell{\textbf{Input}\\\textbf{Tokens (m)}}
      & \makecell{\textbf{Generated}\\\textbf{Tokens (k)}}
      & \makecell{\textbf{Avg.}\\\textbf{Cost (\$)}} \\
    \midrule
    \multirow[c]{5}{*}{\lbHarnessCell{claude.png}{Claude Code}}
      & \lbDeepSeekModel       & 0.0          & 4.8             & 15.0          & 71.8            & 176.9 & 64.4  & 27.6 & 115.6 & 1.4  \\
      & \lbClaudeOpus          & \textbf{14.0} & \textbf{26.3} & \textbf{52.7} & 73.0            & 165.8 & 57.7  & 17.7 & 151.5 & 15.3 \\
      & \lbKimiModel           & \underline{4.0} & \underline{10.1} & \underline{32.5} & \underline{76.1} & 187.9 & 110.7 & 36.3 & 149.9 & 16.9 \\
      & \lbGLMModel            & 2.0          & 4.5             & 20.6          & \textbf{78.3}   & 172.4 & 55.4  & 21.8 & 104.3 & 7.7  \\
      & \lbMiniMaxModel        & 3.0          & 4.0             & 12.2          & 73.5            & 343.4 & 30.2  & 44.6 & 66.0  & 4.2  \\
    \midrule
    \multirow[c]{5}{*}{\lbHarnessCell{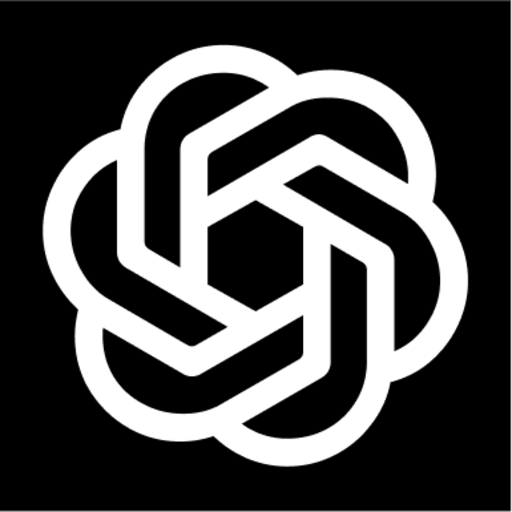}{Codex}}
      & \lbDeepSeekModel        & 0.0          & 3.3             & 18.1          & 72.1            & 197.5 & 56.2 & 22.1 & 125.3 & 0.6  \\
      & \lbKimiModel            & \underline{2.0} & \underline{8.8} & \textbf{32.9} & \textbf{79.3} & 188.3 & 83.5 & 22.2 & 119.3 & 11.5 \\
      & \lbGPT                  & \textbf{3.0} & \textbf{13.6} & 28.6          & 76.0            & 135.2 & 24.9 & 17.5 & 46.4 & 11.7 \\
      & \lbGLMModel             & 1.0          & 4.8             & \underline{32.0} & \underline{78.8} & 204.2 & 48.8 & 22.8 & 136.3 & 6.3  \\
      & \lbMiniMaxModel         & 1.0          & 1.5             & 12.6          & 71.5            & 532.4 & 44.9 & 63.1 & 96.7 & 4.0  \\
    \midrule
    \multirow[c]{6}{*}{\lbHarnessCell{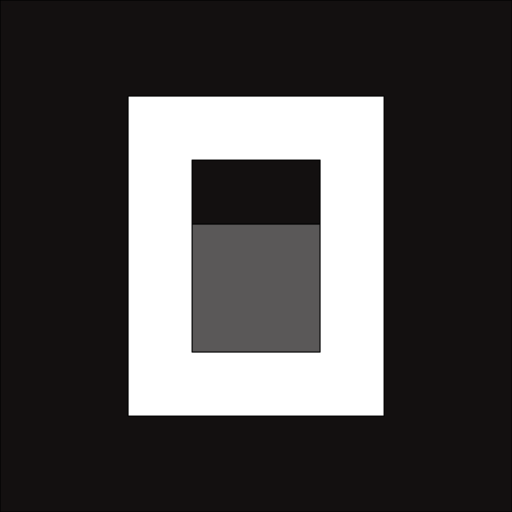}{OpenCode}}
      & \lbDeepSeekModel        & 1.0          & 4.0             & 18.3          & 73.6            & 177.2 & 44.4 & 24.1 & 95.8  & 0.5  \\
      & \lbClaudeOpus           & \textbf{10.0} & \textbf{16.2} & \textbf{37.6} & 65.8            & 148.2 & 77.6 & 30.3 & 140.0 & 18.6 \\
      & \lbKimiModel            & 3.0          & \underline{8.3} & \underline{32.1} & \underline{76.6} & 171.4 & 82.6 & 29.7 & 114.3 & 13.8 \\
      & \lbGPT                  & \underline{5.0} & \underline{8.3} & 15.8          & 43.4            & 53.0  & 82.7 & 8.7  & 28.3  & 5.2  \\
      & \lbGLMModel             & 2.0          & 5.1             & 18.2          & \textbf{80.1}   & 193.4 & 54.9 & 22.7 & 80.2  & 6.7  \\
      & \lbMiniMaxModel         & 0.0          & 1.8             & 12.0          & 72.9            & 373.0 & 37.1 & 54.9 & 71.8  & 3.5  \\
    \midrule
    \multirow[c]{6}{*}{\lbHarnessCell{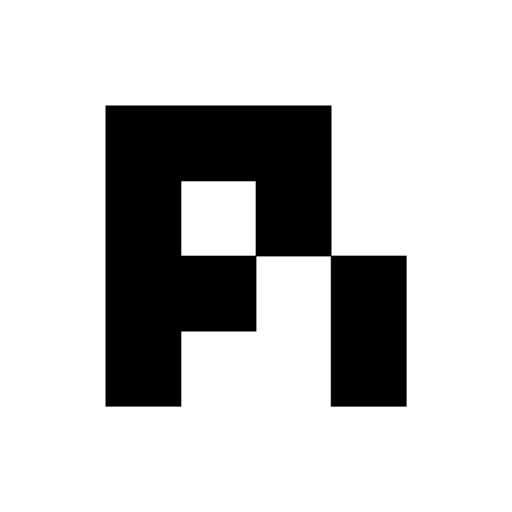}{Pi}}
      & \lbDeepSeekModel        & 1.0          & 3.3             & 28.0          & 73.1            & 182.7 & 51.1  & 23.9 & 105.5 & 0.5  \\
      & \lbClaudeOpus           & \textbf{13.0} & \textbf{21.0} & \textbf{44.5} & \underline{80.2} & 204.4 & 186.0 & 34.1 & 714.0 & 39.7 \\
      & \lbKimiModel            & 1.0          & \underline{8.6} & \underline{33.5} & \textbf{81.1}   & 175.8 & 94.0  & 29.3 & 124.8 & 13.9 \\
      & \lbGPT                  & \underline{3.0} & 7.1             & 22.9          & \underline{80.2} & 72.4  & 18.4  & 5.7  & 26.8  & 4.4  \\
      & \lbGLMModel             & 2.0          & 5.6             & 17.0          & 69.9            & 218.6 & 52.2  & 26.6 & 90.2  & 9.3  \\
      & \lbMiniMaxModel         & 0.0          & 3.0             & 13.7          & 69.9            & 415.5 & 56.6  & 51.3 & 76.5  & 3.3  \\
    \midrule
    \multirow[c]{6}{*}{\lbHarnessCell{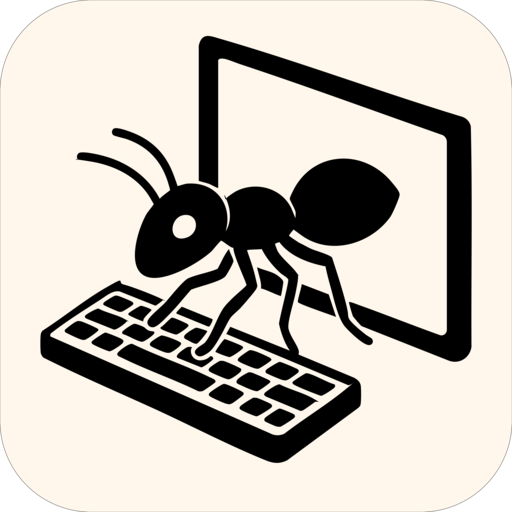}{mini-SWE-agent}}
      & \lbDeepSeekModel        & 0.0          & 3.5             & 14.9          & 69.7            & 212.9 & 57.3  & 27.4 & 105.7 & 0.6  \\
      & \lbClaudeOpus           & \textbf{14.0} & \textbf{20.5} & \textbf{44.7} & 73.9            & 211.1 & 185.5 & 40.7 & 752.0 & 43.6 \\
      & \lbKimiModel            & \underline{5.0} & \underline{11.1} & \underline{34.0} & \textbf{80.3} & 240.0 & 108.4 & 51.5 & 178.1 & 22.4 \\
      & \lbGPT                  & \underline{5.0} & 9.1             & 29.4          & \underline{78.0} & 72.4  & 43.8  & 10.0 & 100.7 & 9.3  \\
      & \lbGLMModel             & 0.0          & 2.5             & 16.9          & 73.9            & 246.3 & 40.7  & 27.8 & 79.8  & 9.0  \\
      & \lbMiniMaxModel         & 1.0          & 2.3             & 11.1          & 70.9            & 473.5 & 54.9  & 66.0 & 84.1  & 4.2  \\
    \bottomrule
  \end{tabular}
  }
  \endgroup
\end{table*}

\section{Evaluation}\label{sec:eval}

\subsection{Experiment Setup}

\textbf{Models and scaffolds.} We evaluate six recent models: DeepSeek V4 Flash~\citep{deepseekv4}, Opus 5~\citep{opus5}, Kimi K3~\citep{kimik3}, GPT-5.6 Sol~\citep{gpt56}, GLM-5.2~\citep{glm52}, and MiniMax M3~\citep{minimaxm3}. We access all models through OpenRouter~\citep{openrouter} and use the second-highest reasoning effort for each model. We pair them with five scaffolds: Claude Code~\citep{claudecode}, Codex~\citep{codex}, OpenCode~\citep{opencode}, Pi~\citep{pi}, and mini-SWE-agent~\citep{minisweagent}. Codex with Opus 5 and Claude Code with GPT-5.6 Sol are unavailable because of provider issues, leaving 28 agents.

\textbf{Environment.} We implement \bench with Harbor~\citep{harbor}, which executes each task in a sandbox. We disable public network access to prevent cheating. Each attempt has a six-hour wall-clock limit and is terminated after one hour without observable progress. We give each agent one attempt at each task, except when provider or infrastructure failures require a retry.

\textbf{Metrics.} \emph{Resolved} is the percentage of tasks accepted by the full task verifier. \emph{Completed} is the percentage of all completed implementable sections over all implementable sections: a section is completed only when all F2P tests mapped to it pass. \emph{F2P Pass} and \emph{P2P Pass} are micro-averaged over all tests. We also report mean turns, wall-clock time, input tokens, generated tokens, and cost.

\subsection{Effectiveness Analysis}

\textbf{Modular development remains challenging.} Table~\ref{tab:overall-lolbench} reports results for all 28 agents. The best agents, Claude Code and mini-SWE-agent with Opus 5, resolve only 14\% of the tasks. Across agents, the mean and median resolved rates are 3.4\% and 2.0\%. Fine-grained progress is also limited: Claude Code with Opus 5 attains the highest completed rate of 26.3\%, whereas the mean and median are 8.0\% and 5.3\%. Thus, the low resolved rates are accompanied by limited section-level progress, rather than arising only from the strict all-or-nothing resolved rates.

\textbf{Model choice can substantially affect effectiveness.} We observe that Opus 5 achieves the highest resolved, completed, and F2P pass rates under every scaffold that supports it. We further analyze the four models available with all scaffolds and measure performance variability using relative standard deviation (RSD). Holding the scaffold fixed, the RSD across models averages 62.0\% for Completed rate and 44.5\% for F2P Pass rate. Holding the model fixed, the RSD across scaffolds averages 24.0\% and 17.3\%, respectively. Relative variability across models is about 2.6 times that across scaffolds for both metrics, indicating that agent effectiveness is highly sensitive to model choice.

\subsection{Efficiency Analysis}

\begin{figure*}[t]
    \centering
    \includegraphics[width=0.57\linewidth]{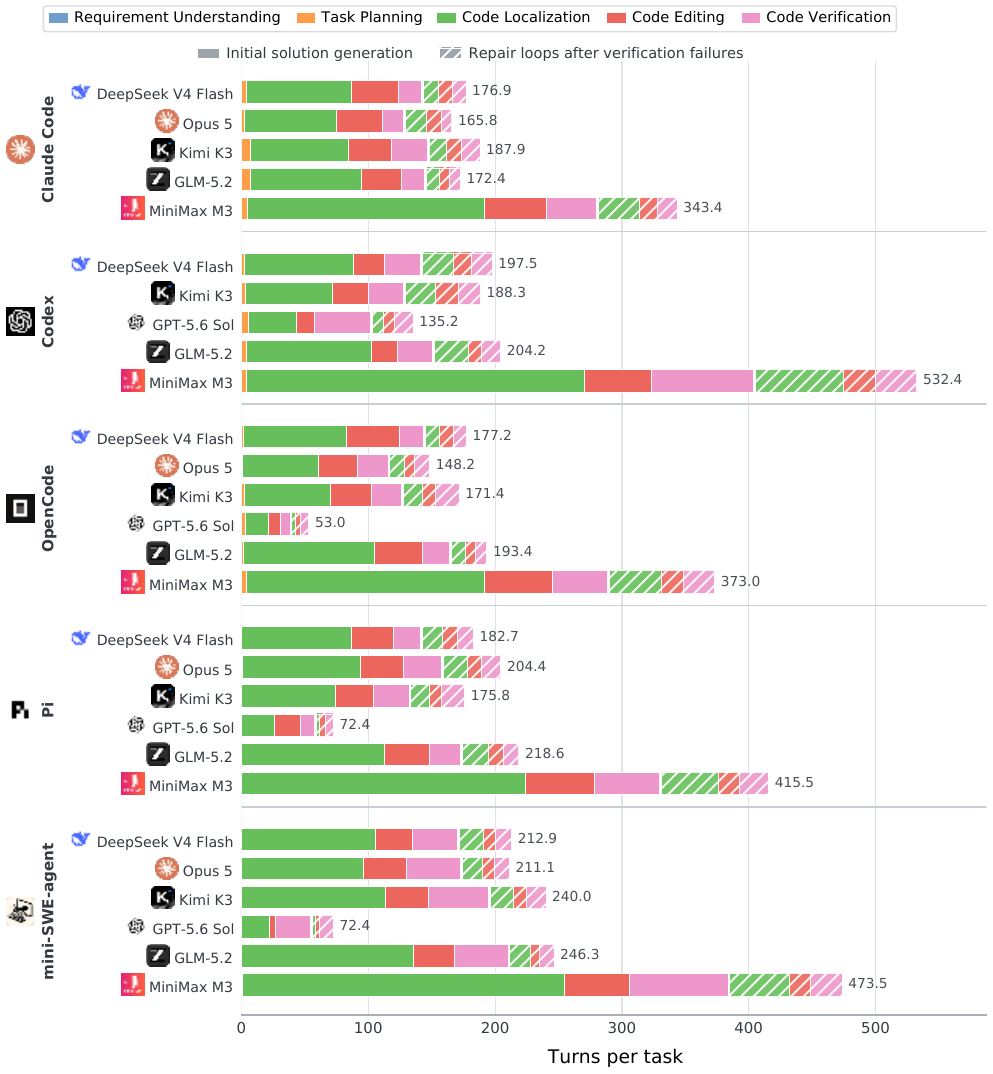}
    \caption{Average agent turns per task across five phases. Solid segments denote initial solution generation, and hatched segments denote repair loops after failed verification.}
    \label{fig:phase}
\end{figure*}

\textbf{Scaffold choice can substantially affect efficiency.} 
With Opus 5, both Claude Code and mini-SWE-agent resolve 14 tasks, but mini-SWE-agent uses 3.2$\times$ the wall-clock time, 5.0$\times$ the generated tokens, and 2.8$\times$ the cost of Claude Code. Comparing the three open-source scaffolds, we find that OpenCode uses fewer turns, generated tokens, and less time than Pi and mini-SWE-agent.

\textbf{Localization dominates agent activities.} We assign agent turns to requirement understanding, task planning, code localization, code editing, or code verification, and we record repair-loop membership separately. As shown in Fig.~\ref{fig:phase}, code localization is the largest phase for 26 of 28 agents during initial solution generation and for 21 during repair. Aggregated over the five phases, it accounts for 60.4\% of initial-generation turns and 44.2\% of repair turns. These results identify repository exploration and localization as the dominant activity for current agents in modular development tasks with high perception complexity. Appendix~\ref{app:effi} adds analysis for phase distribution on token usage.

\subsection{Failure Analysis}

\begin{table*}[t]
  \centering
  \caption{The heat map of failure attribution across the 28 agents on \bench. ``Agent Failure'' indicates failed tasks without analyzable trajectories, and ``Solution Failure'' indicates failed tasks with analyzable trajectories. The 49 agent failures for OpenCode with GPT-5.6 Sol are primarily due to timeouts before submission. The last seven columns show failure attribution across seven phases, and they sum to the ``Solution Failure'' column. Darker backgrounds indicate a larger share of total failure attribution for each agent's solution failures.}
  \label{tab:failure-composition}
  \begingroup
  \definecolor{lbFailureHeat}{HTML}{729ECE}
  \newcommand{\lbModelLogo}[1]{%
    \raisebox{-0.28ex}{\includegraphics[height=1.22em,keepaspectratio]{figures/#1}}}
  \newcommand{\lbAgentLogo}[1]{%
    \raisebox{-0.32ex}{\includegraphics[height=1.38em,keepaspectratio]{figures/#1}}}
  \newcommand{\lbHarnessCell}[2]{%
    \makecell[c]{\lbAgentLogo{#1}\\[-0.15ex]\textsc{#2}}}
  \newcommand{\lbClaudeOpus}{\lbModelLogo{claude.png}\,Opus 5}
  \newcommand{\lbGPT}{\lbModelLogo{openai.png}\,GPT-5.6 Sol}
  \newcommand{\lbKimiModel}{\lbModelLogo{kimi.png}\,Kimi K3}
  \newcommand{\lbGLMModel}{\lbModelLogo{glm.png}\,GLM-5.2}
  \newcommand{\lbMiniMaxModel}{\lbModelLogo{minimax.png}\,MiniMax M3}
  \newcommand{\lbDeepSeekModel}{\lbModelLogo{deepseek.png}\,DeepSeek V4 Flash}
  \small
  \setlength{\tabcolsep}{3.0pt}
  \renewcommand{\arraystretch}{0.96}
  \scalebox{0.6}{%
  \begin{tabular}{@{}clrrrrrrrrr@{}}
    \toprule
    \textbf{Scaffold}
      & \textbf{Model}
      & \makecell{\textbf{Agent}\\\textbf{Failure}}
      & \makecell{\textbf{Solution}\\\textbf{Failure}}
      & \makecell{\textbf{Requirement}\\\textbf{Understanding}}
      & \makecell{\textbf{Task}\\\textbf{Planning}}
      & \makecell{\textbf{Code}\\\textbf{Localization}}
      & \makecell{\textbf{Code}\\\textbf{Editing}}
      & \makecell{\textbf{Code}\\\textbf{Verification}}
      & \textbf{Self-Repair}
      & \textbf{Tool Use} \\
    \midrule
      \multirow[c]{5}{*}{\lbHarnessCell{claude.png}{Claude Code}} & \lbDeepSeekModel & 0 & 100 & \cellcolor{lbFailureHeat!16}11.2 & \cellcolor{lbFailureHeat!15}10.5 & \cellcolor{lbFailureHeat!40}27.8 & \cellcolor{lbFailureHeat!37}25.6 & \cellcolor{lbFailureHeat!20}13.8 & \cellcolor{lbFailureHeat!16}11.0 & \cellcolor{lbFailureHeat!1}0.1 \\
       & \lbClaudeOpus & 6 & 80 & \cellcolor{lbFailureHeat!12}6.9 & \cellcolor{lbFailureHeat!11}6.2 & \cellcolor{lbFailureHeat!45}25.1 & \cellcolor{lbFailureHeat!32}18.1 & \cellcolor{lbFailureHeat!25}13.8 & \cellcolor{lbFailureHeat!18}9.9 & \cellcolor{lbFailureHeat!0}0.0 \\
       & \lbKimiModel & 5 & 91 & \cellcolor{lbFailureHeat!8}5.1 & \cellcolor{lbFailureHeat!7}4.4 & \cellcolor{lbFailureHeat!41}26.4 & \cellcolor{lbFailureHeat!41}25.9 & \cellcolor{lbFailureHeat!15}9.8 & \cellcolor{lbFailureHeat!30}18.8 & \cellcolor{lbFailureHeat!1}0.6 \\
       & \lbGLMModel & 0 & 98 & \cellcolor{lbFailureHeat!12}8.2 & \cellcolor{lbFailureHeat!11}7.7 & \cellcolor{lbFailureHeat!38}25.9 & \cellcolor{lbFailureHeat!46}31.7 & \cellcolor{lbFailureHeat!19}12.9 & \cellcolor{lbFailureHeat!17}11.6 & \cellcolor{lbFailureHeat!0}0.0 \\
       & \lbMiniMaxModel & 1 & 96 & \cellcolor{lbFailureHeat!14}9.4 & \cellcolor{lbFailureHeat!14}9.2 & \cellcolor{lbFailureHeat!36}24.2 & \cellcolor{lbFailureHeat!44}29.3 & \cellcolor{lbFailureHeat!18}12.2 & \cellcolor{lbFailureHeat!17}11.5 & \cellcolor{lbFailureHeat!1}0.2 \\
    \midrule
      \multirow[c]{5}{*}{\lbHarnessCell{codex.png}{Codex}} & \lbDeepSeekModel & 0 & 100 & \cellcolor{lbFailureHeat!6}4.5 & \cellcolor{lbFailureHeat!6}3.9 & \cellcolor{lbFailureHeat!57}39.9 & \cellcolor{lbFailureHeat!35}24.6 & \cellcolor{lbFailureHeat!26}18.3 & \cellcolor{lbFailureHeat!12}8.5 & \cellcolor{lbFailureHeat!1}0.3 \\
       & \lbKimiModel & 0 & 98 & \cellcolor{lbFailureHeat!3}2.1 & \cellcolor{lbFailureHeat!2}1.7 & \cellcolor{lbFailureHeat!61}41.7 & \cellcolor{lbFailureHeat!35}23.8 & \cellcolor{lbFailureHeat!24}16.8 & \cellcolor{lbFailureHeat!17}11.4 & \cellcolor{lbFailureHeat!1}0.5 \\
       & \lbGPT & 1 & 96 & \cellcolor{lbFailureHeat!2}1.2 & \cellcolor{lbFailureHeat!1}1.0 & \cellcolor{lbFailureHeat!53}35.5 & \cellcolor{lbFailureHeat!30}20.5 & \cellcolor{lbFailureHeat!30}19.8 & \cellcolor{lbFailureHeat!27}18.0 & \cellcolor{lbFailureHeat!0}0.0 \\
       & \lbGLMModel & 0 & 99 & \cellcolor{lbFailureHeat!5}3.3 & \cellcolor{lbFailureHeat!4}2.9 & \cellcolor{lbFailureHeat!61}42.1 & \cellcolor{lbFailureHeat!36}25.2 & \cellcolor{lbFailureHeat!24}16.3 & \cellcolor{lbFailureHeat!13}9.2 & \cellcolor{lbFailureHeat!0}0.0 \\
       & \lbMiniMaxModel & 0 & 99 & \cellcolor{lbFailureHeat!9}6.3 & \cellcolor{lbFailureHeat!9}6.4 & \cellcolor{lbFailureHeat!52}36.1 & \cellcolor{lbFailureHeat!34}23.5 & \cellcolor{lbFailureHeat!23}15.8 & \cellcolor{lbFailureHeat!14}9.4 & \cellcolor{lbFailureHeat!2}1.5 \\
    \midrule
      \multirow[c]{6}{*}{\lbHarnessCell{opencode.png}{OpenCode}} & \lbDeepSeekModel & 1 & 98 & \cellcolor{lbFailureHeat!12}8.2 & \cellcolor{lbFailureHeat!12}8.5 & \cellcolor{lbFailureHeat!37}25.6 & \cellcolor{lbFailureHeat!45}30.7 & \cellcolor{lbFailureHeat!19}12.8 & \cellcolor{lbFailureHeat!17}11.8 & \cellcolor{lbFailureHeat!1}0.4 \\
       & \lbClaudeOpus & 11 & 79 & \cellcolor{lbFailureHeat!10}5.7 & \cellcolor{lbFailureHeat!9}5.1 & \cellcolor{lbFailureHeat!49}27.1 & \cellcolor{lbFailureHeat!31}16.9 & \cellcolor{lbFailureHeat!17}9.5 & \cellcolor{lbFailureHeat!27}14.7 & \cellcolor{lbFailureHeat!0}0.0 \\
       & \lbKimiModel & 1 & 96 & \cellcolor{lbFailureHeat!8}5.6 & \cellcolor{lbFailureHeat!9}6.1 & \cellcolor{lbFailureHeat!39}26.5 & \cellcolor{lbFailureHeat!38}25.5 & \cellcolor{lbFailureHeat!20}13.7 & \cellcolor{lbFailureHeat!27}18.1 & \cellcolor{lbFailureHeat!1}0.5 \\
       & \lbGPT & 49 & 46 & \cellcolor{lbFailureHeat!3}0.8 & \cellcolor{lbFailureHeat!2}0.6 & \cellcolor{lbFailureHeat!41}13.2 & \cellcolor{lbFailureHeat!57}18.4 & \cellcolor{lbFailureHeat!20}6.5 & \cellcolor{lbFailureHeat!20}6.5 & \cellcolor{lbFailureHeat!0}0.0 \\
       & \lbGLMModel & 0 & 98 & \cellcolor{lbFailureHeat!11}7.8 & \cellcolor{lbFailureHeat!11}7.5 & \cellcolor{lbFailureHeat!39}26.5 & \cellcolor{lbFailureHeat!47}32.2 & \cellcolor{lbFailureHeat!18}12.6 & \cellcolor{lbFailureHeat!17}11.4 & \cellcolor{lbFailureHeat!0}0.0 \\
       & \lbMiniMaxModel & 1 & 99 & \cellcolor{lbFailureHeat!12}8.3 & \cellcolor{lbFailureHeat!11}7.8 & \cellcolor{lbFailureHeat!37}25.6 & \cellcolor{lbFailureHeat!48}33.0 & \cellcolor{lbFailureHeat!16}11.2 & \cellcolor{lbFailureHeat!19}13.1 & \cellcolor{lbFailureHeat!0}0.0 \\
    \midrule
      \multirow[c]{6}{*}{\lbHarnessCell{pi.png}{Pi}} & \lbDeepSeekModel & 3 & 96 & \cellcolor{lbFailureHeat!15}10.2 & \cellcolor{lbFailureHeat!15}10.4 & \cellcolor{lbFailureHeat!48}32.6 & \cellcolor{lbFailureHeat!36}24.1 & \cellcolor{lbFailureHeat!16}10.5 & \cellcolor{lbFailureHeat!11}7.6 & \cellcolor{lbFailureHeat!1}0.6 \\
       & \lbClaudeOpus & 7 & 80 & \cellcolor{lbFailureHeat!13}7.1 & \cellcolor{lbFailureHeat!12}6.5 & \cellcolor{lbFailureHeat!55}30.7 & \cellcolor{lbFailureHeat!31}17.6 & \cellcolor{lbFailureHeat!15}8.5 & \cellcolor{lbFailureHeat!15}8.2 & \cellcolor{lbFailureHeat!3}1.4 \\
       & \lbKimiModel & 1 & 98 & \cellcolor{lbFailureHeat!6}4.2 & \cellcolor{lbFailureHeat!6}4.3 & \cellcolor{lbFailureHeat!56}38.4 & \cellcolor{lbFailureHeat!42}28.7 & \cellcolor{lbFailureHeat!16}10.7 & \cellcolor{lbFailureHeat!16}11.1 & \cellcolor{lbFailureHeat!1}0.6 \\
       & \lbGPT & 2 & 95 & \cellcolor{lbFailureHeat!2}1.4 & \cellcolor{lbFailureHeat!1}0.7 & \cellcolor{lbFailureHeat!44}29.5 & \cellcolor{lbFailureHeat!60}39.9 & \cellcolor{lbFailureHeat!18}11.7 & \cellcolor{lbFailureHeat!18}11.7 & \cellcolor{lbFailureHeat!1}0.1 \\
       & \lbGLMModel & 1 & 97 & \cellcolor{lbFailureHeat!15}10.4 & \cellcolor{lbFailureHeat!15}10.1 & \cellcolor{lbFailureHeat!40}27.3 & \cellcolor{lbFailureHeat!44}29.9 & \cellcolor{lbFailureHeat!16}10.7 & \cellcolor{lbFailureHeat!12}8.2 & \cellcolor{lbFailureHeat!1}0.4 \\
       & \lbMiniMaxModel & 1 & 99 & \cellcolor{lbFailureHeat!15}10.4 & \cellcolor{lbFailureHeat!14}9.8 & \cellcolor{lbFailureHeat!35}24.5 & \cellcolor{lbFailureHeat!49}34.2 & \cellcolor{lbFailureHeat!12}8.5 & \cellcolor{lbFailureHeat!15}10.4 & \cellcolor{lbFailureHeat!2}1.2 \\
    \midrule
      \multirow[c]{6}{*}{\lbHarnessCell{mini-swe-agent.png}{mini-SWE-agent}} & \lbDeepSeekModel & 1 & 99 & \cellcolor{lbFailureHeat!4}3.0 & \cellcolor{lbFailureHeat!2}1.7 & \cellcolor{lbFailureHeat!56}38.6 & \cellcolor{lbFailureHeat!37}25.7 & \cellcolor{lbFailureHeat!36}24.9 & \cellcolor{lbFailureHeat!7}5.1 & \cellcolor{lbFailureHeat!0}0.0 \\
       & \lbClaudeOpus & 12 & 74 & \cellcolor{lbFailureHeat!1}0.6 & \cellcolor{lbFailureHeat!1}0.6 & \cellcolor{lbFailureHeat!59}30.6 & \cellcolor{lbFailureHeat!27}14.2 & \cellcolor{lbFailureHeat!32}16.6 & \cellcolor{lbFailureHeat!21}10.8 & \cellcolor{lbFailureHeat!1}0.6 \\
       & \lbKimiModel & 1 & 94 & \cellcolor{lbFailureHeat!2}1.1 & \cellcolor{lbFailureHeat!2}1.1 & \cellcolor{lbFailureHeat!57}37.5 & \cellcolor{lbFailureHeat!28}18.5 & \cellcolor{lbFailureHeat!35}23.0 & \cellcolor{lbFailureHeat!18}11.7 & \cellcolor{lbFailureHeat!2}1.1 \\
       & \lbGPT & 0 & 95 & \cellcolor{lbFailureHeat!2}1.5 & \cellcolor{lbFailureHeat!2}1.5 & \cellcolor{lbFailureHeat!54}35.9 & \cellcolor{lbFailureHeat!36}24.0 & \cellcolor{lbFailureHeat!21}13.7 & \cellcolor{lbFailureHeat!27}18.2 & \cellcolor{lbFailureHeat!1}0.2 \\
       & \lbGLMModel & 1 & 99 & \cellcolor{lbFailureHeat!6}4.1 & \cellcolor{lbFailureHeat!5}3.2 & \cellcolor{lbFailureHeat!56}39.0 & \cellcolor{lbFailureHeat!34}23.4 & \cellcolor{lbFailureHeat!29}20.3 & \cellcolor{lbFailureHeat!13}8.8 & \cellcolor{lbFailureHeat!1}0.2 \\
       & \lbMiniMaxModel & 0 & 99 & \cellcolor{lbFailureHeat!11}7.6 & \cellcolor{lbFailureHeat!10}7.1 & \cellcolor{lbFailureHeat!45}31.4 & \cellcolor{lbFailureHeat!34}23.6 & \cellcolor{lbFailureHeat!25}17.0 & \cellcolor{lbFailureHeat!16}11.2 & \cellcolor{lbFailureHeat!2}1.1 \\
    \bottomrule
  \end{tabular}
  }
  \endgroup
\end{table*}

\textbf{Method.} 
We perform a failure analysis on the trajectories of failed tasks for all 28 agents. For each failed task, we attribute its failure to several failure modes in different phases mentioned in Fig.~\ref{fig:phase}. We add self-repair and tool use as extra phases to separate failures that occur outside initial solution generation. We use pre-defined rules to detect observable localization, editing, verification, repair, and tool-use signals, and an outcome-blinded LLM judge to detect semantic requirement understanding and planning issues. For solution failures with multiple modes, we normalize confidence scores, so each solution failure contributes one unit of attribution. Table~\ref{tab:failure-composition} reports phase-level attribution, and Appendix~\ref{app:failure} provides the taxonomy and per-phase failure mode results.

\textbf{Incomplete cross-module context is the dominant failure mode.} Averaged over the 28 agents, code localization accounts for the largest 30.9 units of failure attribution, followed by code editing at 25.3 units. These values are fractional attribution weights aggregated across solution failures. Table~\ref{tab:failure-code-localization} further shows that nearly all localization attribution, averaging 30.0 units, comes from \texttt{cross\_module\_context\_missing}: the submitted solution reaches some target files and modules but also misses some. The phase attribution also varies with the scaffold. Among the four models shared by all scaffolds, Codex assigns 11.8--16.2 more attribution units to code localization than Claude Code, while mini-SWE-agent assigns 7.2--13.1 more. Code localization is the largest phase for 19 agents, while editing is the largest for the remaining nine. This shows that modular development tasks on large software systems require comprehensive perception and implementation capabilities.

\subsection{Perception Capability Analysis}

\begin{figure*}[t]
  \centering
  \includegraphics[width=0.7\textwidth]{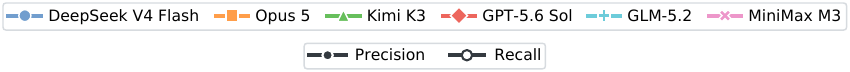}
  \vspace{-0.4em}

  \begin{subfigure}[t]{0.24\textwidth}
    \centering
    \includegraphics[width=\linewidth,height=0.72\linewidth,keepaspectratio]{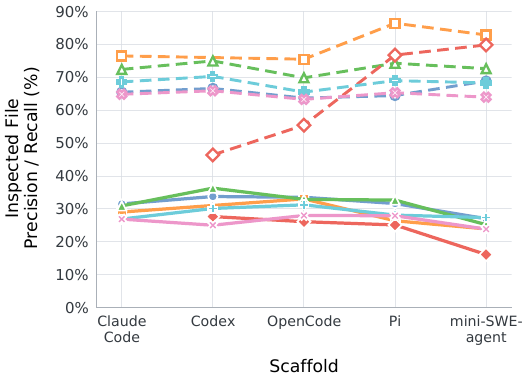}
    \caption{Inspected files.}
    \label{fig:localization-inspected-file}
  \end{subfigure}\hfill
  \begin{subfigure}[t]{0.24\textwidth}
    \centering
    \includegraphics[width=\linewidth,height=0.72\linewidth,keepaspectratio]{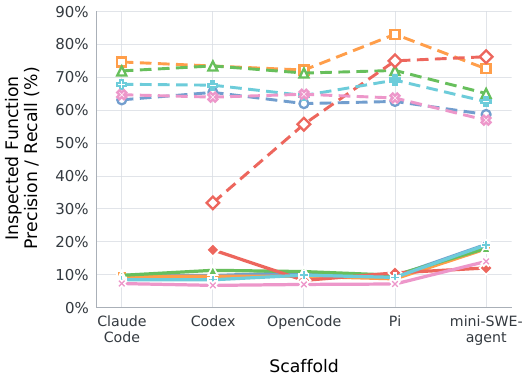}
    \caption{Inspected functions.}
    \label{fig:localization-inspected-function}
  \end{subfigure}\hfill
  \begin{subfigure}[t]{0.24\textwidth}
    \centering
    \includegraphics[width=\linewidth,height=0.72\linewidth,keepaspectratio]{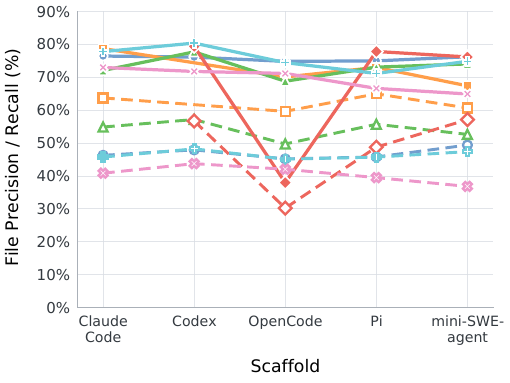}
    \caption{Modified files.}
    \label{fig:localization-file}
  \end{subfigure}\hfill
  \begin{subfigure}[t]{0.24\textwidth}
    \centering
    \includegraphics[width=\linewidth,height=0.72\linewidth,keepaspectratio]{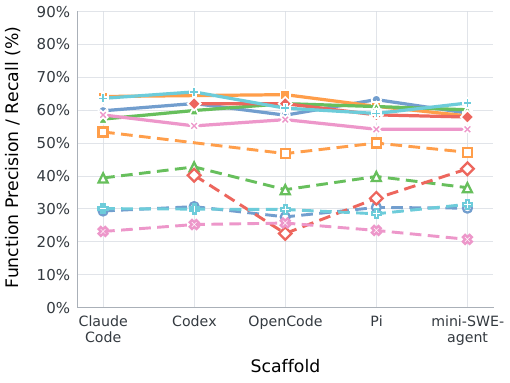}
    \caption{Modified functions.}
    \label{fig:localization-function}
  \end{subfigure}

  \caption{Macro-averaged precision and recall for code localization based on inspected (left) and modified (right) files and functions.}
  \label{fig:localization}
\end{figure*}

To further evaluate the perception capability of current agents in modular development tasks, we compare the source files and functions each agent inspects or modifies with those modified by the reference implementation. We exclude newly added entities because they have no pre-existing location to recover. Figure~\ref{fig:localization} reports task-level macro averages.

\textbf{Current agents struggle to pinpoint all related files and functions.} Figure~\ref{fig:localization} shows incomplete code localization coverage at both the inspection and modification stages. Across all 28 agents, inspection recall is 46.4--86.4\% for files and 31.9--83.1\% for functions, indicating that some reference edit locations remain uninspected. Inspection precision is only 16.1--36.3\% for files and 6.8--19.2\% for functions, reflecting broad exploration beyond the entities modified by the reference implementation. Agents select some reference edit locations more precisely when making changes: across the 27 agents other than OpenCode with GPT-5.6 Sol, modification precision is 64.9--80.4\% for files and 54.2--65.6\% for functions. However, modification recall remains limited to 36.9--65.0\% for files and 20.8--53.4\% for functions. These results show that broad repository exploration does not ensure comprehensive coverage of edit locations when implementing abstract EPs.

\textbf{Current agents perform substantially better on specifications than on abstract proposals.} We follow the instruction format in NL2Repo-Bench~\citep{nl2repo} and reformulate EPs as specifications by appending a file tree and API specification derived from the reference solution. We evaluate three selected agents on the specifications and report the paired results in Table~\ref{tab:api-lolbench}. Compared with EPs, the specifications increase resolved rate by 16--22 percentage points, completed rate by 13.7--19.2 points, and F2P pass rate by 13.3--27.6 points. For Claude Code with Opus 5, F2P pass rate rises from 52.7\% to 72.7\%. These paired improvements show that explicit implementation guidance in specifications can remove much of the difficulty, while agents need perception capability to convert abstract EPs into detailed specifications.

\begin{table*}[t]
  \centering
  \caption{Results of three agents on \bench with file trees and API specification in EPs. All columns are calculated the same way in Table~\ref{tab:overall-lolbench}. Parenthesized values report signed absolute changes computed by subtracting the baseline values in Table~\ref{tab:overall-lolbench} from the values here.}
  \label{tab:api-lolbench}
  \begingroup
  \newcommand{\lbModelLogo}[1]{%
    \raisebox{-0.28ex}{\includegraphics[height=1.22em,keepaspectratio]{figures/#1}}}
  \newcommand{\lbAgentLogo}[1]{%
    \raisebox{-0.32ex}{\includegraphics[height=1.38em,keepaspectratio]{figures/#1}}}
  \newcommand{\lbModelHarness}[4]{%
    \lbModelLogo{#1}\,#2\;{+}\;\lbAgentLogo{#3}\,\textsc{#4}}
  \newcommand{\lbClaudeCodeOpus}{%
    \lbModelHarness{claude.png}{Opus 5}{claude.png}{Claude Code}}
  \newcommand{\lbCodexGPT}{%
    \lbModelHarness{openai.png}{GPT-5.6 Sol}{codex.png}{Codex}}
  \newcommand{\lbOpenCodeDeepSeek}{%
    \lbModelHarness{deepseek.png}{DeepSeek V4 Flash}{opencode.png}{OpenCode}}
  \newcommand{\lbValueDelta}[2]{%
    \makecell{#1\\[-0.2ex]{\scriptsize\color{black!55}(#2)}}}
  \small
  \setlength{\tabcolsep}{3.2pt}
  \renewcommand{\arraystretch}{0.96}
  \scalebox{0.7}{%
  \begin{tabular}{@{}lrrrrrrrr@{}}
    \toprule
    \textbf{Agent}
      & \makecell{\textbf{Resolved}\\\textbf{(\%)}}
      & \makecell{\textbf{Completed}\\\textbf{(\%)}}
      & \makecell{\textbf{F2P Pass}\\\textbf{(\%)}}
      & \makecell{\textbf{P2P Pass}\\\textbf{(\%)}}
      & \makecell{\textbf{Avg.}\\\textbf{Turns}}
      & \makecell{\textbf{Avg. Time}\\\textbf{(min)}}
      & \makecell{\textbf{Generated}\\\textbf{Tokens (k)}}
      & \makecell{\textbf{Avg.}\\\textbf{Cost (\$)}} \\
    \midrule
    \lbClaudeCodeOpus
      & \lbValueDelta{\textbf{34.0}}{+20.0}
      & \lbValueDelta{\textbf{42.9}}{+16.6}
      & \lbValueDelta{\textbf{72.7}}{+20.0}
      & \lbValueDelta{\underline{81.1}}{+8.1}
      & \lbValueDelta{168.0}{+2.2}
      & \lbValueDelta{50.8}{-6.9}
      & \lbValueDelta{141.5}{-10.0}
      & \lbValueDelta{15.2}{-0.1} \\
    \lbCodexGPT
      & \lbValueDelta{\underline{25.0}}{+22.0}
      & \lbValueDelta{\underline{32.8}}{+19.2}
      & \lbValueDelta{\underline{56.2}}{+27.6}
      & \lbValueDelta{\textbf{85.7}}{+9.7}
      & \lbValueDelta{123.8}{-11.4}
      & \lbValueDelta{27.4}{+2.5}
      & \lbValueDelta{47.8}{+1.4}
      & \lbValueDelta{10.8}{-0.9} \\
    \lbOpenCodeDeepSeek
      & \lbValueDelta{17.0}{+16.0}
      & \lbValueDelta{17.7}{+13.7}
      & \lbValueDelta{31.6}{+13.3}
      & \lbValueDelta{69.4}{-4.2}
      & \lbValueDelta{232.0}{+54.8}
      & \lbValueDelta{55.1}{+10.7}
      & \lbValueDelta{128.7}{+32.9}
      & \lbValueDelta{0.9}{+0.4} \\
    \bottomrule
  \end{tabular}
  }
  \endgroup
\end{table*}

\section{Conclusion and Future Work}

We introduce \bench, a multilingual benchmark of 100 modular development tasks across 29 large software systems. Built from human-written enhancement proposals, \bench evaluates the full process of grounding user intent and high-level design in an existing codebase and producing verified implementations. Its tasks combine perception complexity with implementation complexity. Across 28 agents, the highest resolved rate is only 14\%. Our analyses identify incomplete cross-module context as the major failure mode. Providing reference-derived implementation guidance improves resolved rates by 16--22 percentage points across three agents, highlighting the challenge of translating proposals into concrete changes on large software systems. Future work should focus on improving the perception capability of coding agents to handle practical modular development tasks on large software systems.

\bibliography{iclr2027_conference}
\bibliographystyle{iclr2027_conference}

\appendix
\section{Statistics of Projects in \bench}
\label{sec:project-statistics}

\begin{table*}[t]
  \centering
  \caption{Statistics of projects in \bench. ``Language'' is assigned at the project level and applies to every task from that project. ``Repo Size'' is the mean source code lines excluding comments and blank lines. ``GitHub Stars'' are as of September 18, 2026, and ``Maintenance Days'' is the number of days elapsed from each repository's creation on GitHub to September 18, 2026.}
  \label{tab:lolbench-projects}
  \begingroup
  \small
  \setlength{\tabcolsep}{5.0pt}
  \renewcommand{\arraystretch}{0.94}
  \scalebox{0.9}{%
  \begin{tabular}{@{}llcrrrr@{}}
    \toprule
    \textbf{Domain}
      & \textbf{Project}
      & \textbf{Language}
      & \textbf{\#Tasks}
      & \makecell{\textbf{Repo Size}\\\textbf{(kLoC)}}
      & \makecell{\textbf{GitHub}\\\textbf{Stars}}
      & \makecell{\textbf{Maintenance}\\\textbf{Days}} \\
    \midrule
    \multirow[c]{9}{*}{Compilers}
      & CPython    & Python     & 18 & 1,344.4 & 77,201  & 3,507 \\
      & Cargo      & Rust       & 1  & 230.1 & 15,489  & 4,581 \\
      & Mypy       & Python     & 1  & 213.2 & 20,644  & 5,033 \\
      & OpenJDK    & Java       & 14 & 7,744.6 & 23,358  & 2,923 \\
      & PHP        & C          & 4  & 2,615.7 & 40,393  & 5,573 \\
      & Roslyn     & C\#        & 1  & 5,981.0 & 20,672  & 4,268 \\
      & Ruff       & Rust       & 1  & 297.9 & 49,679  & 1,501 \\
      & Rust       & Rust       & 4  & 2,390.0 & 118,921 & 5,938 \\
      & TypeScript & TypeScript & 4  & 2,379.0 & 111,099 & 4,476 \\
    \midrule
    \multirow[c]{8}{*}{Data Science}
      & Apache Arrow      & C++    & 2  & 906.2 & 17,136 & 3,866 \\
      & Apache DataFusion & Rust   & 2  & 475.2 & 9,323  & 1,980 \\
      & Apache Flink      & Java   & 13 & 1,824.0 & 26,342 & 4,486 \\
      & Apache Iceberg    & Java   & 1  & 635.7 & 9,248  & 2,860 \\
      & Apache Kafka      & Java   & 7  & 778.2 & 33,749 & 5,513 \\
      & NumPy             & Python & 2  & 335.2 & 32,765 & 5,849 \\
      & Pandas            & Python & 1  & 421.1 & 49,734 & 5,869 \\
      & Scikit-learn      & Python & 1  & 218.2 & 67,288 & 5,876 \\
    \midrule
    \multirow[c]{6}{*}{Distributed Systems}
      & cert-manager & Go & 1 & 166.0 & 14,081  & 3,404 \\
      & gRPC-Go      & Go & 1 & 208.8 & 23,069  & 4,302 \\
      & Kubernetes   & Go & 7 & 3,805.3 & 127,805 & 4,487 \\
      & Kueue        & Go & 1 & 1,947.1 & 2,984   & 1,675 \\
      & OpenTofu     & Go & 4 & 438.7 & 30,212  & 1,129 \\
      & Prometheus   & Go & 3 & 250.7 & 66,114  & 5,046 \\
    \midrule
    \multirow[c]{3}{*}{Databases}
      & DuckDB & C++  & 1 & 1,080.6 & 41,492 & 3,006 \\
      & Presto & Java & 1 & 897.8 & 16,739 & 5,153 \\
      & Vitess & Go   & 1 & 1,043.9 & 21,344 & 4,831 \\
    \midrule
    \multirow[c]{3}{*}{Web}
      & ASP.NET Core & C\#    & 1 & 1,821.5 & 38,445 & 4,574 \\
      & Django       & Python & 1 & 246.3 & 91,134 & 5,256 \\
      & FastAPI      & Python & 1 & 52.5 & 102,419 & 2,841 \\
    \bottomrule
  \end{tabular}%
  }
  \endgroup
\end{table*}

Table~\ref{tab:lolbench-projects} details the project composition of the 100 tasks in \bench.  The benchmark spans 29 projects in five domains: compilers contribute 48 tasks, data science systems 29, distributed systems 17, and databases and web frameworks three each.  It also covers eight programming languages, with 36 Java, 25 Python, 18 Go, 8 Rust, 3 C++, 4 C, 4 TypeScript, and 2 C\# tasks. 

The projects we selected are well-maintained large software systems. Based on the per-project means, all repositories exceed 50k source LoC, and 12 exceed one million. Mean repository size ranges from 52.5k source LoC for FastAPI to 7.7 million for OpenJDK.  As of September 18, 2026, their GitHub stars range from 2,984 to 127,805, and their elapsed times since repository creation on GitHub range from 1,129 to 5,938 days. This indicates that these software systems have a significant impact on the open-source community and have been maintained for a long time. Therefore, \bench combines diverse languages and architectural styles while consistently requiring agents to navigate complex software systems.

\section{Details of EP--PR Matching}\label{sec:quality}

This section specifies the quality-control pipeline used in the EP--PR matching phase in Sec.~\ref{sec:meth}.  For each candidate, the mapping artifact contains (1)~a Section Classification Summary (SCS) covering every EP section, (2)~a PR File Summary (PFS), (3)~one body block for each mapped implementable section, with separate tables for direct modifications and associated support changes, and (4)~tables for unmapped EP sections and PR files.  Source-code rows are additionally resolved to changed classes and functions. A candidate must pass the static and semantic validation gates before it is quality-ranked.

\subsection{Static validation}
\label{app:static-validation}

The 28 static rules are deterministic checks over the EP, the PR, and the mapping artifact. They are:

\begin{enumerate}
  \setlength{\itemsep}{1pt}
  \setlength{\parskip}{0pt}
  \item \textbf{PR-file completeness.} Every changed PR file is assigned to exactly one role: directly mapped, associated, or unmapped. The three-way total equals the PR file count.
  \item \textbf{File validity.} Every file named by the mapping occurs in the PR diff.
  \item \textbf{EP-section completeness.} Every Markdown or reStructuredText heading in the EP occurs in the SCS and is placed consistently: mapped body block, unmapped implementable table, or SCS-only contextual entry.
  \item \textbf{Controlled vocabularies.} Section, PR-file, and unmapped-reason labels belong to their predefined vocabularies, and the relevant tables use the required schemas.
  \item \textbf{SCS consistency.} The SCS has the required columns, implementable sections use implementation/evaluation categories, non-implementable sections use knowledge/contextual/process categories, and its mapping flags agree with the body blocks.
  \item \textbf{Document order.} Mapped sections are numbered and occur in the same order as in the EP.
  \item \textbf{Per-section proportions.} Each body block reports direct, associated, and accounted file counts and percentages that agree with its tables.
  \item \textbf{Global arithmetic.} The reported EP-section coverage and PR file coverage are arithmetically correct.
  \item \textbf{Test-plan coverage.} When the EP and PR contain test sections and test files, the sections are implementable evaluation sections, and the test files are assigned to them.
  \item \textbf{Implementability--category agreement.} The implementability flag and section category form a valid pair.
  \item \textbf{PFS consistency.} The PFS has the required columns, direct and associated roles are mutually exclusive, and every PFS assignment agrees with the body or unmapped-file table.
  \item \textbf{Unmapped-section consistency.} The unmapped-section table contains exactly the SCS rows marked implementable but unmapped, with the same identifiers and a permitted reason.
  \item \textbf{No redundant section lists.} Contextual and process sections appear only in the SCS, rather than in additional stand-alone lists.
  \item \textbf{Multi-section accounting.} If one file contributes to several sections, the PFS lists every owning or associated section.
  \item \textbf{Statistics consistency.} Regenerated per-instance statistics agree with the mapping, including category totals and coverage values.
  \item \textbf{Unique SCS entries.} SCS identifiers and titles are unique, and code fragments are not mistaken for section headings.
  \item \textbf{Unique PFS entries.} Each PR file has exactly one PFS row.
  \item \textbf{Sequential numbering.} SCS and PFS rows are consecutively numbered, and mapped body blocks are in ascending order.
  \item \textbf{SCS--body correspondence.} Each body block corresponds to one SCS row marked implementable and mapped, and conversely.
  \item \textbf{File-level entries.} Mapping tables name individual files, not directories, globs, or aggregate paths.
  \item \textbf{Unmapped-count arithmetic.} Unmapped EP sections and PR files equal their respective totals minus mapped/accounted items.
  \item \textbf{Verbatim evidence.} Every mapped body block quotes text that occurs verbatim in the appropriate EP section and includes a concise requirement summary.
  \item \textbf{No synthetic sections.} Every SCS title is backed by an actual EP heading.
  \item \textbf{Modification-summary coverage.} Every directly mapped file has exactly one corresponding summary bullet, with no extra bullets.
  \item \textbf{Non-overlapping quotations.} A section quotation neither absorbs a mapped child section nor overlaps the quotation of another body block.
  \item \label{rule:self-contained-ep}\textbf{Self-contained EP.} The EP contains substantive requirement text rather than only a link to an external design document, and ambiguous cases are forwarded to semantic rule S8.
  \item \textbf{Touched-scope coverage.} For each source file, every changed class/function pair recovered from the pre- and post-PR syntax trees occurs in at least one direct or associated row.  Pure file-level changes are exempt.
  \item \textbf{Unique scope ownership.} Each changed
  (file, class, function) tuple has one canonical assignment, even when its file contributes to multiple EP sections.
\end{enumerate}

The controlled vocabularies in rule 4 are: \{knowledge, implementation, evaluation, contextual, process\} for EP sections, \{source, test, test-data, data, build, documentation, generated, vendor\} for PR files, and \{deferred, pre-existing, out-of-scope, documentation-only, runtime-behavior, external-dependency, test-only, insufficient-context\} for an unmapped implementable section.

Rules 27--28 use syntax-tree scopes from both sides of each diff: added lines are attributed using the post-PR tree and deleted lines using the pre-PR tree. This avoids treating a multi-purpose file as an indivisible unit while still requiring every changed code entity to have a unique semantic owner.

\subsection{Semantic validation}

Static checks cannot decide whether an EP section matches a code change in the PR.  We therefore run eight focused LLM audits. Each audit receives the complete EP and mapping, as well as the PR patches and the extracted class/function scopes. We use separate prompts to reduce interference between criteria.  The common prompt wrapper is:

\begin{quote}\small
\emph{You are a quality auditor for LoLBench EP--PR mapping files. Read the provided EP, PR data, and mapping. Apply only rule S$k$. Report clear, actionable errors, identifying the affected section, file, or scope. Return a
JSON list of issue strings; return \texttt{[]} when no issue is found (except for the verdict format specified by S8). Do not report borderline stylistic judgments.}
\end{quote}

The rule-specific instructions appended to this wrapper are:

\begin{enumerate}
  \renewcommand{\labelenumi}{S\arabic{enumi}.}
  \setlength{\itemsep}{2pt}
  \setlength{\parskip}{0pt}
  \item \textbf{Section coverage.} Extract all Markdown and
  reStructuredText headings, and report headings missing from the SCS, SCS entries unsupported by the EP, and ordering errors.
  \item \textbf{Section classification.} Read each section and verify both its implementability and its category: implementation/evaluation for concrete behavior or test requirements, and knowledge/contextual/process for background, rationale, and other contextual descriptions.
  \item \textbf{Test mapping.} Verify that changed test files are assigned to the finest available test subsection and that corresponding test sections
  are implementable evaluation sections, rather than assigning tests to an unrelated implementation section.
  \item \textbf{PR-file category.} Classify each patch as source, test, test-data, data, build, documentation, generated, or vendor. Inspect patch content in ambiguous cases instead of relying only on paths or suffixes.
  \item \textbf{Mapping semantics.} For every mapped section, verify that its direct files/scopes implement the stated behavior, that the most specific matching section owns them, and that associated rows have genuine support relationships.  Report clear cross-layer or cross-feature mismatches.
  \item \textbf{Citation and summary fidelity.} Verify that each quotation is exact text from the section named by its path, that any truncation is a faithful prefix, and that the one-to-three-sentence summary is accurate and contains no fabricated claim.
  \item \textbf{Direct versus associated changes.} Put only changes explicitly required by the section in the direct table.  Treat tests, generated files,
  dependencies, vendored code, build/CI files, and documentation as associated unless the section explicitly requires them.  Also verify that each direct
  file's modification summary accurately explains its contribution.
  \item \textbf{External-link sufficiency.} Enumerate external links and decide whether each is incidental background or contains essential specification
  text.  Essential accessible content must be inlined.  If essential content is inaccessible, return \texttt{invalid:<reason>} and reject the candidate;
  otherwise return \texttt{complete} or \texttt{inlined} as appropriate.
\end{enumerate}

The semantic audits are run through Claude Code~\citep{claudecode} as independent subagents in batches of five.  For S1--S7, an empty issue list is a pass.  Rule S8 is applied to EPs containing external links, especially those flagged by Appendix~\ref{app:static-validation}, Rule~\ref{rule:self-contained-ep}. A curator reviews the reported evidence, regenerates a faulty mapping rather than locally patching it, and reruns both validation stages.  Only candidates
with no unresolved static or semantic findings proceed to quality ranking.

\subsection{Quality ranking}

The validation rules are hard constraints, but ranking measures the quality of a valid mapping.  The LLM judge considers only semantic alignment between the EP and
PR, not whether the upstream implementation itself is a good engineering solution.  It grades every implementable section on the five dimensions in Table~\ref{tab:mapping-quality-dimensions}, and non-implementable sections are
excluded in this process.

\begin{table*}[t]
  \centering
  \caption{Dimensions and score anchors used to evaluate EP-to-PR mappings. Each dimension is scored from 1 to 10, and the final section score is their weighted sum.}
  \label{tab:mapping-quality-dimensions}
  \footnotesize
  \setlength{\tabcolsep}{5pt}
  \renewcommand{\arraystretch}{1.08}
  \begin{tabularx}{\textwidth}{@{}l c X@{}}
    \toprule
    \textbf{Dimension} & \textbf{Weight} & \textbf{Score anchors} \\
    \midrule
    Requirement coverage
      & 40\%
      & \textbf{9--10:} all concrete requirements are mapped;
        \textbf{7--8:} one minor sub-requirement or edge case is missing or
        indirect; \textbf{5--6:} an important requirement is only partly
        covered; \textbf{3--4:} several central requirements are missing or
        mapped only to support; \textbf{1--2:} little valid coverage. \\
    \midrule
    Mapping precision
      & 25\%
      & \textbf{9--10:} no meaningful unrelated behavior;
        \textbf{7--8:} only incidental support or shared helpers;
        \textbf{5--6:} substantial unrelated behavior; \textbf{3--4:} many
        loosely related scopes that belong elsewhere or should be associated;
        \textbf{1--2:} mostly unrelated scopes. \\
    \midrule
    Scope correctness
      & 15\%
      & \textbf{9--10:} correct section, direct/associated role, category, and
        unique scope ownership; \textbf{7--8:} a few minor helper or category
        errors; \textbf{5--6:} several incorrect assignments, but still
        usable; \textbf{3--4:} assignment is mostly based on file proximity;
        \textbf{1--2:} ownership is largely incorrect. \\
    \midrule
    Granularity
      & 10\%
      & \textbf{9--10:} a fine-grained semantic boundary;
        \textbf{7--8:} a few broad rows, but coherent ownership;
        \textbf{5--6:} a large mixed set of scopes; \textbf{3--4:} the section
        is mostly a catch-all; \textbf{1--2:} no useful granularity. \\
    \midrule
    Requirement specificity
      & 10\%
      & \textbf{9--10:} concrete APIs, syntax, semantics, configuration, or
        deliverables; \textbf{7--8:} only routine details are implicit;
        \textbf{5--6:} substantial inference is required; \textbf{3--4:} very
        short or vague; \textbf{1--2:} too vague to justify the mapping without
        external information. \\
    \bottomrule
  \end{tabularx}
\end{table*}

\paragraph{Judge prompt.}
Bracketed fields below are replaced with the full candidate artifacts.

\begin{quote}\small
\emph{You are grading the semantic quality of an EP-to-PR mapping, not the
quality of the PR implementation. Read [EP], [PR patches and changed
class/function scopes], and [mapping]. For every implementable EP section:
(1) extract its concrete requirement claims; (2) determine whether its direct
files/scopes cover all claims without importing unrelated behavior; (3) verify
file, class, function, and direct-versus-associated ownership; (4) assess
whether the mapping is sufficiently fine-grained; and (5) assess whether the EP
text is specific enough to justify the mapping. Score requirement coverage,
mapping precision, scope/function correctness, granularity, and requirement
specificity from 1 to 10 using the supplied rubric. Assign a semantic
importance weight independently of the number of mapped files. Return one row
per implementable section containing the five scores, importance weight,
weighted section score, and a concise evidence-based reason. Then return the
weighted instance score and its main reason. Do not apply an additional
instance-level adjustment.}
\end{quote}

Let $d_{ij}$ be the score of section $i$ on dimension $j$.  Its quality score
is
\[
q_i = 0.40d_{i,\mathrm{cov}} + 0.25d_{i,\mathrm{prec}}
    + 0.15d_{i,\mathrm{scope}} + 0.10d_{i,\mathrm{gran}}
    + 0.10d_{i,\mathrm{spec}}.
\]

The judge separately assigns each section an importance weight $w_i$ using
Table~\ref{tab:section-importance-weights}.

\begin{table*}[t]
  \centering
  \caption{Semantic importance weights for EP sections. Modifiers are applied
  independently after choosing the base weight. Implementable section weights
  are clipped to $[0.1,1.0]$; non-implementable sections always receive zero.}
  \label{tab:section-importance-weights}
  \small
  \setlength{\tabcolsep}{6pt}
  \renewcommand{\arraystretch}{1.08}
  \begin{tabularx}{\textwidth}{@{}l c X@{}}
    \toprule
    \textbf{Type} & \textbf{Weight} & \textbf{Section semantics or condition} \\
    \midrule
    Base & 1.0 & Core public API, syntax, protocol/schema, runtime semantics,
      or principal feature behavior. \\
    Base & 0.8 & Important internal mechanism required by the main behavior. \\
    Base & 0.6 & Compatibility, migration, configuration, rollout, error
      handling, or deprecation behavior. \\
    Base & 0.4 & Explicit test, validation, benchmark, or performance
      requirement. \\
    Base & 0.2 & Explicit documentation, build, generated-output, fixture, or
      support requirement. \\
    Base & 0.0 & Non-implementable knowledge, context, or process section. \\
    \midrule
    Modifier & $+0.1$ & Strong normative language such as \emph{must},
      \emph{shall}, or \emph{required}, or an explicitly named goal. \\
    Modifier & $+0.1$ & Many other implementable sections conceptually depend
      on this section. \\
    Modifier & $-0.1$ & Primarily illustrative or example-based. \\
    Modifier & $-0.1$ & Vague with weak requirement force, but still
      implementable. \\
    Modifier & $-0.2$ & Optional, deferred, exploratory, or nice-to-have. \\
    \bottomrule
  \end{tabularx}
\end{table*}

The number of mapped files never affects \mbox{$w_i$.} Broad or weakly justified
mappings are instead penalized through precision, granularity, and requirement
specificity.

The candidate score is the importance-weighted mean
\begin{equation*}
Q = \frac{\displaystyle\sum_i w_i\,q_i}{\displaystyle\sum_i w_i}.
\end{equation*}
Section and candidate scores are rounded to one decimal place.  We retain a candidate only if it has passed both validation gates, receives $Q \geq 8.0$, and passes final expert review.

\section{Details of Task Enhancement}\label{app:enhance}

\subsection{Incorrect solution mutant generation}
\label{app:mutant-generation}

We construct incorrect solution mutants to identify requirement-level
deficiencies in the original F2P suites. Each mutant is a complete candidate solution that intentionally violates a section in the EP. An accepted mutant must preserve the behavior exercised by all P2P tests and have a documented functional difference from the reference solution.

\paragraph{EP mutation.}
We do not directly mutate solutions. Instead, we mutate one implementable section in the EP and first generate EP mutants. We apply three methods for EP mutations:
\begin{enumerate}
    \setlength{\itemsep}{2pt}
    \setlength{\parskip}{0pt}
    \item \textbf{Section Revert.} Remove or reverse behavior attributed to a section, yielding a partial implementation that omits a
    required functionality while retaining the support needed by the remaining implementation.
    \item \textbf{Requirement Mismatch.} Introduce a plausible semantic error relative to an explicit requirement clause. Examples include an incorrect default or validation boundary, omitted parameter propagation, an incorrect enumeration or option mapping, altered error behavior, or a disabled feature flag.
    \item \textbf{Semantic Mutation.} Alter the required structure or execution flow while preserving a plausible implementation. Examples include bypassing a new dispatch path, calling the old implementation, dropping a required method call, returning a default, empty, or routing execution to an incorrect handler.
\end{enumerate}

\textbf{Solution Mutation.} Given the EP mutants and the reference solutions, we then ask LLMs to modify the reference solutions based on the EP mutants. For the generated solution candidates, we verify their semantic non-equivalence with the original reference solutions. Every candidate must identify the exact mapped section and requirement it violates, describe the reference and mutated behaviors, and specify a concrete scenario under which they diverge. The candidate also includes a non-equivalence rationale explaining how this difference
violates the requirement. These records must allow a reviewer to assess the functional difference without relying on an F2P failure. We then validate the candidates on P2P tests and filter out candidates that cannot pass. Finally, we combine LLM judges and expert reviewers to review the remaining candidates and select those with distinct functionality as solution mutants. Table~\ref{tab:mutant-distribution} summarizes the solution mutants generated by three mutation methods.

\begin{table}[t]
  \centering
  \caption{Distribution of retained solution mutants, excluding retired or specification-equivalent candidates.}
  \label{tab:mutant-distribution}
  \begingroup
  \small
  \setlength{\tabcolsep}{10pt}
  \renewcommand{\arraystretch}{1.08}
  \begin{tabular}{@{}lrr@{}}
    \toprule
    \textbf{Mutation method} & \textbf{Count} & \textbf{Share (\%)} \\
    \midrule
    Section Revert       & 158   & 9.36 \\
    Requirement Mismatch & 1,205 & 71.39 \\
    Semantic Mutation    & 325   & 19.25 \\
    \midrule
    \textbf{Total} & \textbf{1,688} & \textbf{100.00} \\
    \bottomrule
  \end{tabular}
  \endgroup
\end{table}

\subsection{F2P test augmentation}
\label{app:test-augmentation}

\textbf{Mutation-guided test generation.} We run all original F2P tests on generated solution mutants. When the original F2P tests cannot kill certain solution mutants, we use Codex with GPT-5.6 Sol and Claude Code with Opus 4.8~\citep{opus48} to add tests that distinguish these mutants from the reference solution. Each candidate targets one or more surviving mutants and asserts an EP-specified outcome through system-level or end-to-end behavior. The EP defines the expected outcome, while the mutant's documented distinguishing scenario guides the input choice.

\textbf{Coverage-guided test generation.} We identify tasks whose original F2P test suite has line coverage lower than 50\%. For these tasks, we direct Codex with GPT-5.6 Sol and Claude Code with Opus 4.8 to add new F2P tests targeting uncovered reference-patch behavior.

\paragraph{Test audit.}
We audit every generated F2P test before inclusion. We inspect the selected test body and the source symbols it references. Tests that invoke or otherwise depend on internal symbols are rejected. We also reject tests that directly use newly introduced source symbols whose names are absent from the EP. Rejected candidates must be rewritten to exercise permitted public behavior.
We deduplicate candidates, discard unstable tests, and retain the added tests separately from the original suite. Table~\ref{tab:test-augmentation} summarizes the statistics of added F2P tests. On all 1,688 solution mutants, the original tests kill 1,031 (61.08\%), while the combined original and added tests kill 1,620 (95.97\%), leaving only 68 surviving mutants. The combined test suite also improves the line coverage from 60.07\% to 78.50\%.

\begin{table}[t]
  \centering
  \caption{F2P test augmentation. ``Overall'' combines original and added tests. Coverage is the mean per-task coverage of F2P tests. Mutation scores use the 1,688 retained mutants in Table~\ref{tab:mutant-distribution}.}
  \label{tab:test-augmentation}
  \begingroup
  \small
  \setlength{\tabcolsep}{9pt}
  \renewcommand{\arraystretch}{1.08}
  \begin{tabular}{@{}lrr@{}}
    \toprule
    \textbf{Metric} & \textbf{Original} & \textbf{Overall} \\
    \midrule
    F2P tests & 1,480 & 2,234 \\
    \quad Added: mutant-killing & --- & 622 \\
    \quad Added: coverage & --- & 132 \\
    \midrule
    Mutants killed & 1,031 & 1,620 \\
    Mutation score (\%) & 61.08 & 95.97 \\
    \midrule
    Mean F2P coverage (\%) & 60.07 & 78.50 \\
    \bottomrule
  \end{tabular}
  \endgroup
\end{table}

\paragraph{Mutation-guided generation prompts.}
The following prompt templates summarize the mutation-guided F2P generation instructions. Angle brackets denote task-specific inputs, including the EP, evaluation environment, original tests, reference solution, and surviving-mutant records.

\noindent\textbf{System prompt.}
\begin{quote}\small
\emph{You generate additional fail-to-pass (F2P) tests for a software task described by an enhancement proposal (EP). Your goal is to expose requirement violations in solution mutants that survive the original F2P suite. Use the EP as the behavioral oracle and the mutants' distinguishing scenarios to identify missing checks.}

\emph{Follow these constraints:}
\begin{enumerate}
    \setlength{\itemsep}{2pt}
    \setlength{\parskip}{0pt}
    \item \emph{Exercise system-level or end-to-end behavior through permitted public entry points. Assert the outcome required by the EP, rather than merely asserting that the result differs from a mutant.}
    \item \emph{Do not invoke, import, construct, or assert directly on internal source symbols. Do not directly use a newly introduced source symbol unless its exact name appears in the EP. Generated tests are audited, and candidates that violate these rules are rejected.}
    \item \emph{Use the project's existing test framework and execution conventions. Add tests separately from the original suite; do not modify the reference solution or original tests.}
    \item \emph{Identify the EP clause, input, expected outcome, public entry point, and targeted mutant IDs for each candidate. Prefer a parameterized test when several mutants can be distinguished through the same requirement and entry point.}
    \item \emph{An accepted candidate must fail or error before implementation, pass with the reference solution, and fail on every targeted mutant. Remove duplicates and reject unstable candidates.}
    \item \emph{If no permitted public behavior distinguishes a mutant, report that limitation for review. Do not bypass the restriction by calling an internal symbol.}
\end{enumerate}
\end{quote}

\noindent\textbf{Task prompt.}
\begin{quote}\small
\emph{Generate new F2P test cases for the following task.}

\emph{EP: \textless{}enhancement proposal and section identifiers\textgreater{}}

\emph{Repository and evaluation environment: \textless{}base revision, source tree, test framework, and execution commands\textgreater{}}

\emph{Reference solution: \textless{}reference implementation patch\textgreater{}}

\emph{Original F2P suite: \textless{}test sources, test selectors, and mutant execution results\textgreater{}}

\emph{Surviving mutants: \textless{}mutant IDs, patches, violated requirement clauses, distinguishing scenarios, and expected differences from the reference solution\textgreater{}}

\emph{Test-surface constraints: \textless{}permitted public entry points, internal-symbol restrictions, and newly introduced source symbols\textgreater{}}

\emph{For each surviving mutant, translate its distinguishing scenario into a test that asserts the EP-required outcome through permitted public behavior. Return the added test patch and executable test selectors. For each candidate, report its EP clause, targeted mutant IDs, input, expected outcome, and public entry point. Explain why its imports, calls, constructions, and assertions satisfy the symbol restrictions.}

\emph{Run each candidate before implementation, with the reference solution, and with its targeted mutants; report the observed results. Audit the selected test body for internal-symbol dependencies and newly introduced symbols absent from the EP. Rewrite or discard any candidate that fails this audit or the execution checks. Report any surviving mutant for which no admissible test could be produced.}
\end{quote}

\paragraph{Coverage-guided generation prompts.}
The following templates summarize the coverage-guided F2P generation instructions. They generate extra F2P test cases for tasks whose combined existing F2P and P2P suites have line coverage below 50\%.

\noindent\textbf{System prompt.}
\begin{quote}\small
\emph{You generate additional fail-to-pass (F2P) tests for a software task described by an enhancement proposal (EP). Your goal is to exercise uncovered executable lines in the reference implementation through system-level or end-to-end behavior, while asserting outcomes required by the EP.}

\emph{Follow these constraints:}
\begin{enumerate}
    \setlength{\itemsep}{2pt}
    \setlength{\parskip}{0pt}
    \item \emph{Use the coverage report to identify executable reference-patch lines not reached by the union of the current F2P and P2P suites. Prioritize large connected uncovered spans and relate each target to an implementable EP section using the EP--implementation mapping.}
    \item \emph{Identify a permitted public entry point and an input that reaches each target span. Assert the EP-required observable outcome, rather than an internal implementation detail or a coverage count alone.}
    \item \emph{Do not invoke, import, construct, or assert directly on internal source symbols. Do not directly use a newly introduced source symbol unless its exact name appears in the EP. Audit each candidate for these restrictions.}
    \item \emph{Use the project's existing test framework and execution conventions. Add tests separately from the original suite; do not modify the reference implementation or original tests. Deduplicate candidates and reject unstable tests.}
    \item \emph{Each retained candidate must fail or error before implementation because the required behavior is absent, pass with the reference implementation, and exercise previously uncovered target lines. A test that passes before implementation is not an F2P test, even if it increases coverage.}
    \item \emph{Rerun coverage after adding accepted tests and report newly covered reference-patch lines and the updated F2P--P2P union coverage using the same executable-line denominator. Record the EP clause, public entry point, input, expected outcome, test selector, and validation results for each candidate.}
    \item \emph{If an uncovered span cannot be reached through permitted public behavior, document the span and the reason for review. Do not bypass the restrictions with an internal-symbol test or weaken F2P validity to meet a coverage target.}
\end{enumerate}
\end{quote}

\noindent\textbf{Task prompt.}
\begin{quote}\small
\emph{Generate new F2P tests for the uncovered requirement behavior in the following task.}

\emph{EP and implementation mapping: \textless{}enhancement proposal, section identifiers, and mapped implementation files and scopes\textgreater{}}

\emph{Repository and evaluation environment: \textless{}base revision, source tree, test framework, test commands, and coverage commands\textgreater{}}

\emph{Reference solution: \textless{}reference implementation patch\textgreater{}}

\emph{Current test suites: \textless{}original and previously accepted augmented F2P tests, P2P tests, and executable selectors\textgreater{}}

\emph{Coverage report: \textless{}executable reference-patch lines, per-file F2P and P2P covered-line sets, their union, and baseline union coverage\textgreater{}}

\emph{Test-surface constraints: \textless{}permitted public entry points, internal-symbol restrictions, and newly introduced source symbols\textgreater{}}

\emph{Identify uncovered executable spans and their EP requirements. For each reachable target, design a public-behavior test with an EP-derived expected outcome. Return the added test patch and executable selectors, together with each candidate's target span, EP clause, entry point, input, expected outcome, and symbol-restriction audit.}

\emph{Run each candidate before implementation and with the reference solution, then rerun the augmented F2P and P2P suites with the reference solution and collect coverage. Report the observed results, newly covered target lines, and before/after union coverage. Retain only valid, stable F2P tests with verified coverage gains; report remaining gaps and explain any that cannot be exercised through permitted public behavior.}
\end{quote}

\section{Effectiveness on Different Programming Languages}\label{app:pl}

\begin{figure}[p]
  \captionsetup[subfigure]{font=footnotesize,skip=0pt}
  \centering
  \includegraphics[width=\textwidth]{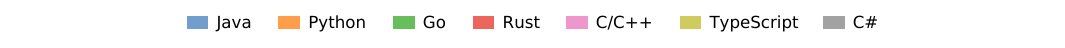}
  \par\smallskip

  \begin{subfigure}[t]{0.66\textwidth}
    \centering
    \includegraphics[width=\linewidth]{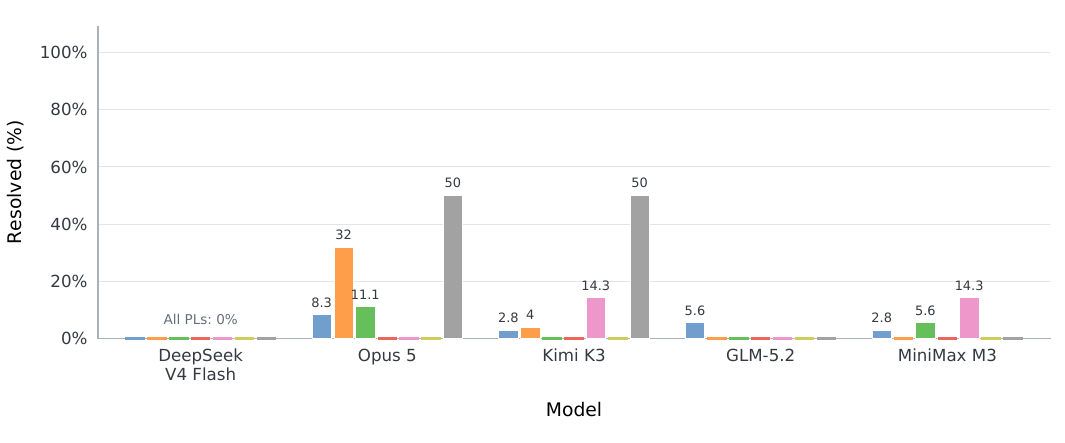}
    \caption{Claude Code}
    \label{fig:pl-resolved-claude-code}
  \end{subfigure}\par
  \begin{subfigure}[t]{0.66\textwidth}
    \centering
    \includegraphics[width=\linewidth]{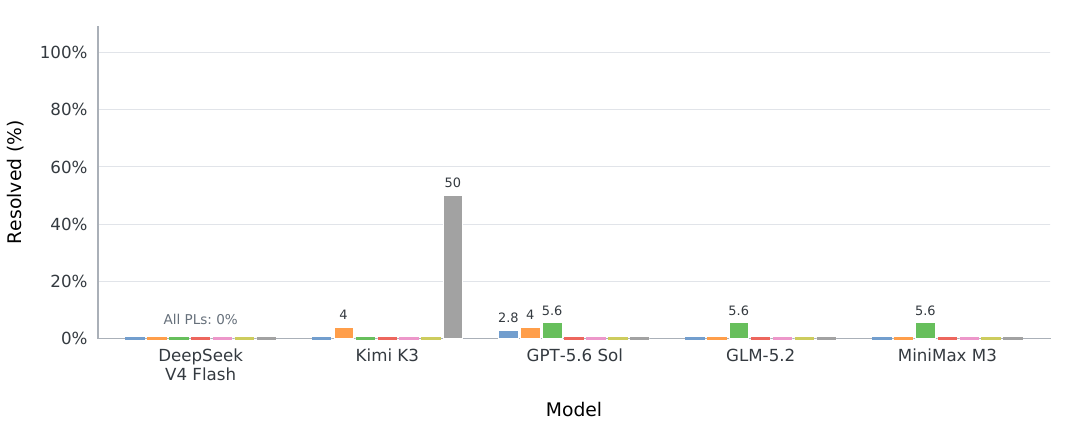}
    \caption{Codex}
    \label{fig:pl-resolved-codex}
  \end{subfigure}\par

  \begin{subfigure}[t]{0.66\textwidth}
    \centering
    \includegraphics[width=\linewidth]{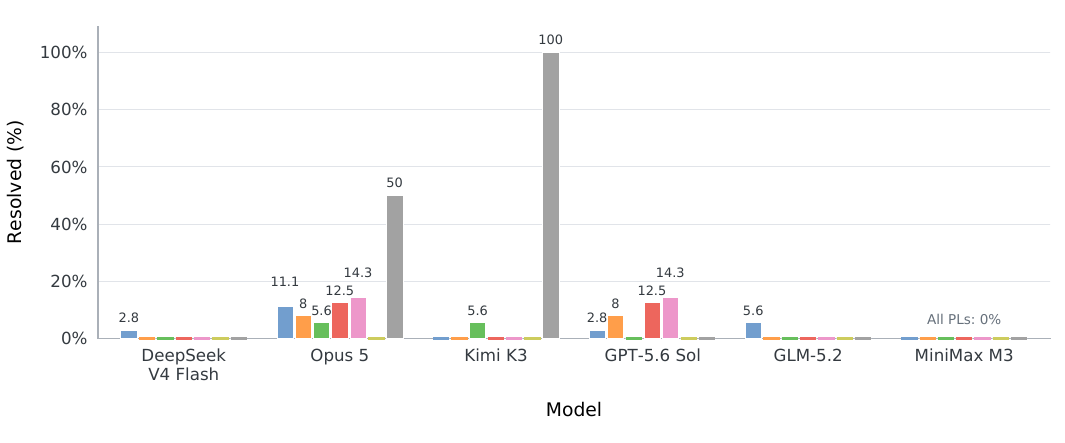}
    \caption{OpenCode}
    \label{fig:pl-resolved-opencode}
  \end{subfigure}\par
  \begin{subfigure}[t]{0.66\textwidth}
    \centering
    \includegraphics[width=\linewidth]{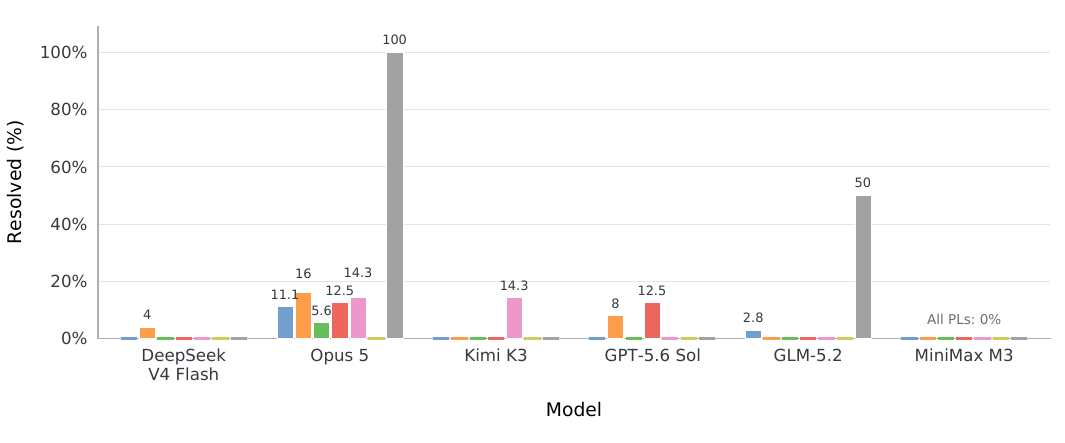}
    \caption{Pi}
    \label{fig:pl-resolved-pi}
  \end{subfigure}\par

  \begin{subfigure}[t]{0.66\textwidth}
    \centering
    \includegraphics[width=\linewidth]{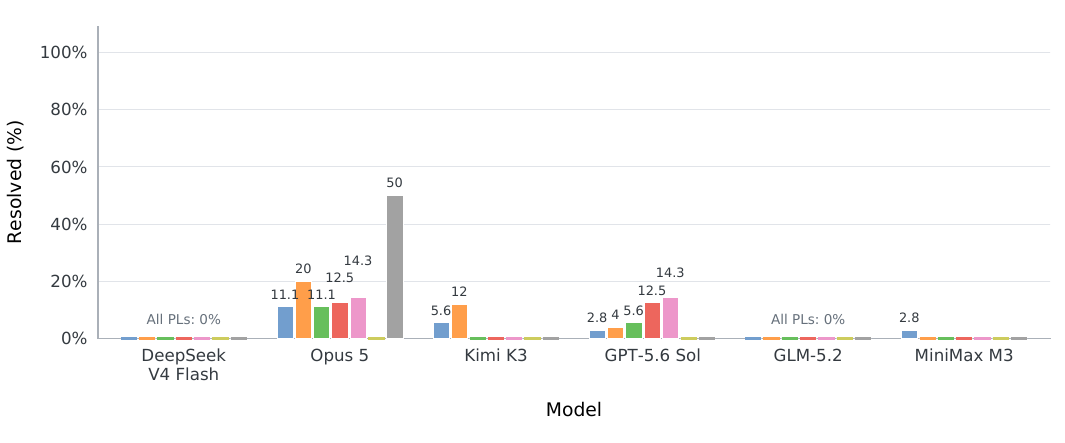}
    \caption{mini-SWE-agent}
    \label{fig:pl-resolved-mini-swe-agent}
  \end{subfigure}\par

  \caption{Resolved\% by programming language.}
  \label{fig:pl-resolved}
\end{figure}

\begin{figure}[p]
  \captionsetup[subfigure]{font=footnotesize,skip=0pt}
  \centering
  \includegraphics[width=\textwidth]{figures/pl_legend.pdf}
  \par\smallskip

  \begin{subfigure}[t]{0.66\textwidth}
    \centering
    \includegraphics[width=\linewidth]{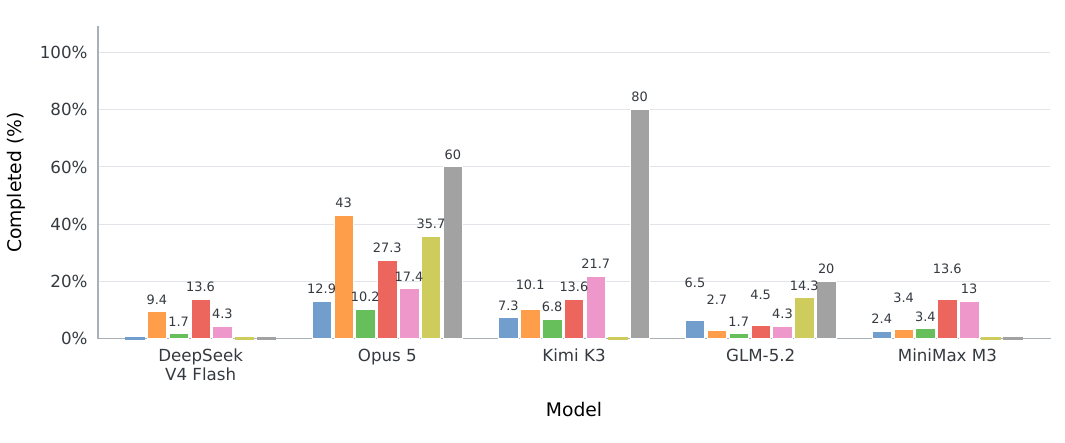}
    \caption{Claude Code}
    \label{fig:pl-completed-claude-code}
  \end{subfigure}\par
  \begin{subfigure}[t]{0.66\textwidth}
    \centering
    \includegraphics[width=\linewidth]{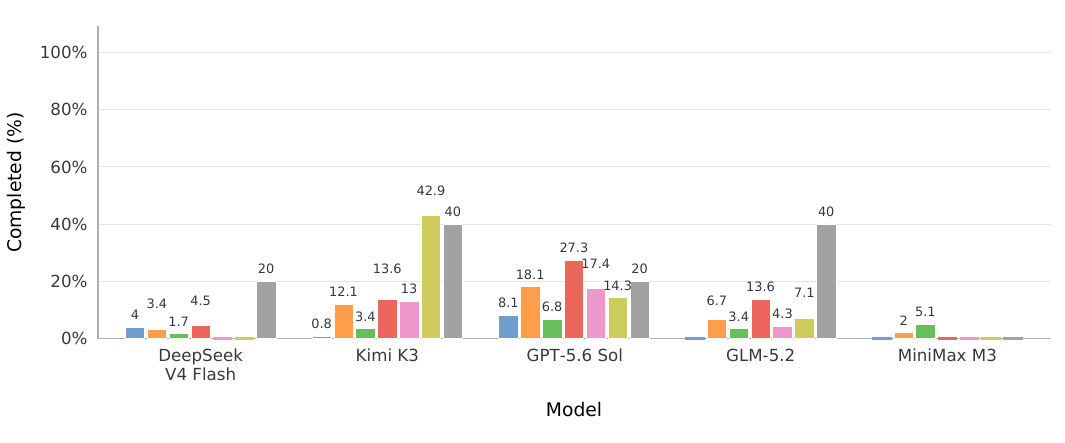}
    \caption{Codex}
    \label{fig:pl-completed-codex}
  \end{subfigure}\par

  \begin{subfigure}[t]{0.66\textwidth}
    \centering
    \includegraphics[width=\linewidth]{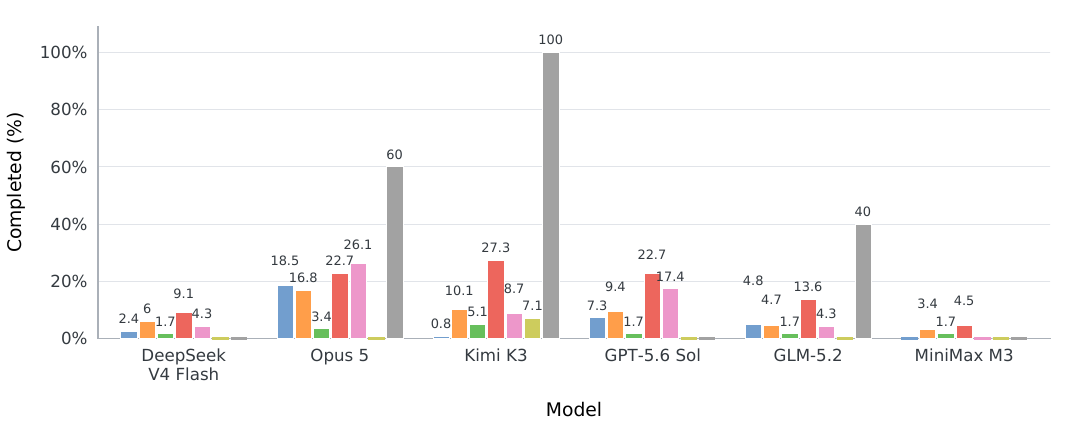}
    \caption{OpenCode}
    \label{fig:pl-completed-opencode}
  \end{subfigure}\par
  \begin{subfigure}[t]{0.66\textwidth}
    \centering
    \includegraphics[width=\linewidth]{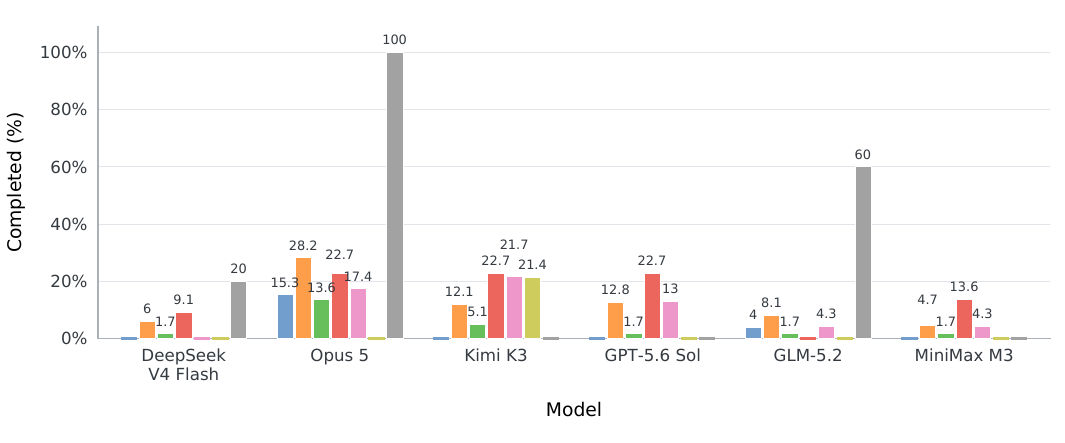}
    \caption{Pi}
    \label{fig:pl-completed-pi}
  \end{subfigure}\par

  \begin{subfigure}[t]{0.66\textwidth}
    \centering
    \includegraphics[width=\linewidth]{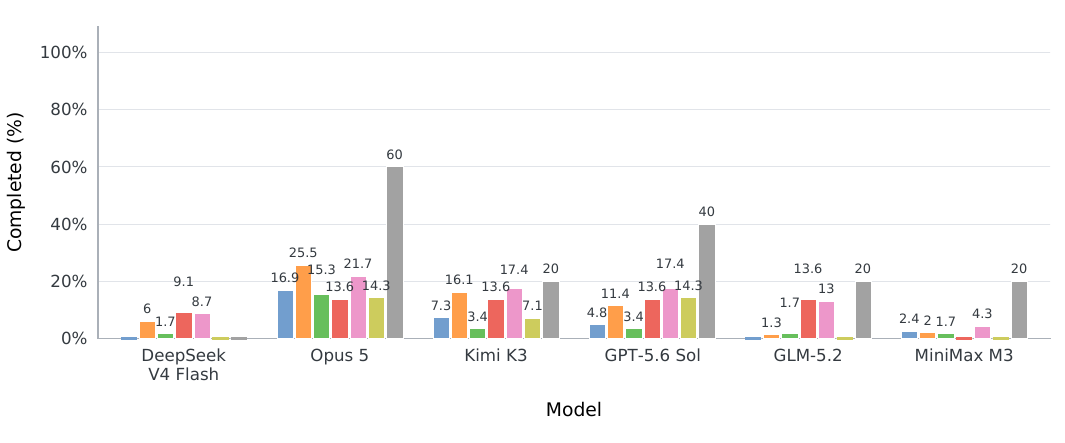}
    \caption{mini-SWE-agent}
    \label{fig:pl-completed-mini-swe-agent}
  \end{subfigure}\par

  \caption{Completed\% by programming language.}
  \label{fig:pl-completed}
\end{figure}

\begin{figure}[p]
  \captionsetup[subfigure]{font=footnotesize,skip=0pt}
  \centering
  \includegraphics[width=\textwidth]{figures/pl_legend.pdf}
  \par\smallskip

  \begin{subfigure}[t]{0.66\textwidth}
    \centering
    \includegraphics[width=\linewidth]{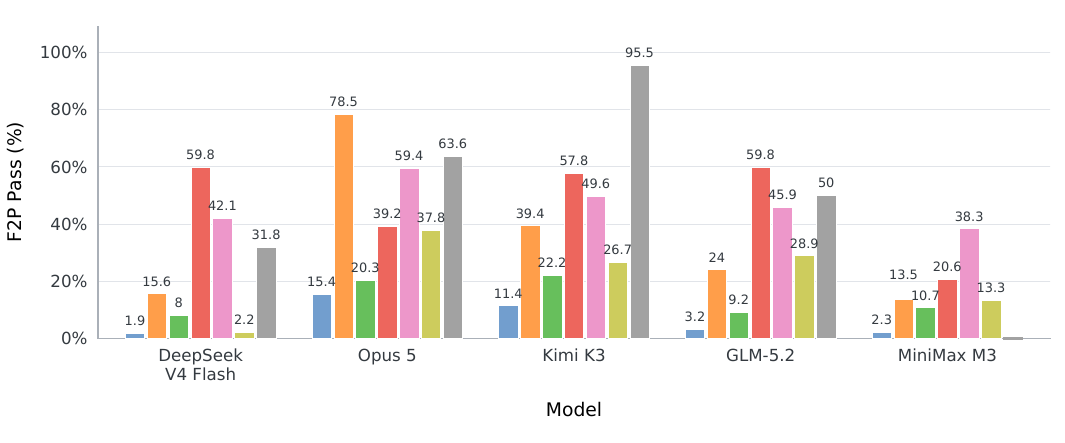}
    \caption{Claude Code}
    \label{fig:pl-f2p-claude-code}
  \end{subfigure}\par
  \begin{subfigure}[t]{0.66\textwidth}
    \centering
    \includegraphics[width=\linewidth]{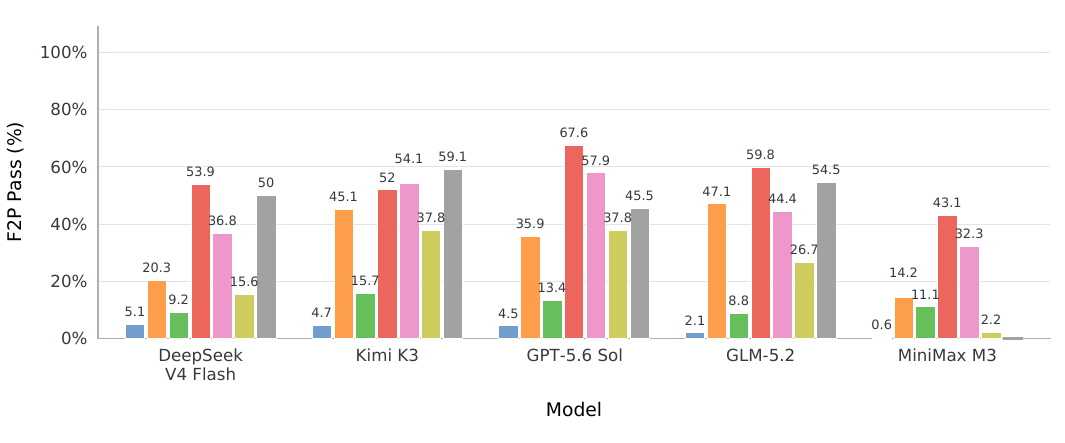}
    \caption{Codex}
    \label{fig:pl-f2p-codex}
  \end{subfigure}\par

  \begin{subfigure}[t]{0.66\textwidth}
    \centering
    \includegraphics[width=\linewidth]{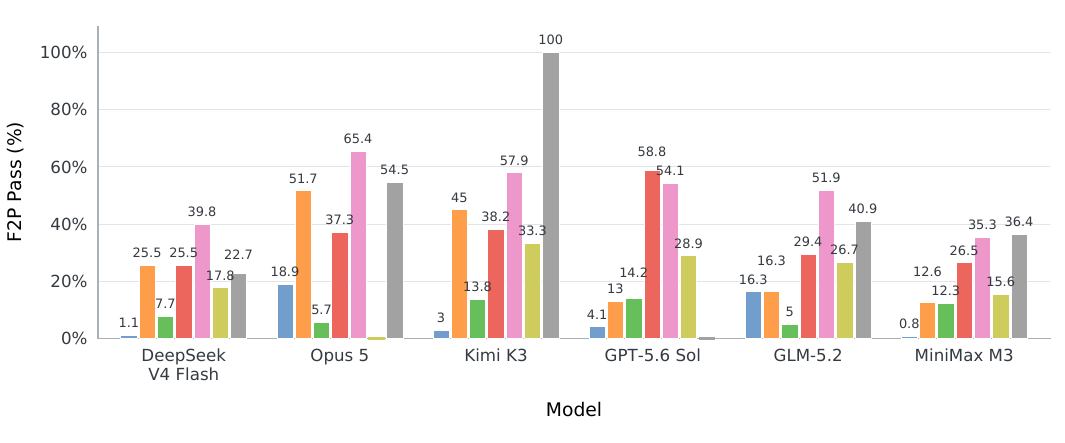}
    \caption{OpenCode}
    \label{fig:pl-f2p-opencode}
  \end{subfigure}\par
  \begin{subfigure}[t]{0.66\textwidth}
    \centering
    \includegraphics[width=\linewidth]{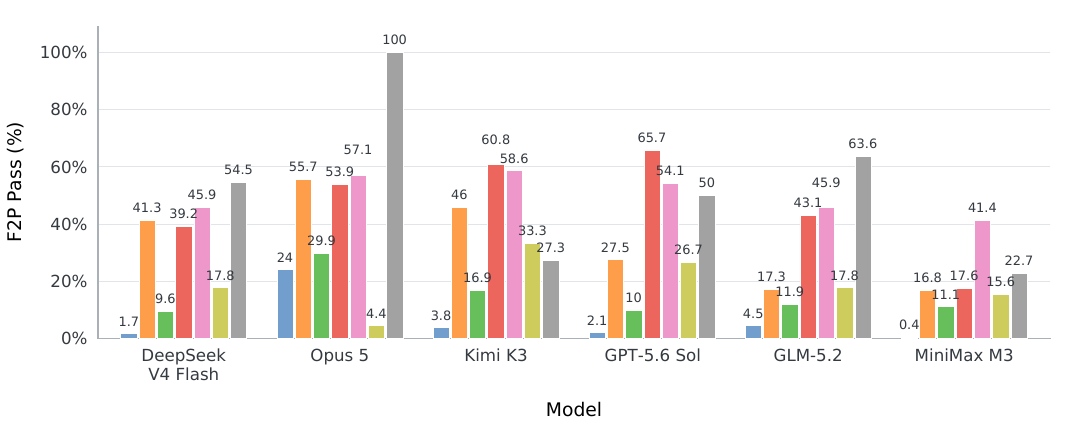}
    \caption{Pi}
    \label{fig:pl-f2p-pi}
  \end{subfigure}\par

  \begin{subfigure}[t]{0.66\textwidth}
    \centering
    \includegraphics[width=\linewidth]{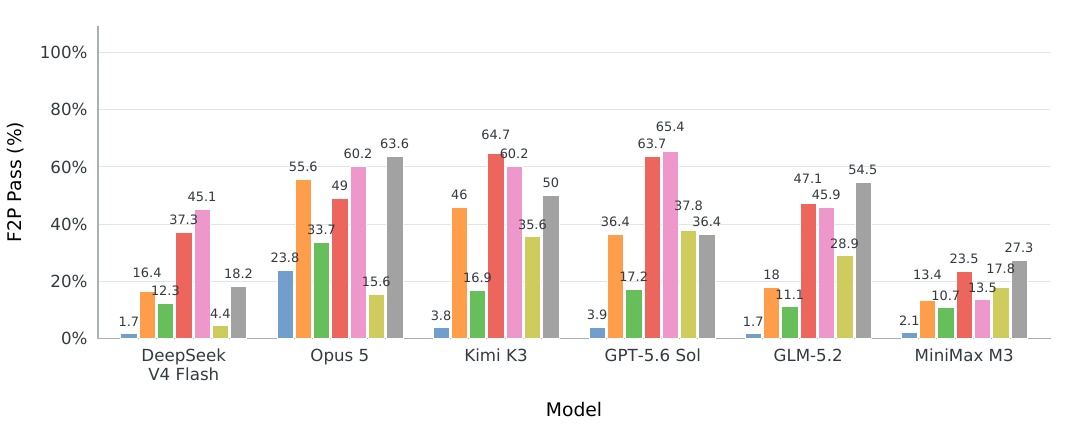}
    \caption{mini-SWE-agent}
    \label{fig:pl-f2p-mini-swe-agent}
  \end{subfigure}\par

  \caption{F2P Pass\% by programming language.}
  \label{fig:pl-f2p}
\end{figure}

\begin{figure}[p]
  \captionsetup[subfigure]{font=footnotesize,skip=0pt}
  \centering
  \includegraphics[width=\textwidth]{figures/pl_legend.pdf}
  \par\smallskip

  \begin{subfigure}[t]{0.66\textwidth}
    \centering
    \includegraphics[width=\linewidth]{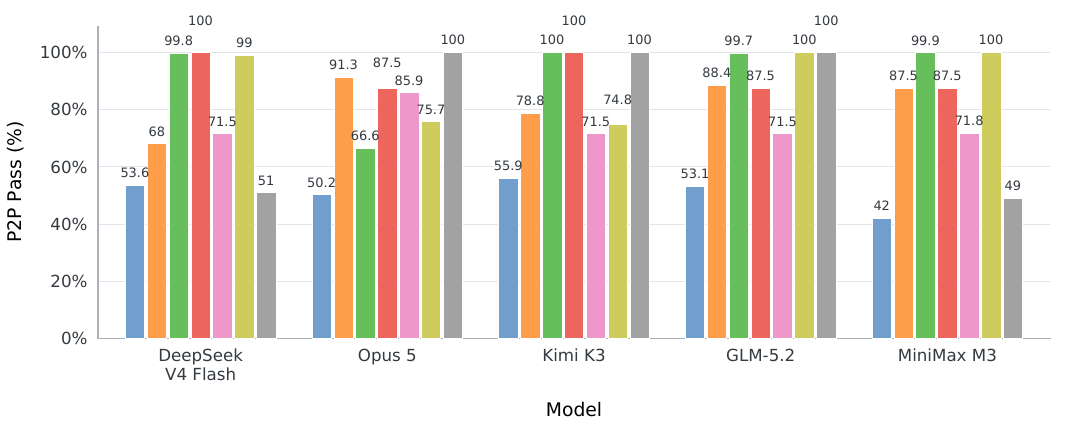}
    \caption{Claude Code}
    \label{fig:pl-p2p-claude-code}
  \end{subfigure}\par
  \begin{subfigure}[t]{0.66\textwidth}
    \centering
    \includegraphics[width=\linewidth]{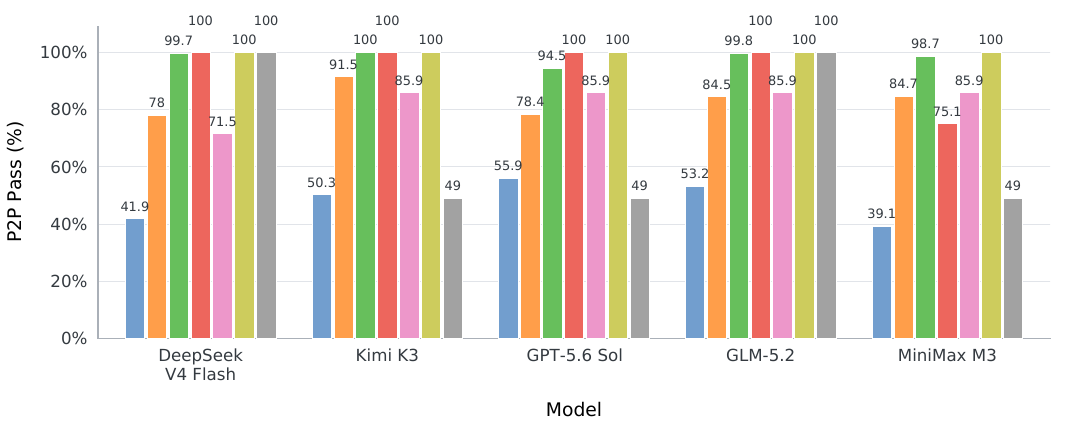}
    \caption{Codex}
    \label{fig:pl-p2p-codex}
  \end{subfigure}\par

  \begin{subfigure}[t]{0.66\textwidth}
    \centering
    \includegraphics[width=\linewidth]{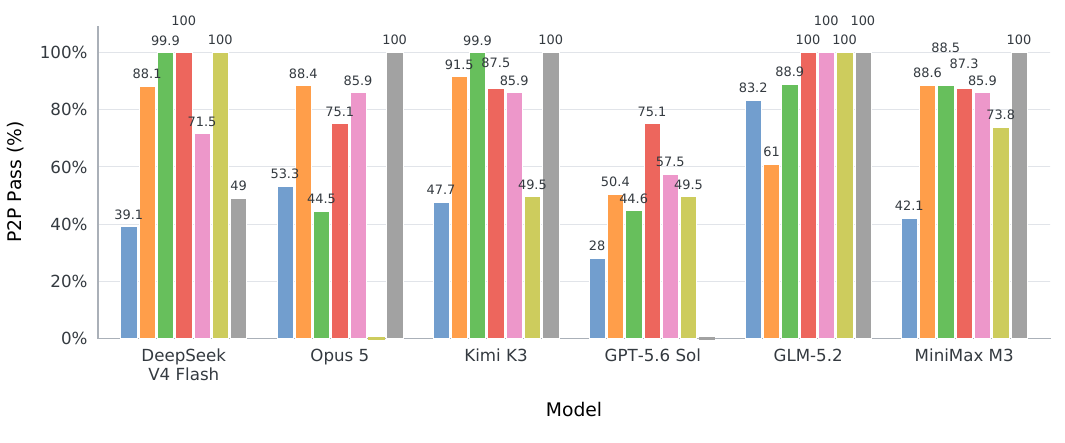}
    \caption{OpenCode}
    \label{fig:pl-p2p-opencode}
  \end{subfigure}\par
  \begin{subfigure}[t]{0.66\textwidth}
    \centering
    \includegraphics[width=\linewidth]{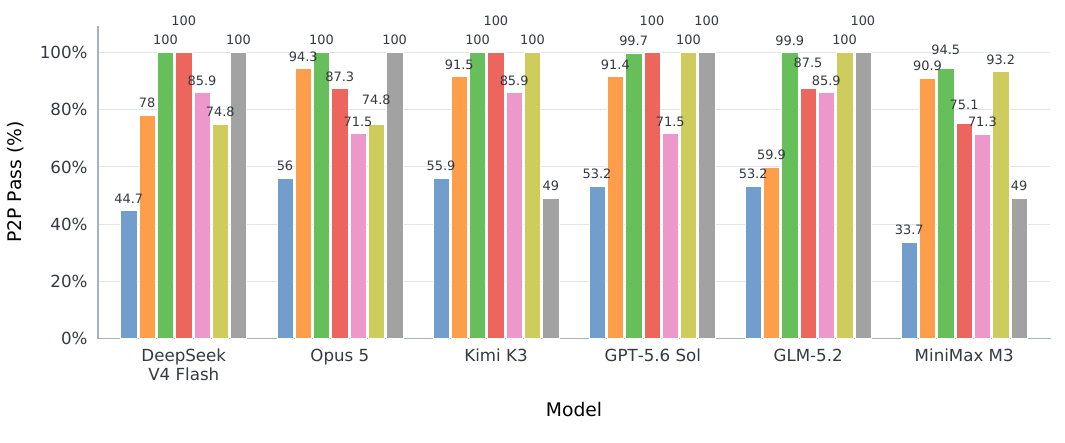}
    \caption{Pi}
    \label{fig:pl-p2p-pi}
  \end{subfigure}\par

  \begin{subfigure}[t]{0.66\textwidth}
    \centering
    \includegraphics[width=\linewidth]{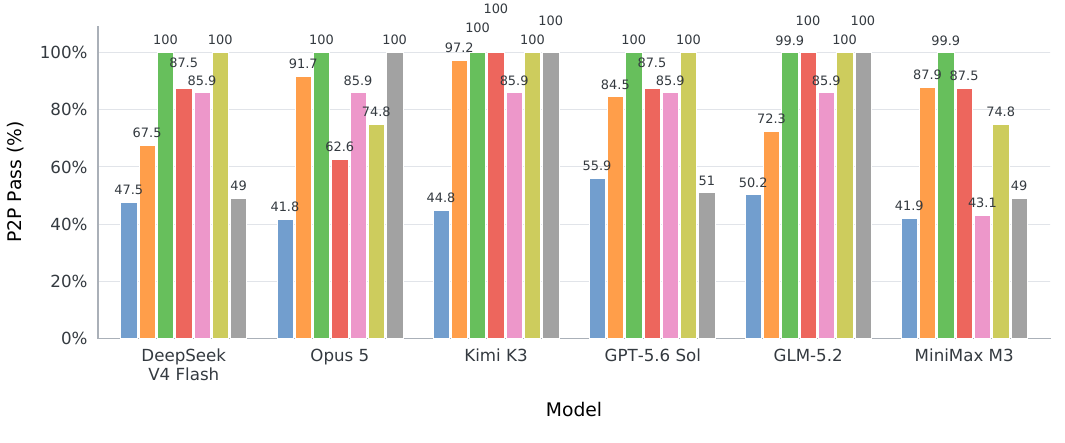}
    \caption{mini-SWE-agent}
    \label{fig:pl-p2p-mini-swe-agent}
  \end{subfigure}\par

  \caption{P2P Pass\% by programming language.}
  \label{fig:pl-p2p}
\end{figure}

Figures~\ref{fig:pl-resolved}--\ref{fig:pl-p2p} break down the four effectiveness metrics by programming language across the five scaffolds. Each figure uses the same language colors and reports the models evaluated with each scaffold. Tasks use the project languages listed in Table~\ref{tab:lolbench-projects}. C and C++ tasks are grouped as C/C++.

\textbf{Modular development tasks remain hard for each programming language.}
Fig.~\ref{fig:pl-resolved} shows that the median resolved rate across the 28 agents is zero in six of the seven language groups. Even the strongest agent resolves only 32.0\% of Python tasks, while the best rates for Java, Go, Rust, and C/C++ range from 11.1\% to 14.3\%, and no agent resolves a TypeScript task. Section-level progress is also limited: completed rates, averaged over the 28 agents within each language group, range from 4.0\% to 13.8\% across the groups other than C\# in Fig.~\ref{fig:pl-completed}. Although some agents resolve all C\# tasks, this subset contains only two tasks, and 20 of the 28 agents resolve neither. Therefore, the difficulty of \bench extends across programming languages.

\textbf{Current agents perform better on Python, C/C++, and Rust.}
This advantage is most evident in Fig.~\ref{fig:pl-f2p}. Averaging the language-specific F2P pass rates equally over all 28 agents yields 31.7\% for Python, 48.5\% for C/C++, and 46.3\% for Rust, compared with 6.0\% for Java, 13.5\% for Go, and 21.7\% for TypeScript. This comparison shows that current agents are more capable of handling modular development tasks in Python, C/C++, and Rust, rather than Java, Go, and TypeScript. 

\textbf{Scaffold rankings vary across different languages.} In Figures~\ref{fig:pl-resolved} and~\ref{fig:pl-f2p}, we find that with Opus 5, Claude Code achieves a Python resolved rate of 32.0\% and an F2P pass rate of 78.5\%, but OpenCode achieves only 8.0\% and 51.7\%, respectively. The ordering reverses for C/C++: OpenCode resolves 14.3\% of tasks and passes 65.4\% of F2P tests, while Claude Code resolves none and passes 59.4\% of F2P tests. Thus, for Opus 5, OpenCode achieves higher resolved and F2P pass rates on C/C++ tasks, while Claude Code achieves higher rates on Python tasks. The same model can therefore benefit from different scaffolds across these tasks with different languages.

\clearpage

\section{Phase Distribution on Token Usage}\label{app:effi}

\begin{figure}[t]
    \centering
    \includegraphics[width=\textwidth]{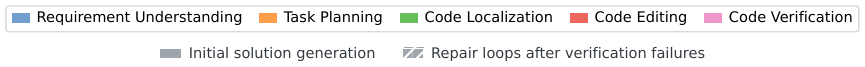}
    \par\smallskip
    \begin{subfigure}[t]{0.49\textwidth}
        \centering
        \includegraphics[width=\linewidth]{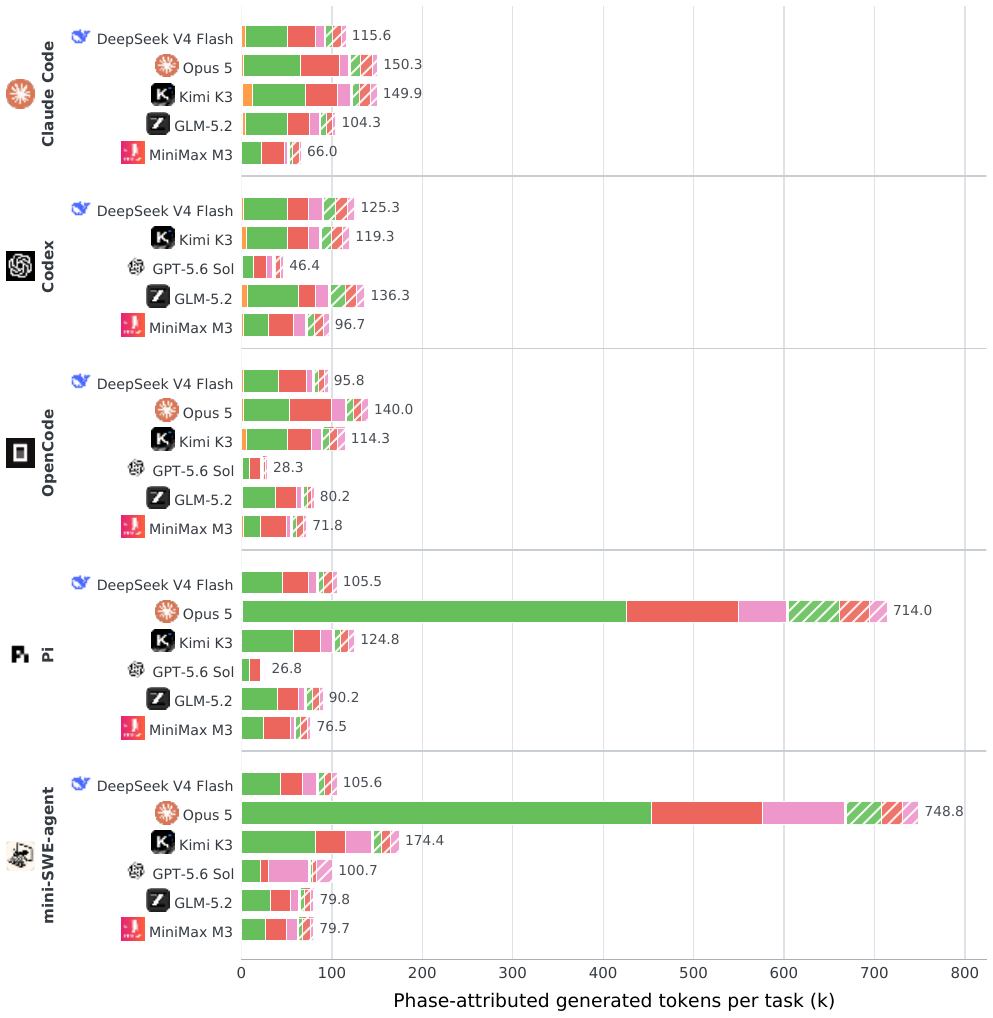}
        \caption{Generated tokens (thousands).}
        \label{fig:generated-token-distribution}
    \end{subfigure}\hfill
    \begin{subfigure}[t]{0.49\textwidth}
        \centering
        \includegraphics[width=\linewidth]{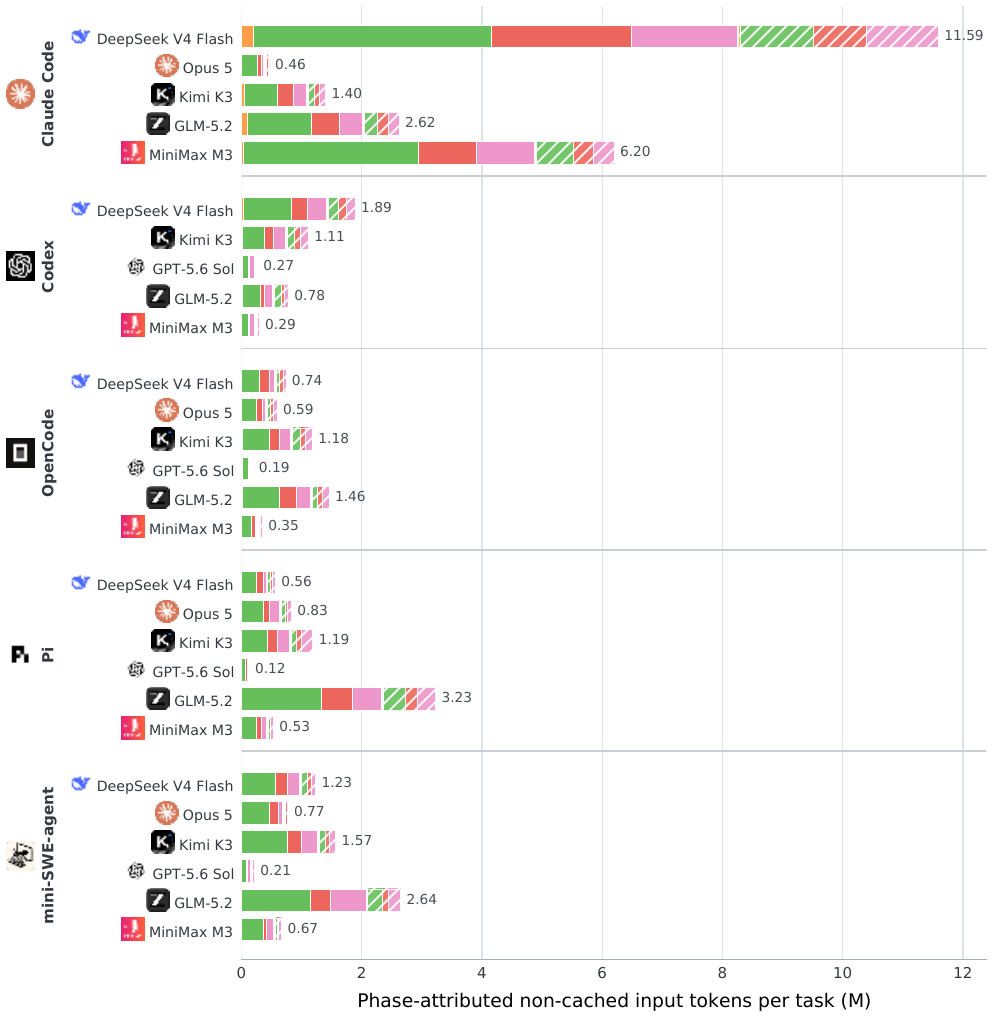}
        \caption{Non-cached input tokens (millions).}
        \label{fig:input-token-distribution}
    \end{subfigure}
    \caption{Phase-attributed token usage per attempt across 28 agents. Non-cached input excludes cache reads but includes cache writes. Bar labels sum the five colored phases and omit usage without phase attribution. Colors denote phases, solid segments denote initial solution generation, and hatched segments denote repair loops after verification failures.}
    \label{fig:token-distribution}
\end{figure}

\textbf{Current agents consume about 11$\times$ as many non-cached input tokens as generated tokens.}
We report the non-cached input token usage instead of overall input token usage to avoid duplicate counts of input contexts such as the system prompts. Averaging over all attempts with recorded usage, each attempt costs 1.60 million non-cached input tokens and 145.8 thousand generated tokens, yielding a ratio of 11.0$\times$. Computing the ratio of mean non-cached input tokens to mean generated tokens separately for each agent yields a median of 7.9$\times$ across the 28 agents. Given the complex nature of tasks in \bench, current agents typically spend more tokens understanding and exploring the EP and codebase than editing the code.

\textbf{Code localization dominates non-cached input and generated token usage.}
Code localization has the largest macro-averaged shares, accounting for 52.4\% of non-cached input tokens and 45.4\% of generated tokens in Fig.~\ref{fig:token-distribution}. It is the largest phase in non-cached input token usage for 26 of the 28 agents and in generated token usage for 19. Code editing accounts for smaller macro-averaged ratios of 19.8\% and 35.1\%, respectively. Within repair loops, code localization remains the largest phase for non-cached input tokens at 39.5\%. The overall localization shares reinforce the major efficiency bottleneck identified in Fig.~\ref{fig:phase}.

\textbf{Code verification is the second-largest phase of non-cached input token usage.}
Code verification has a macro-averaged share of 26.4\% of non-cached input tokens, exceeding code editing at 19.8\% in Fig.~\ref{fig:input-token-distribution}.
It ranks second for 18 of the 28 agents and first for two. Its macro-averaged share of non-cached input usage rises from 23.5\% during initial solution generation to 36.6\% during repair loops, indicating the frequent need for code verification during revision. This demonstrates the difficulty of modular development tasks, in which current agents have to spend a large token budget on verifying their solutions against the complex dependencies in large software systems.

\begin{table*}[t]
  \centering
  \caption{Failure phases, failure modes, and detection methods for the 24 modes observed across the 28 evaluated agents on \bench.}
  \label{tab:failure-phase-rootcause}
  \begingroup
  \footnotesize
  \setlength{\tabcolsep}{4pt}
  \renewcommand{\arraystretch}{1.0}
  \begin{tabular}{@{}p{0.17\textwidth}p{0.28\textwidth}p{0.29\textwidth}p{0.17\textwidth}@{}}
    \toprule
    \textbf{Failure Phase}
      & \textbf{Failure Mode}
      & \textbf{Description}
      & \textbf{Detection Method} \\
    \midrule
    \multirow[t]{4}{0.17\textwidth}{Requirement Understanding}
      & \texttt{misread\_\allowbreak{}requirement}
      & Interprets the stated requirement incorrectly.
      & LLM-based \\
      & \texttt{missed\_\allowbreak{}constraint}
      & Omits an explicit constraint from the reasoning or solution.
      & LLM-based \\
      & \texttt{incomplete\_\allowbreak{}requirement\_\allowbreak{}coverage}
      & Addresses only part of the requested behavior.
      & LLM-based \\
      & \texttt{scope\_\allowbreak{}misjudged}
      & Chooses a boundary for the change that is too narrow or too broad.
      & LLM-based \\
    \midrule
    \multirow[t]{2}{0.17\textwidth}{Task Planning}
      & \texttt{flawed\_\allowbreak{}plan}
      & Forms a plan that cannot fully satisfy the requirement.
      & LLM-based \\
      & \texttt{abandoned\_\allowbreak{}plan}
      & Stops following the stated plan before completing it.
      & LLM-based \\
    \midrule
    \multirow[t]{3}{0.17\textwidth}{Code Localization}
      & \texttt{missed\_\allowbreak{}relevant\_\allowbreak{}file}
      & Fails to inspect or modify a file needed for the change.
      & Pre-defined rules \\
      & \texttt{wrong\_\allowbreak{}file\_\allowbreak{}localization}
      & Focuses edits on files unrelated to the required change.
      & Pre-defined rules \\
      & \texttt{cross\_\allowbreak{}module\_\allowbreak{}context\_\allowbreak{}missing}
      & Misses required context or dependencies across modules.
      & Pre-defined rules \\
    \midrule
    \multirow[t]{4}{0.17\textwidth}{Code Editing}
      & \texttt{incorrect\_\allowbreak{}patch}
      & Changes relevant code, but implements the required behavior incorrectly.
      & Pre-defined rules \\
      & \texttt{relevant\_\allowbreak{}change\_\allowbreak{}omitted}
      & Leaves out a required change in relevant code.
      & Pre-defined rules \\
      & \texttt{no\_\allowbreak{}patch\_\allowbreak{}produced}
      & Completes the attempt without producing a source patch.
      & Pre-defined rules \\
      & \texttt{gave\_\allowbreak{}up\_\allowbreak{}early}
      & Stops the attempt before producing an adequate fix.
      & Pre-defined rules \\
    \midrule
    \multirow[t]{4}{0.17\textwidth}{Code Verification}
      & \texttt{no\_\allowbreak{}validation\_\allowbreak{}attempted}
      & Runs no recognized validation of the patch.
      & Pre-defined rules \\
      & \texttt{no\_\allowbreak{}test\_\allowbreak{}after\_\allowbreak{}final\_\allowbreak{}edit}
      & Makes a final edit after the last test and does not retest.
      & Pre-defined rules \\
      & \texttt{authored\_\allowbreak{}test\_\allowbreak{}never\_\allowbreak{}run}
      & Writes or modifies a test but never executes it.
      & Pre-defined rules \\
      & \texttt{test\_\allowbreak{}result\_\allowbreak{}misinterpreted}
      & Interprets the test output or status incorrectly.
      & Pre-defined rules \\
    \midrule
    \multirow[t]{4}{0.17\textwidth}{Self-Repair}
      & \texttt{repair\_\allowbreak{}not\_\allowbreak{}reverified}
      & Makes a repair after a failure but does not verify it.
      & Pre-defined rules \\
      & \texttt{repeated\_\allowbreak{}ineffective\_\allowbreak{}attempt}
      & Repeats repair attempts that do not resolve the failure.
      & Pre-defined rules \\
      & \texttt{retry\_\allowbreak{}without\_\allowbreak{}change}
      & Retries a failed action without an intervening change.
      & Pre-defined rules \\
      & \texttt{reverted\_\allowbreak{}own\_\allowbreak{}change}
      & Undoes an earlier edit while attempting to repair the solution.
      & Pre-defined rules \\
    \midrule
    \multirow[t]{3}{0.17\textwidth}{Tool Use}
      & \texttt{tool\_\allowbreak{}failure\_\allowbreak{}not\_\allowbreak{}retried}
      & Leaves a failed tool action unretried.
      & Pre-defined rules \\
      & \texttt{repeated\_\allowbreak{}patch\_\allowbreak{}apply\_\allowbreak{}failure}
      & Repeatedly submits a patch that the tool cannot apply.
      & Pre-defined rules \\
      & \texttt{hallucinated\_\allowbreak{}path\_\allowbreak{}unrecovered}
      & Uses a nonexistent path and never recovers from the error.
      & Pre-defined rules \\
    \bottomrule
  \end{tabular}
  \endgroup
\end{table*}

\section{Failure Mode Analysis}\label{app:failure}

\subsection{Method}

\textbf{Failure taxonomy.}
The analysis decomposes failures into seven phases and their associated failure modes. Table~\ref{tab:failure-phase-rootcause} presents the mappings of failure phases and failure modes. To ensure a stable and mostly objective analysis, we implement predefined rules to detect 18 failure modes. For the 6 remaining failure modes that require semantic understanding, we use an LLM judge.

\textbf{Objective evidence extraction.}
We normalize scaffold-specific logs into a common sequence of events such as messages, tool calls, and results. We then analyze the sequences using pre-defined deterministic rules and identify potential evidence. This evidence does not directly establish a failure mode; instead, it is only a suspect candidate pending further analysis. For example, a file in the reference solution that is neither read nor edited provides evidence of a localization miss, whereas a file that is read but left unmodified supports a relevant-change omission in Code Editing. Editing only part of the reference file set supports a cross-module localization failure, and editing all reference files while still failing verification supports an incorrect-patch failure. We collect all evidence identified by pre-defined rules for further analysis.

\textbf{Semantic judgments.}
Requirement understanding and task planning failures require semantic understanding, so we assess them with an outcome-blinded LLM judge using a closed set of labels. The judge receives the task and visible process evidence, but neither the reference solution nor the outcome, and must cite supporting trajectory steps. We admit only medium- or high-confidence judgments with valid evidence citations, and we exclude low-confidence and unparseable judgments. We use GPT-5.6 Luna~\citep{gpt56luna} with the pro reasoning mode (OpenRouter id \texttt{openai/gpt-5.6-luna-pro}) and temperature $0$ to perform the semantic judgment.

\paragraph{Judge prompts.}
The fixed system prompt below includes the complete label set and response format.


\begin{quote}\small

\emph{You are a rigorous evaluator of an AI software-engineering agent's PROCESS on a
bug-fix / feature task. You assess exactly two capabilities:}

\begin{itemize}
  \setlength{\itemsep}{2pt}
  \setlength{\parskip}{0pt}
  \item \emph{\texttt{requirement\_understanding}: did the agent correctly grasp WHAT
    the issue asks for?}
  \item \emph{\texttt{task\_planning}: did the agent choose a viable APPROACH / sequence
    to get there?}
\end{itemize}

\emph{For each capability, choose exactly ONE label from this closed set (or
\texttt{none}):}

\emph{\texttt{requirement\_understanding}:}

\begin{itemize}
  \setlength{\itemsep}{2pt}
  \setlength{\parskip}{0pt}
  \item \emph{\texttt{misread\_requirement}: Misunderstood what the issue is actually
    asking for.}
  \item \emph{\texttt{missed\_constraint}: Overlooked an explicit constraint, edge case,
    or requirement stated in the issue.}
  \item \emph{\texttt{incomplete\_requirement\_coverage}: Understood and addressed only
    part of a multi-part requirement.}
  \item \emph{\texttt{scope\_misjudged}: Tackled a meaningfully broader or narrower
    problem than the issue asks.}
  \item \emph{\texttt{none}: the agent's understanding looks adequate from the
    evidence.}
\end{itemize}

\emph{\texttt{task\_planning}:}

\begin{itemize}
  \setlength{\itemsep}{2pt}
  \setlength{\parskip}{0pt}
  \item \emph{\texttt{flawed\_plan}: Adopted an approach that cannot satisfy the stated
    requirement.}
  \item \emph{\texttt{abandoned\_plan}: Stated a reasonable plan, then diverged from it
    without justification.}
  \item \emph{\texttt{none}: the agent's planning looks adequate from the evidence.}
\end{itemize}

\emph{CRITICAL RULES:}

\begin{enumerate}
  \setlength{\itemsep}{2pt}
  \setlength{\parskip}{0pt}
  \item \emph{You are NOT told whether the task ultimately passed or failed, and you are
    NOT shown the reference solution. Judge ONLY the agent's process from the evidence
    below. Do not try to guess the outcome.}
  \item \emph{Only flag an error you can support with SPECIFIC step numbers from the
    evidence. If you cannot point to evidence, answer \texttt{none}.}
  \item \emph{Code \emph{correctness} is judged by a separate system. Do NOT label a
    misunderstanding just because you suspect the final code is wrong. A
    different-but-valid approach is NOT an error.}
  \item \emph{Some agents' internal reasoning is hidden/encrypted (noted below). When
    reasoning is hidden, do NOT infer that the agent ``failed to consider'' something from
    its absence --- judge from visible actions and messages, and lower your confidence.}
  \item \emph{Be conservative. \texttt{none} is the right answer whenever the agent's
    comprehension/plan looks adequate. Reserve \texttt{low} confidence for genuinely
    ambiguous cases.}
  \item \emph{The requested scope is the whole TASK REQUIREMENT unless that text
    explicitly narrows it. If the agent says the requested work is too large/infeasible
    and therefore plans or submits only a subset, stubs, documentation, or a ``minimal''
    approximation, that is direct evidence of a narrower scope
    (\texttt{scope\_misjudged}) and usually a plan that cannot satisfy the task
    (\texttt{flawed\_plan}). Do not excuse deliberate partial implementation merely
    because the agent describes its resource limits.}
  \item \emph{Conversely, ordinary sequencing language such as ``first'', ``focus on X'', or
    ``start with X'' is not scope reduction when the agent continues toward the remaining
    requirements.}
  \item \emph{The user message may contain a HIGHLIGHTED SCOPE/PLAN RISK section.
    Classify every highlighted statement as either temporary sequencing or a
    submission-scope reduction. A coherent, well-tested partial implementation is still
    a \texttt{flawed\_plan} because it cannot satisfy the full task. Returning
    \texttt{none} despite an explicit partial/subset/stub decision is allowed only when
    later cited evidence shows that the agent rescinded that decision and completed the
    remaining requested scope. State that disposition in the rationale; ``reasonable
    given time/constraints'' is not a valid excuse.}
\end{enumerate}

\emph{Respond with a SINGLE JSON object and nothing else, in exactly this shape:}

{\fontencoding{T1}\ttfamily\small\raggedright
\{\\
\quad "requirement\_understanding": \{\\
\qquad "error\_type": "\textless{}label or none\textgreater{}",\\
\qquad "confidence": "high\textbar{}medium\textbar{}low",\\
\qquad "evidence\_steps": [\textless{}int\textgreater{}, ...],\\
\qquad "rationale": "\textless{}=240 chars"\\
\quad \},\\
\quad "task\_planning": \{\\
\qquad "error\_type": "\textless{}label or none\textgreater{}",\\
\qquad "confidence": "high\textbar{}medium\textbar{}low",\\
\qquad "evidence\_steps": [\textless{}int\textgreater{}, ...],\\
\qquad "rationale": "\textless{}=240 chars"\\
\quad \}\\
\}\par}

\end{quote}

The accompanying user-message template is shown below. Angle-bracketed fields are replaced with the task specification, visible trajectory evidence, and coverage metadata; the step rows are repeated for the selected evidence. The benchmark-contract block, hidden-reasoning note, and highlighted scope/plan statements are included only when applicable.


\begin{quote}\small

\emph{TASK REQUIREMENT (the work requested):}
\par
\emph{\textless{}task\_requirement\textgreater{}}

\emph{TASK COVERAGE: \textless{}task\_coverage\_mode\textgreater{} --- shown
\textless{}shown\_characters\textgreater{} of \textless{}total\_characters\textgreater{}
characters.}

\emph{BENCHMARK CONTRACT (not the feature specification):}
\par
\emph{\textless{}benchmark\_instructions\textgreater{}}

\emph{AGENT PROCESS EVIDENCE (numbered steps; cite these step numbers):}
\par
{\raggedright\emph{[\textless{}step\_id\textgreater{}]
\textless{}message\_or\_action\_type\textgreater{}
\textless{}tool\_if\_any\textgreater{}: \textless{}visible\_process\_text\textgreater{}}\par}

\emph{NOTE: this agent's internal reasoning is HIDDEN/encrypted. Judge only from the
visible messages and actions above, and lower confidence accordingly.}

\emph{HIGHLIGHTED SCOPE/PLAN RISK STATEMENTS (also present above):}
\par
{\raggedright\emph{[\textless{}risk\_step\_id\textgreater{}]
\textless{}message\_or\_action\_type\textgreater{}:
\textless{}scope\_or\_plan\_statement\textgreater{}}\par}
\par
\emph{For each highlighted statement, decide whether it is temporary sequencing or a
deliberate reduction of the submitted scope. If you return \texttt{none}, cite later
evidence that the agent rescinded and completed each reduction.}

\emph{PROCESS COVERAGE: \textless{}process\_coverage\_mode\textgreater{} --- shown
\textless{}shown\_rows\textgreater{} of \textless{}total\_eligible\_rows\textgreater{}
eligible rows. Rows containing explicit scope-reduction language were prioritized. Do
not infer a missing plan solely from omitted rows.}

\emph{Return the JSON verdict now.}

\end{quote}

\textbf{Confidence assignment.}
Each candidate first receives a categorical confidence label: high, medium, or low. For candidates identified by deterministic rules, the triggering predicate specifies the label from the observed evidence. For semantic judgments, we use the confidence label supplied by the single judge response, together with its error label and supporting trajectory steps. The judge is instructed to lower confidence when reasoning is hidden and to use low confidence for ambiguous cases. An error judgment without citations to steps visible in the judge input is downgraded to low confidence before aggregation. Only high- and medium-confidence candidates enter the attribution calculation, with confidence scores of $1.0$ and $0.6$, respectively. A failed task can exhibit multiple failure modes. For task $t$, let $E_t$ denote its admitted candidates and $c_{ti}$ the confidence score of candidate $i$. We normalize these scores within each task to obtain each candidate's fractional attribution $w_{ti}$:
\begin{equation}
    w_{ti} = \frac{c_{ti}}
    {\sum_{j\in E_t} c_{tj}},
    \qquad \sum_{i\in E_t} w_{ti}=1.
    \label{eq:failure-attribution}
\end{equation}
Normalization allocates exactly one unit of attribution per failed task, irrespective of the number of detected modes.

\textbf{Aggregation.}
Let $F_a$ be the evaluable failed tasks for agent $a$, and let $p_i$ and $m_i$ denote the phase and failure mode of candidate $i$. The attribution of failure mode
$m$ in phase $p$, and the total attribution to that phase, are
\begin{equation}
    M_{a,p,m} = \sum_{t\in F_a}
    \sum_{\substack{i\in E_t:\,p_i=p,\,m_i=m}} w_{ti},
    \qquad A_{a,p}=\sum_m M_{a,p,m}.
    \label{eq:failure-aggregation}
\end{equation}
The per-phase failure mode tables report $M_{a,p,m}$, while the phase summary reports $A_{a,p}$. Thus, failure mode attribution values sum to their phase total, and $\sum_p A_{a,p}=|F_a|$. We use largest-remainder rounding to one decimal place, first preserving each agent's total across phases and then each displayed phase total across failure modes. These values are fractional allocations of failed tasks, not counts of tasks with a particular error flag. The resulting decomposition provides a reproducible diagnosis of observable failure patterns.

\subsection{Results}

\textbf{Requirement Understanding.}
Table~\ref{tab:failure-requirement-understanding} shows that scope misjudgment receives the largest or joint-largest attribution for 27 agents at the requirement understanding phase.
For Claude Code with DeepSeek V4 Flash, it accounts for 8.9 of the phase's 11.2 attribution units. Missed constraints and incomplete requirement coverage are secondary contributors, while direct requirement misreading receives at most 1.0 units per agent. This pattern suggests that the major difficulty is determining the full extent of the requested change, including its constraints and affected functionality.

\textbf{Task Planning.}
Table~\ref{tab:failure-task-planning} shows that nearly all planning attribution comes from flawed plans. For example, this mode accounts for all 10.5 planning attribution units for Claude Code with DeepSeek V4 Flash and all 10.1 units for Pi with GLM-5.2. Abandoned plans receive at most 0.8 units per agent. These judgments suggest that agents often formulate strategies that cannot fully satisfy the requirements. Effective planning therefore requires checking whether the proposed steps collectively cover the requested behavior and its implementation dependencies.

\textbf{Code Localization.}
Missing cross-module context is the largest localization failure mode for all 28 agents in Table~\ref{tab:failure-code-localization}. It accounts for 41.4 of 41.7 attribution units for Codex with Kimi K3, whereas missed relevant files and wrong-file localization receive at most 2.1 and 0.5 units, respectively, across agents. The dominant pattern is incomplete coverage of the files and modules involved in the reference solutions. This suggests that agents can reach part of the relevant code but struggle to identify the dependencies needed for a complete implementation.

\textbf{Code Editing.}
Table~\ref{tab:failure-code-editing} identifies incorrect patches as the largest editing contributor for 22 agents and omitted relevant changes as the largest for the remaining six. Both can contribute substantially for a single agent: Pi with GPT-5.6 Sol receives 19.4 units for incorrect patches and 20.5 for omitted changes, together accounting for all 39.9 editing units. Missing patches and early abandonment contribute comparatively little overall. These results highlight two implementation difficulties: correctly modifying relevant code and carrying identified requirements through to all necessary edits.

\textbf{Code Verification.}
The absence of recognized validation is the leading verification failure mode for all 28 agents in Table~\ref{tab:failure-code-verification}. For mini-SWE-agent with DeepSeek V4 Flash, it accounts for 17.5 of 24.9 attribution units at the verification phase, followed by 6.3 units for missing tests after the final edit. Authored tests that are never executed and misinterpreted test results receive smaller aggregate attribution. The distribution highlights gaps in executing validation and ensuring that its results apply to the final submitted solution.

\textbf{Self Repair.}
Table~\ref{tab:failure-self-repair-loop} shows that unverified repairs and repeated ineffective attempts dominate repair attribution. Their relative importance varies across agents: repeated ineffective attempts are the largest contributor for all six agents using Pi, while unverified repairs account for 14.6 of 18.2 repair units for mini-SWE-agent with GPT-5.6 Sol. Self-reversions and unchanged retries contribute less overall. These patterns suggest that effective repair requires both revising the implementation strategy in response to failure evidence and verifying whether subsequent changes actually resolve the observed problem.

\textbf{Tool Use.}
Tool-use attribution ranges from 0.0 to 1.5 units per agent in Table~\ref{tab:failure-tool-use}. Unrecovered invalid paths receive the largest aggregate attribution, although other mechanical errors dominate particular agents. Unretried tool failures account for 1.3 of 1.4 units for Pi with Opus 5, while repeated patch-application failures account for 0.9 of 1.5 units for Codex with MiniMax M3. These results show that failures to recover from tool errors contribute relatively little to the overall failure attribution.

\begin{table*}[t]
  \centering
  \caption{Failure mode distribution for Requirement Understanding failures across the 28 agents. The phase column reproduces the corresponding failure attribution in Table~\ref{tab:failure-composition} and equals the sum of failure mode values in every row. Darker backgrounds indicate a larger share of total failure attribution for each agent's solution failures.}
  \label{tab:failure-requirement-understanding}
  \begingroup
  \definecolor{lbFailureHeat}{HTML}{729ECE}
  \newcommand{\lbModelLogo}[1]{%
    \raisebox{-0.28ex}{\includegraphics[height=1.22em,keepaspectratio]{figures/#1}}}
  \newcommand{\lbAgentLogo}[1]{%
    \raisebox{-0.32ex}{\includegraphics[height=1.38em,keepaspectratio]{figures/#1}}}
  \newcommand{\lbHarnessCell}[2]{%
    \makecell[c]{\lbAgentLogo{#1}\\[-0.15ex]\textsc{#2}}}
  \newcommand{\lbClaudeOpus}{\lbModelLogo{claude.png}\,Opus 5}
  \newcommand{\lbGPT}{\lbModelLogo{openai.png}\,GPT-5.6 Sol}
  \newcommand{\lbKimiModel}{\lbModelLogo{kimi.png}\,Kimi K3}
  \newcommand{\lbGLMModel}{\lbModelLogo{glm.png}\,GLM-5.2}
  \newcommand{\lbMiniMaxModel}{\lbModelLogo{minimax.png}\,MiniMax M3}
  \newcommand{\lbDeepSeekModel}{\lbModelLogo{deepseek.png}\,DeepSeek V4 Flash}
  \small
  \setlength{\tabcolsep}{3.0pt}
  \renewcommand{\arraystretch}{0.96}
  \scalebox{0.8}{%
  \begin{tabular}{@{}clr@{\hspace{7pt}}rrrr@{}}
    \toprule
    \multirow{2}{*}{\textbf{Scaffold}}
      & \multirow{2}{*}{\textbf{Model}}
      & \multicolumn{1}{c}{\textbf{Phase}}
      & \multicolumn{4}{c}{\textbf{Failure Modes}} \\
    \cmidrule(lr){3-3}\cmidrule(lr){4-7}
      &
      & \makecell{\textbf{Requirement}\\\textbf{Understanding}}
      & \makecell{\textbf{Scope}\\\textbf{Misjudged}}
      & \makecell{\textbf{Missed}\\\textbf{Constraint}}
      & \makecell{\textbf{Incomplete}\\\textbf{Coverage}}
      & \makecell{\textbf{Misread}\\\textbf{Requirement}} \\
    \midrule
      \multirow[c]{5}{*}{\lbHarnessCell{claude.png}{Claude Code}} & \lbDeepSeekModel & \cellcolor{lbFailureHeat!11}11.2 & \cellcolor{lbFailureHeat!9}8.9 & \cellcolor{lbFailureHeat!1}1.1 & \cellcolor{lbFailureHeat!1}1.0 & \cellcolor{lbFailureHeat!1}0.2 \\
       & \lbClaudeOpus & \cellcolor{lbFailureHeat!9}6.9 & \cellcolor{lbFailureHeat!5}3.7 & \cellcolor{lbFailureHeat!2}1.8 & \cellcolor{lbFailureHeat!2}1.4 & \cellcolor{lbFailureHeat!0}0.0 \\
       & \lbKimiModel & \cellcolor{lbFailureHeat!6}5.1 & \cellcolor{lbFailureHeat!3}2.8 & \cellcolor{lbFailureHeat!1}1.2 & \cellcolor{lbFailureHeat!1}0.8 & \cellcolor{lbFailureHeat!1}0.3 \\
       & \lbGLMModel & \cellcolor{lbFailureHeat!8}8.2 & \cellcolor{lbFailureHeat!7}6.4 & \cellcolor{lbFailureHeat!1}0.7 & \cellcolor{lbFailureHeat!1}1.1 & \cellcolor{lbFailureHeat!0}0.0 \\
       & \lbMiniMaxModel & \cellcolor{lbFailureHeat!10}9.4 & \cellcolor{lbFailureHeat!7}6.6 & \cellcolor{lbFailureHeat!1}0.5 & \cellcolor{lbFailureHeat!2}1.6 & \cellcolor{lbFailureHeat!1}0.7 \\
    \midrule
      \multirow[c]{5}{*}{\lbHarnessCell{codex.png}{Codex}} & \lbDeepSeekModel & \cellcolor{lbFailureHeat!5}4.5 & \cellcolor{lbFailureHeat!3}2.9 & \cellcolor{lbFailureHeat!1}0.7 & \cellcolor{lbFailureHeat!1}0.7 & \cellcolor{lbFailureHeat!1}0.2 \\
       & \lbKimiModel & \cellcolor{lbFailureHeat!2}2.1 & \cellcolor{lbFailureHeat!1}1.0 & \cellcolor{lbFailureHeat!1}0.3 & \cellcolor{lbFailureHeat!1}0.8 & \cellcolor{lbFailureHeat!0}0.0 \\
       & \lbGPT & \cellcolor{lbFailureHeat!1}1.2 & \cellcolor{lbFailureHeat!1}0.2 & \cellcolor{lbFailureHeat!1}0.2 & \cellcolor{lbFailureHeat!1}0.8 & \cellcolor{lbFailureHeat!0}0.0 \\
       & \lbGLMModel & \cellcolor{lbFailureHeat!3}3.3 & \cellcolor{lbFailureHeat!2}2.1 & \cellcolor{lbFailureHeat!0}0.0 & \cellcolor{lbFailureHeat!1}1.2 & \cellcolor{lbFailureHeat!0}0.0 \\
       & \lbMiniMaxModel & \cellcolor{lbFailureHeat!6}6.3 & \cellcolor{lbFailureHeat!6}5.6 & \cellcolor{lbFailureHeat!1}0.7 & \cellcolor{lbFailureHeat!0}0.0 & \cellcolor{lbFailureHeat!0}0.0 \\
    \midrule
      \multirow[c]{6}{*}{\lbHarnessCell{opencode.png}{OpenCode}} & \lbDeepSeekModel & \cellcolor{lbFailureHeat!8}8.2 & \cellcolor{lbFailureHeat!6}6.2 & \cellcolor{lbFailureHeat!1}0.8 & \cellcolor{lbFailureHeat!1}1.1 & \cellcolor{lbFailureHeat!1}0.1 \\
       & \lbClaudeOpus & \cellcolor{lbFailureHeat!7}5.7 & \cellcolor{lbFailureHeat!4}3.0 & \cellcolor{lbFailureHeat!2}1.7 & \cellcolor{lbFailureHeat!1}1.0 & \cellcolor{lbFailureHeat!0}0.0 \\
       & \lbKimiModel & \cellcolor{lbFailureHeat!6}5.6 & \cellcolor{lbFailureHeat!3}3.2 & \cellcolor{lbFailureHeat!1}0.9 & \cellcolor{lbFailureHeat!2}1.5 & \cellcolor{lbFailureHeat!0}0.0 \\
       & \lbGPT & \cellcolor{lbFailureHeat!2}0.8 & \cellcolor{lbFailureHeat!1}0.6 & \cellcolor{lbFailureHeat!0}0.0 & \cellcolor{lbFailureHeat!0}0.0 & \cellcolor{lbFailureHeat!1}0.2 \\
       & \lbGLMModel & \cellcolor{lbFailureHeat!8}7.8 & \cellcolor{lbFailureHeat!6}6.3 & \cellcolor{lbFailureHeat!1}0.5 & \cellcolor{lbFailureHeat!1}1.0 & \cellcolor{lbFailureHeat!0}0.0 \\
       & \lbMiniMaxModel & \cellcolor{lbFailureHeat!8}8.3 & \cellcolor{lbFailureHeat!6}5.6 & \cellcolor{lbFailureHeat!1}1.3 & \cellcolor{lbFailureHeat!1}1.1 & \cellcolor{lbFailureHeat!1}0.3 \\
    \midrule
      \multirow[c]{6}{*}{\lbHarnessCell{pi.png}{Pi}} & \lbDeepSeekModel & \cellcolor{lbFailureHeat!11}10.2 & \cellcolor{lbFailureHeat!9}8.7 & \cellcolor{lbFailureHeat!1}0.3 & \cellcolor{lbFailureHeat!1}1.0 & \cellcolor{lbFailureHeat!1}0.2 \\
       & \lbClaudeOpus & \cellcolor{lbFailureHeat!9}7.1 & \cellcolor{lbFailureHeat!6}5.0 & \cellcolor{lbFailureHeat!2}1.5 & \cellcolor{lbFailureHeat!1}0.6 & \cellcolor{lbFailureHeat!0}0.0 \\
       & \lbKimiModel & \cellcolor{lbFailureHeat!4}4.2 & \cellcolor{lbFailureHeat!3}2.9 & \cellcolor{lbFailureHeat!1}0.3 & \cellcolor{lbFailureHeat!1}1.0 & \cellcolor{lbFailureHeat!0}0.0 \\
       & \lbGPT & \cellcolor{lbFailureHeat!1}1.4 & \cellcolor{lbFailureHeat!1}0.6 & \cellcolor{lbFailureHeat!1}0.5 & \cellcolor{lbFailureHeat!1}0.3 & \cellcolor{lbFailureHeat!0}0.0 \\
       & \lbGLMModel & \cellcolor{lbFailureHeat!11}10.4 & \cellcolor{lbFailureHeat!8}7.9 & \cellcolor{lbFailureHeat!2}1.9 & \cellcolor{lbFailureHeat!1}0.6 & \cellcolor{lbFailureHeat!0}0.0 \\
       & \lbMiniMaxModel & \cellcolor{lbFailureHeat!11}10.4 & \cellcolor{lbFailureHeat!6}6.1 & \cellcolor{lbFailureHeat!2}2.3 & \cellcolor{lbFailureHeat!2}1.6 & \cellcolor{lbFailureHeat!1}0.4 \\
    \midrule
      \multirow[c]{6}{*}{\lbHarnessCell{mini-swe-agent.png}{mini-SWE-agent}} & \lbDeepSeekModel & \cellcolor{lbFailureHeat!3}3.0 & \cellcolor{lbFailureHeat!1}1.3 & \cellcolor{lbFailureHeat!1}1.0 & \cellcolor{lbFailureHeat!1}0.4 & \cellcolor{lbFailureHeat!1}0.3 \\
       & \lbClaudeOpus & \cellcolor{lbFailureHeat!1}0.6 & \cellcolor{lbFailureHeat!1}0.2 & \cellcolor{lbFailureHeat!1}0.2 & \cellcolor{lbFailureHeat!1}0.2 & \cellcolor{lbFailureHeat!0}0.0 \\
       & \lbKimiModel & \cellcolor{lbFailureHeat!1}1.1 & \cellcolor{lbFailureHeat!1}1.1 & \cellcolor{lbFailureHeat!0}0.0 & \cellcolor{lbFailureHeat!0}0.0 & \cellcolor{lbFailureHeat!0}0.0 \\
       & \lbGPT & \cellcolor{lbFailureHeat!2}1.5 & \cellcolor{lbFailureHeat!1}1.2 & \cellcolor{lbFailureHeat!1}0.1 & \cellcolor{lbFailureHeat!1}0.2 & \cellcolor{lbFailureHeat!0}0.0 \\
       & \lbGLMModel & \cellcolor{lbFailureHeat!4}4.1 & \cellcolor{lbFailureHeat!2}2.3 & \cellcolor{lbFailureHeat!2}1.6 & \cellcolor{lbFailureHeat!1}0.2 & \cellcolor{lbFailureHeat!0}0.0 \\
       & \lbMiniMaxModel & \cellcolor{lbFailureHeat!8}7.6 & \cellcolor{lbFailureHeat!6}5.6 & \cellcolor{lbFailureHeat!1}0.4 & \cellcolor{lbFailureHeat!1}0.6 & \cellcolor{lbFailureHeat!1}1.0 \\
    \bottomrule
  \end{tabular}
  }
  \endgroup
\end{table*}

\begin{table*}[t]
  \centering
  \caption{Failure mode distribution for Task Planning failures across the 28 agents. The phase column reproduces the corresponding failure attribution in Table~\ref{tab:failure-composition} and equals the sum of failure mode values in every row. Darker backgrounds indicate a larger share of total failure attribution for each agent's solution failures.}
  \label{tab:failure-task-planning}
  \begingroup
  \definecolor{lbFailureHeat}{HTML}{729ECE}
  \newcommand{\lbModelLogo}[1]{%
    \raisebox{-0.28ex}{\includegraphics[height=1.22em,keepaspectratio]{figures/#1}}}
  \newcommand{\lbAgentLogo}[1]{%
    \raisebox{-0.32ex}{\includegraphics[height=1.38em,keepaspectratio]{figures/#1}}}
  \newcommand{\lbHarnessCell}[2]{%
    \makecell[c]{\lbAgentLogo{#1}\\[-0.15ex]\textsc{#2}}}
  \newcommand{\lbClaudeOpus}{\lbModelLogo{claude.png}\,Opus 5}
  \newcommand{\lbGPT}{\lbModelLogo{openai.png}\,GPT-5.6 Sol}
  \newcommand{\lbKimiModel}{\lbModelLogo{kimi.png}\,Kimi K3}
  \newcommand{\lbGLMModel}{\lbModelLogo{glm.png}\,GLM-5.2}
  \newcommand{\lbMiniMaxModel}{\lbModelLogo{minimax.png}\,MiniMax M3}
  \newcommand{\lbDeepSeekModel}{\lbModelLogo{deepseek.png}\,DeepSeek V4 Flash}
  \small
  \setlength{\tabcolsep}{3.0pt}
  \renewcommand{\arraystretch}{0.96}
  \scalebox{0.8}{%
  \begin{tabular}{@{}clr@{\hspace{7pt}}rr@{}}
    \toprule
    \multirow{2}{*}{\textbf{Scaffold}}
      & \multirow{2}{*}{\textbf{Model}}
      & \multicolumn{1}{c}{\textbf{Phase}}
      & \multicolumn{2}{c}{\textbf{Failure Modes}} \\
    \cmidrule(lr){3-3}\cmidrule(lr){4-5}
      &
      & \makecell{\textbf{Task}\\\textbf{Planning}}
      & \makecell{\textbf{Flawed}\\\textbf{Plan}}
      & \makecell{\textbf{Abandoned}\\\textbf{Plan}} \\
    \midrule
      \multirow[c]{5}{*}{\lbHarnessCell{claude.png}{Claude Code}} & \lbDeepSeekModel & \cellcolor{lbFailureHeat!10}10.5 & \cellcolor{lbFailureHeat!10}10.5 & \cellcolor{lbFailureHeat!0}0.0 \\
       & \lbClaudeOpus & \cellcolor{lbFailureHeat!8}6.2 & \cellcolor{lbFailureHeat!8}6.2 & \cellcolor{lbFailureHeat!0}0.0 \\
       & \lbKimiModel & \cellcolor{lbFailureHeat!5}4.4 & \cellcolor{lbFailureHeat!5}4.4 & \cellcolor{lbFailureHeat!0}0.0 \\
       & \lbGLMModel & \cellcolor{lbFailureHeat!8}7.7 & \cellcolor{lbFailureHeat!8}7.7 & \cellcolor{lbFailureHeat!0}0.0 \\
       & \lbMiniMaxModel & \cellcolor{lbFailureHeat!10}9.2 & \cellcolor{lbFailureHeat!10}9.2 & \cellcolor{lbFailureHeat!0}0.0 \\
    \midrule
      \multirow[c]{5}{*}{\lbHarnessCell{codex.png}{Codex}} & \lbDeepSeekModel & \cellcolor{lbFailureHeat!4}3.9 & \cellcolor{lbFailureHeat!4}3.9 & \cellcolor{lbFailureHeat!0}0.0 \\
       & \lbKimiModel & \cellcolor{lbFailureHeat!2}1.7 & \cellcolor{lbFailureHeat!2}1.7 & \cellcolor{lbFailureHeat!0}0.0 \\
       & \lbGPT & \cellcolor{lbFailureHeat!1}1.0 & \cellcolor{lbFailureHeat!1}1.0 & \cellcolor{lbFailureHeat!0}0.0 \\
       & \lbGLMModel & \cellcolor{lbFailureHeat!3}2.9 & \cellcolor{lbFailureHeat!3}2.9 & \cellcolor{lbFailureHeat!0}0.0 \\
       & \lbMiniMaxModel & \cellcolor{lbFailureHeat!7}6.4 & \cellcolor{lbFailureHeat!7}6.4 & \cellcolor{lbFailureHeat!0}0.0 \\
    \midrule
      \multirow[c]{6}{*}{\lbHarnessCell{opencode.png}{OpenCode}} & \lbDeepSeekModel & \cellcolor{lbFailureHeat!9}8.5 & \cellcolor{lbFailureHeat!9}8.5 & \cellcolor{lbFailureHeat!0}0.0 \\
       & \lbClaudeOpus & \cellcolor{lbFailureHeat!6}5.1 & \cellcolor{lbFailureHeat!6}5.1 & \cellcolor{lbFailureHeat!0}0.0 \\
       & \lbKimiModel & \cellcolor{lbFailureHeat!6}6.1 & \cellcolor{lbFailureHeat!6}6.1 & \cellcolor{lbFailureHeat!0}0.0 \\
       & \lbGPT & \cellcolor{lbFailureHeat!1}0.6 & \cellcolor{lbFailureHeat!1}0.6 & \cellcolor{lbFailureHeat!0}0.0 \\
       & \lbGLMModel & \cellcolor{lbFailureHeat!8}7.5 & \cellcolor{lbFailureHeat!8}7.5 & \cellcolor{lbFailureHeat!0}0.0 \\
       & \lbMiniMaxModel & \cellcolor{lbFailureHeat!8}7.8 & \cellcolor{lbFailureHeat!8}7.8 & \cellcolor{lbFailureHeat!0}0.0 \\
    \midrule
      \multirow[c]{6}{*}{\lbHarnessCell{pi.png}{Pi}} & \lbDeepSeekModel & \cellcolor{lbFailureHeat!11}10.4 & \cellcolor{lbFailureHeat!11}10.4 & \cellcolor{lbFailureHeat!0}0.0 \\
       & \lbClaudeOpus & \cellcolor{lbFailureHeat!8}6.5 & \cellcolor{lbFailureHeat!8}6.5 & \cellcolor{lbFailureHeat!0}0.0 \\
       & \lbKimiModel & \cellcolor{lbFailureHeat!4}4.3 & \cellcolor{lbFailureHeat!4}4.3 & \cellcolor{lbFailureHeat!0}0.0 \\
       & \lbGPT & \cellcolor{lbFailureHeat!1}0.7 & \cellcolor{lbFailureHeat!1}0.7 & \cellcolor{lbFailureHeat!0}0.0 \\
       & \lbGLMModel & \cellcolor{lbFailureHeat!10}10.1 & \cellcolor{lbFailureHeat!10}10.1 & \cellcolor{lbFailureHeat!0}0.0 \\
       & \lbMiniMaxModel & \cellcolor{lbFailureHeat!10}9.8 & \cellcolor{lbFailureHeat!9}9.0 & \cellcolor{lbFailureHeat!1}0.8 \\
    \midrule
      \multirow[c]{6}{*}{\lbHarnessCell{mini-swe-agent.png}{mini-SWE-agent}} & \lbDeepSeekModel & \cellcolor{lbFailureHeat!2}1.7 & \cellcolor{lbFailureHeat!1}1.5 & \cellcolor{lbFailureHeat!1}0.2 \\
       & \lbClaudeOpus & \cellcolor{lbFailureHeat!1}0.6 & \cellcolor{lbFailureHeat!1}0.6 & \cellcolor{lbFailureHeat!0}0.0 \\
       & \lbKimiModel & \cellcolor{lbFailureHeat!1}1.1 & \cellcolor{lbFailureHeat!1}1.1 & \cellcolor{lbFailureHeat!0}0.0 \\
       & \lbGPT & \cellcolor{lbFailureHeat!2}1.5 & \cellcolor{lbFailureHeat!2}1.5 & \cellcolor{lbFailureHeat!0}0.0 \\
       & \lbGLMModel & \cellcolor{lbFailureHeat!3}3.2 & \cellcolor{lbFailureHeat!3}3.2 & \cellcolor{lbFailureHeat!0}0.0 \\
       & \lbMiniMaxModel & \cellcolor{lbFailureHeat!7}7.1 & \cellcolor{lbFailureHeat!7}6.8 & \cellcolor{lbFailureHeat!1}0.3 \\
    \bottomrule
  \end{tabular}
  }
  \endgroup
\end{table*}

\begin{table*}[t]
  \centering
  \caption{Failure mode distribution for Code Localization failures across the 28 agents. The phase column reproduces the corresponding failure attribution in Table~\ref{tab:failure-composition} and equals the sum of failure mode values in every row. Darker backgrounds indicate a larger share of total failure attribution for each agent's solution failures.}
  \label{tab:failure-code-localization}
  \begingroup
  \definecolor{lbFailureHeat}{HTML}{729ECE}
  \newcommand{\lbModelLogo}[1]{%
    \raisebox{-0.28ex}{\includegraphics[height=1.22em,keepaspectratio]{figures/#1}}}
  \newcommand{\lbAgentLogo}[1]{%
    \raisebox{-0.32ex}{\includegraphics[height=1.38em,keepaspectratio]{figures/#1}}}
  \newcommand{\lbHarnessCell}[2]{%
    \makecell[c]{\lbAgentLogo{#1}\\[-0.15ex]\textsc{#2}}}
  \newcommand{\lbClaudeOpus}{\lbModelLogo{claude.png}\,Opus 5}
  \newcommand{\lbGPT}{\lbModelLogo{openai.png}\,GPT-5.6 Sol}
  \newcommand{\lbKimiModel}{\lbModelLogo{kimi.png}\,Kimi K3}
  \newcommand{\lbGLMModel}{\lbModelLogo{glm.png}\,GLM-5.2}
  \newcommand{\lbMiniMaxModel}{\lbModelLogo{minimax.png}\,MiniMax M3}
  \newcommand{\lbDeepSeekModel}{\lbModelLogo{deepseek.png}\,DeepSeek V4 Flash}
  \small
  \setlength{\tabcolsep}{3.0pt}
  \renewcommand{\arraystretch}{0.96}
  \scalebox{0.8}{%
  \begin{tabular}{@{}clr@{\hspace{7pt}}rrr@{}}
    \toprule
    \multirow{2}{*}{\textbf{Scaffold}}
      & \multirow{2}{*}{\textbf{Model}}
      & \multicolumn{1}{c}{\textbf{Phase}}
      & \multicolumn{3}{c}{\textbf{Failure Modes}} \\
    \cmidrule(lr){3-3}\cmidrule(lr){4-6}
      &
      & \makecell{\textbf{Code}\\\textbf{Localization}}
      & \makecell{\textbf{Cross-Module}\\\textbf{Context Missing}}
      & \makecell{\textbf{Missed}\\\textbf{Relevant File}}
      & \makecell{\textbf{Wrong File}\\\textbf{Localization}} \\
    \midrule
      \multirow[c]{5}{*}{\lbHarnessCell{claude.png}{Claude Code}} & \lbDeepSeekModel & \cellcolor{lbFailureHeat!28}27.8 & \cellcolor{lbFailureHeat!27}26.8 & \cellcolor{lbFailureHeat!1}0.7 & \cellcolor{lbFailureHeat!1}0.3 \\
       & \lbClaudeOpus & \cellcolor{lbFailureHeat!31}25.1 & \cellcolor{lbFailureHeat!30}24.3 & \cellcolor{lbFailureHeat!1}0.8 & \cellcolor{lbFailureHeat!0}0.0 \\
       & \lbKimiModel & \cellcolor{lbFailureHeat!29}26.4 & \cellcolor{lbFailureHeat!29}26.0 & \cellcolor{lbFailureHeat!1}0.4 & \cellcolor{lbFailureHeat!0}0.0 \\
       & \lbGLMModel & \cellcolor{lbFailureHeat!26}25.9 & \cellcolor{lbFailureHeat!24}23.9 & \cellcolor{lbFailureHeat!2}1.6 & \cellcolor{lbFailureHeat!1}0.4 \\
       & \lbMiniMaxModel & \cellcolor{lbFailureHeat!25}24.2 & \cellcolor{lbFailureHeat!24}23.4 & \cellcolor{lbFailureHeat!1}0.5 & \cellcolor{lbFailureHeat!1}0.3 \\
    \midrule
      \multirow[c]{5}{*}{\lbHarnessCell{codex.png}{Codex}} & \lbDeepSeekModel & \cellcolor{lbFailureHeat!40}39.9 & \cellcolor{lbFailureHeat!39}39.3 & \cellcolor{lbFailureHeat!1}0.6 & \cellcolor{lbFailureHeat!0}0.0 \\
       & \lbKimiModel & \cellcolor{lbFailureHeat!43}41.7 & \cellcolor{lbFailureHeat!42}41.4 & \cellcolor{lbFailureHeat!1}0.3 & \cellcolor{lbFailureHeat!0}0.0 \\
       & \lbGPT & \cellcolor{lbFailureHeat!37}35.5 & \cellcolor{lbFailureHeat!37}35.2 & \cellcolor{lbFailureHeat!1}0.3 & \cellcolor{lbFailureHeat!0}0.0 \\
       & \lbGLMModel & \cellcolor{lbFailureHeat!42}42.1 & \cellcolor{lbFailureHeat!40}39.6 & \cellcolor{lbFailureHeat!2}2.1 & \cellcolor{lbFailureHeat!1}0.4 \\
       & \lbMiniMaxModel & \cellcolor{lbFailureHeat!36}36.1 & \cellcolor{lbFailureHeat!35}34.6 & \cellcolor{lbFailureHeat!1}1.3 & \cellcolor{lbFailureHeat!1}0.2 \\
    \midrule
      \multirow[c]{6}{*}{\lbHarnessCell{opencode.png}{OpenCode}} & \lbDeepSeekModel & \cellcolor{lbFailureHeat!26}25.6 & \cellcolor{lbFailureHeat!26}25.0 & \cellcolor{lbFailureHeat!1}0.6 & \cellcolor{lbFailureHeat!0}0.0 \\
       & \lbClaudeOpus & \cellcolor{lbFailureHeat!34}27.1 & \cellcolor{lbFailureHeat!33}25.8 & \cellcolor{lbFailureHeat!2}1.3 & \cellcolor{lbFailureHeat!0}0.0 \\
       & \lbKimiModel & \cellcolor{lbFailureHeat!28}26.5 & \cellcolor{lbFailureHeat!27}26.3 & \cellcolor{lbFailureHeat!1}0.2 & \cellcolor{lbFailureHeat!0}0.0 \\
       & \lbGPT & \cellcolor{lbFailureHeat!29}13.2 & \cellcolor{lbFailureHeat!29}13.2 & \cellcolor{lbFailureHeat!0}0.0 & \cellcolor{lbFailureHeat!0}0.0 \\
       & \lbGLMModel & \cellcolor{lbFailureHeat!27}26.5 & \cellcolor{lbFailureHeat!25}24.7 & \cellcolor{lbFailureHeat!2}1.5 & \cellcolor{lbFailureHeat!1}0.3 \\
       & \lbMiniMaxModel & \cellcolor{lbFailureHeat!26}25.6 & \cellcolor{lbFailureHeat!24}24.2 & \cellcolor{lbFailureHeat!1}1.1 & \cellcolor{lbFailureHeat!1}0.3 \\
    \midrule
      \multirow[c]{6}{*}{\lbHarnessCell{pi.png}{Pi}} & \lbDeepSeekModel & \cellcolor{lbFailureHeat!34}32.6 & \cellcolor{lbFailureHeat!32}30.7 & \cellcolor{lbFailureHeat!1}1.4 & \cellcolor{lbFailureHeat!1}0.5 \\
       & \lbClaudeOpus & \cellcolor{lbFailureHeat!38}30.7 & \cellcolor{lbFailureHeat!38}30.5 & \cellcolor{lbFailureHeat!1}0.2 & \cellcolor{lbFailureHeat!0}0.0 \\
       & \lbKimiModel & \cellcolor{lbFailureHeat!39}38.4 & \cellcolor{lbFailureHeat!38}37.7 & \cellcolor{lbFailureHeat!1}0.7 & \cellcolor{lbFailureHeat!0}0.0 \\
       & \lbGPT & \cellcolor{lbFailureHeat!31}29.5 & \cellcolor{lbFailureHeat!31}29.2 & \cellcolor{lbFailureHeat!1}0.3 & \cellcolor{lbFailureHeat!0}0.0 \\
       & \lbGLMModel & \cellcolor{lbFailureHeat!28}27.3 & \cellcolor{lbFailureHeat!27}26.0 & \cellcolor{lbFailureHeat!1}0.9 & \cellcolor{lbFailureHeat!1}0.4 \\
       & \lbMiniMaxModel & \cellcolor{lbFailureHeat!25}24.5 & \cellcolor{lbFailureHeat!24}23.7 & \cellcolor{lbFailureHeat!1}0.7 & \cellcolor{lbFailureHeat!1}0.1 \\
    \midrule
      \multirow[c]{6}{*}{\lbHarnessCell{mini-swe-agent.png}{mini-SWE-agent}} & \lbDeepSeekModel & \cellcolor{lbFailureHeat!39}38.6 & \cellcolor{lbFailureHeat!38}38.0 & \cellcolor{lbFailureHeat!1}0.6 & \cellcolor{lbFailureHeat!0}0.0 \\
       & \lbClaudeOpus & \cellcolor{lbFailureHeat!41}30.6 & \cellcolor{lbFailureHeat!41}30.2 & \cellcolor{lbFailureHeat!1}0.4 & \cellcolor{lbFailureHeat!0}0.0 \\
       & \lbKimiModel & \cellcolor{lbFailureHeat!40}37.5 & \cellcolor{lbFailureHeat!39}37.0 & \cellcolor{lbFailureHeat!1}0.5 & \cellcolor{lbFailureHeat!0}0.0 \\
       & \lbGPT & \cellcolor{lbFailureHeat!38}35.9 & \cellcolor{lbFailureHeat!37}35.4 & \cellcolor{lbFailureHeat!1}0.5 & \cellcolor{lbFailureHeat!0}0.0 \\
       & \lbGLMModel & \cellcolor{lbFailureHeat!39}39.0 & \cellcolor{lbFailureHeat!38}37.6 & \cellcolor{lbFailureHeat!1}1.2 & \cellcolor{lbFailureHeat!1}0.2 \\
       & \lbMiniMaxModel & \cellcolor{lbFailureHeat!32}31.4 & \cellcolor{lbFailureHeat!31}30.6 & \cellcolor{lbFailureHeat!1}0.6 & \cellcolor{lbFailureHeat!1}0.2 \\
    \bottomrule
  \end{tabular}
  }
  \endgroup
\end{table*}

\begin{table*}[t]
  \centering
  \caption{Failure mode distribution for Code Editing failures across the 28 agents. The phase column reproduces the corresponding failure attribution in Table~\ref{tab:failure-composition} and equals the sum of failure mode values in every row. Darker backgrounds indicate a larger share of total failure attribution for each agent's solution failures.}
  \label{tab:failure-code-editing}
  \begingroup
  \definecolor{lbFailureHeat}{HTML}{729ECE}
  \newcommand{\lbModelLogo}[1]{%
    \raisebox{-0.28ex}{\includegraphics[height=1.22em,keepaspectratio]{figures/#1}}}
  \newcommand{\lbAgentLogo}[1]{%
    \raisebox{-0.32ex}{\includegraphics[height=1.38em,keepaspectratio]{figures/#1}}}
  \newcommand{\lbHarnessCell}[2]{%
    \makecell[c]{\lbAgentLogo{#1}\\[-0.15ex]\textsc{#2}}}
  \newcommand{\lbClaudeOpus}{\lbModelLogo{claude.png}\,Opus 5}
  \newcommand{\lbGPT}{\lbModelLogo{openai.png}\,GPT-5.6 Sol}
  \newcommand{\lbKimiModel}{\lbModelLogo{kimi.png}\,Kimi K3}
  \newcommand{\lbGLMModel}{\lbModelLogo{glm.png}\,GLM-5.2}
  \newcommand{\lbMiniMaxModel}{\lbModelLogo{minimax.png}\,MiniMax M3}
  \newcommand{\lbDeepSeekModel}{\lbModelLogo{deepseek.png}\,DeepSeek V4 Flash}
  \small
  \setlength{\tabcolsep}{3.0pt}
  \renewcommand{\arraystretch}{0.96}
  \scalebox{0.8}{%
  \begin{tabular}{@{}clr@{\hspace{7pt}}rrrr@{}}
    \toprule
    \multirow{2}{*}{\textbf{Scaffold}}
      & \multirow{2}{*}{\textbf{Model}}
      & \multicolumn{1}{c}{\textbf{Phase}}
      & \multicolumn{4}{c}{\textbf{Failure Modes}} \\
    \cmidrule(lr){3-3}\cmidrule(lr){4-7}
      &
      & \makecell{\textbf{Code}\\\textbf{Editing}}
      & \makecell{\textbf{Incorrect}\\\textbf{Patch}}
      & \makecell{\textbf{Relevant Change}\\\textbf{Omitted}}
      & \makecell{\textbf{No Patch}\\\textbf{Produced}}
      & \makecell{\textbf{Gave Up}\\\textbf{Early}} \\
    \midrule
      \multirow[c]{5}{*}{\lbHarnessCell{claude.png}{Claude Code}} & \lbDeepSeekModel & \cellcolor{lbFailureHeat!26}25.6 & \cellcolor{lbFailureHeat!15}14.6 & \cellcolor{lbFailureHeat!11}11.0 & \cellcolor{lbFailureHeat!0}0.0 & \cellcolor{lbFailureHeat!0}0.0 \\
       & \lbClaudeOpus & \cellcolor{lbFailureHeat!23}18.1 & \cellcolor{lbFailureHeat!14}11.0 & \cellcolor{lbFailureHeat!9}7.1 & \cellcolor{lbFailureHeat!0}0.0 & \cellcolor{lbFailureHeat!0}0.0 \\
       & \lbKimiModel & \cellcolor{lbFailureHeat!28}25.9 & \cellcolor{lbFailureHeat!15}13.9 & \cellcolor{lbFailureHeat!12}11.2 & \cellcolor{lbFailureHeat!1}0.8 & \cellcolor{lbFailureHeat!0}0.0 \\
       & \lbGLMModel & \cellcolor{lbFailureHeat!32}31.7 & \cellcolor{lbFailureHeat!15}14.5 & \cellcolor{lbFailureHeat!17}16.7 & \cellcolor{lbFailureHeat!1}0.5 & \cellcolor{lbFailureHeat!0}0.0 \\
       & \lbMiniMaxModel & \cellcolor{lbFailureHeat!31}29.3 & \cellcolor{lbFailureHeat!14}13.1 & \cellcolor{lbFailureHeat!17}16.2 & \cellcolor{lbFailureHeat!0}0.0 & \cellcolor{lbFailureHeat!0}0.0 \\
    \midrule
      \multirow[c]{5}{*}{\lbHarnessCell{codex.png}{Codex}} & \lbDeepSeekModel & \cellcolor{lbFailureHeat!25}24.6 & \cellcolor{lbFailureHeat!22}21.5 & \cellcolor{lbFailureHeat!3}3.1 & \cellcolor{lbFailureHeat!0}0.0 & \cellcolor{lbFailureHeat!0}0.0 \\
       & \lbKimiModel & \cellcolor{lbFailureHeat!24}23.8 & \cellcolor{lbFailureHeat!21}20.9 & \cellcolor{lbFailureHeat!3}2.9 & \cellcolor{lbFailureHeat!0}0.0 & \cellcolor{lbFailureHeat!0}0.0 \\
       & \lbGPT & \cellcolor{lbFailureHeat!21}20.5 & \cellcolor{lbFailureHeat!21}19.8 & \cellcolor{lbFailureHeat!1}0.7 & \cellcolor{lbFailureHeat!0}0.0 & \cellcolor{lbFailureHeat!0}0.0 \\
       & \lbGLMModel & \cellcolor{lbFailureHeat!25}25.2 & \cellcolor{lbFailureHeat!21}21.1 & \cellcolor{lbFailureHeat!4}4.1 & \cellcolor{lbFailureHeat!0}0.0 & \cellcolor{lbFailureHeat!0}0.0 \\
       & \lbMiniMaxModel & \cellcolor{lbFailureHeat!24}23.5 & \cellcolor{lbFailureHeat!21}20.7 & \cellcolor{lbFailureHeat!3}2.8 & \cellcolor{lbFailureHeat!0}0.0 & \cellcolor{lbFailureHeat!0}0.0 \\
    \midrule
      \multirow[c]{6}{*}{\lbHarnessCell{opencode.png}{OpenCode}} & \lbDeepSeekModel & \cellcolor{lbFailureHeat!31}30.7 & \cellcolor{lbFailureHeat!16}15.4 & \cellcolor{lbFailureHeat!16}15.3 & \cellcolor{lbFailureHeat!0}0.0 & \cellcolor{lbFailureHeat!0}0.0 \\
       & \lbClaudeOpus & \cellcolor{lbFailureHeat!21}16.9 & \cellcolor{lbFailureHeat!12}9.8 & \cellcolor{lbFailureHeat!8}6.6 & \cellcolor{lbFailureHeat!1}0.5 & \cellcolor{lbFailureHeat!0}0.0 \\
       & \lbKimiModel & \cellcolor{lbFailureHeat!27}25.5 & \cellcolor{lbFailureHeat!14}13.0 & \cellcolor{lbFailureHeat!10}9.6 & \cellcolor{lbFailureHeat!2}1.9 & \cellcolor{lbFailureHeat!1}1.0 \\
       & \lbGPT & \cellcolor{lbFailureHeat!40}18.4 & \cellcolor{lbFailureHeat!22}9.9 & \cellcolor{lbFailureHeat!18}8.2 & \cellcolor{lbFailureHeat!1}0.3 & \cellcolor{lbFailureHeat!0}0.0 \\
       & \lbGLMModel & \cellcolor{lbFailureHeat!33}32.2 & \cellcolor{lbFailureHeat!15}14.6 & \cellcolor{lbFailureHeat!17}16.6 & \cellcolor{lbFailureHeat!1}1.0 & \cellcolor{lbFailureHeat!0}0.0 \\
       & \lbMiniMaxModel & \cellcolor{lbFailureHeat!33}33.0 & \cellcolor{lbFailureHeat!13}12.8 & \cellcolor{lbFailureHeat!20}20.0 & \cellcolor{lbFailureHeat!1}0.2 & \cellcolor{lbFailureHeat!0}0.0 \\
    \midrule
      \multirow[c]{6}{*}{\lbHarnessCell{pi.png}{Pi}} & \lbDeepSeekModel & \cellcolor{lbFailureHeat!25}24.1 & \cellcolor{lbFailureHeat!17}16.0 & \cellcolor{lbFailureHeat!8}8.1 & \cellcolor{lbFailureHeat!0}0.0 & \cellcolor{lbFailureHeat!0}0.0 \\
       & \lbClaudeOpus & \cellcolor{lbFailureHeat!22}17.6 & \cellcolor{lbFailureHeat!15}12.4 & \cellcolor{lbFailureHeat!5}4.2 & \cellcolor{lbFailureHeat!1}1.0 & \cellcolor{lbFailureHeat!0}0.0 \\
       & \lbKimiModel & \cellcolor{lbFailureHeat!29}28.7 & \cellcolor{lbFailureHeat!20}19.2 & \cellcolor{lbFailureHeat!10}9.5 & \cellcolor{lbFailureHeat!0}0.0 & \cellcolor{lbFailureHeat!0}0.0 \\
       & \lbGPT & \cellcolor{lbFailureHeat!42}39.9 & \cellcolor{lbFailureHeat!20}19.4 & \cellcolor{lbFailureHeat!22}20.5 & \cellcolor{lbFailureHeat!0}0.0 & \cellcolor{lbFailureHeat!0}0.0 \\
       & \lbGLMModel & \cellcolor{lbFailureHeat!31}29.9 & \cellcolor{lbFailureHeat!16}15.4 & \cellcolor{lbFailureHeat!14}13.9 & \cellcolor{lbFailureHeat!1}0.6 & \cellcolor{lbFailureHeat!0}0.0 \\
       & \lbMiniMaxModel & \cellcolor{lbFailureHeat!35}34.2 & \cellcolor{lbFailureHeat!15}14.4 & \cellcolor{lbFailureHeat!18}17.9 & \cellcolor{lbFailureHeat!1}0.9 & \cellcolor{lbFailureHeat!1}1.0 \\
    \midrule
      \multirow[c]{6}{*}{\lbHarnessCell{mini-swe-agent.png}{mini-SWE-agent}} & \lbDeepSeekModel & \cellcolor{lbFailureHeat!26}25.7 & \cellcolor{lbFailureHeat!21}21.2 & \cellcolor{lbFailureHeat!5}4.5 & \cellcolor{lbFailureHeat!0}0.0 & \cellcolor{lbFailureHeat!0}0.0 \\
       & \lbClaudeOpus & \cellcolor{lbFailureHeat!19}14.2 & \cellcolor{lbFailureHeat!15}11.2 & \cellcolor{lbFailureHeat!4}3.0 & \cellcolor{lbFailureHeat!0}0.0 & \cellcolor{lbFailureHeat!0}0.0 \\
       & \lbKimiModel & \cellcolor{lbFailureHeat!20}18.5 & \cellcolor{lbFailureHeat!18}16.6 & \cellcolor{lbFailureHeat!2}1.9 & \cellcolor{lbFailureHeat!0}0.0 & \cellcolor{lbFailureHeat!0}0.0 \\
       & \lbGPT & \cellcolor{lbFailureHeat!25}24.0 & \cellcolor{lbFailureHeat!21}20.2 & \cellcolor{lbFailureHeat!4}3.8 & \cellcolor{lbFailureHeat!0}0.0 & \cellcolor{lbFailureHeat!0}0.0 \\
       & \lbGLMModel & \cellcolor{lbFailureHeat!24}23.4 & \cellcolor{lbFailureHeat!20}20.0 & \cellcolor{lbFailureHeat!3}3.4 & \cellcolor{lbFailureHeat!0}0.0 & \cellcolor{lbFailureHeat!0}0.0 \\
       & \lbMiniMaxModel & \cellcolor{lbFailureHeat!24}23.6 & \cellcolor{lbFailureHeat!18}17.5 & \cellcolor{lbFailureHeat!6}5.8 & \cellcolor{lbFailureHeat!1}0.3 & \cellcolor{lbFailureHeat!0}0.0 \\
    \bottomrule
  \end{tabular}
  }
  \endgroup
\end{table*}

\begin{table*}[t]
  \centering
  \caption{Failure mode distribution for Code Verification failures across the 28 agents. The phase column reproduces the corresponding failure attribution in Table~\ref{tab:failure-composition} and equals the sum of failure mode values in every row. Darker backgrounds indicate a larger share of total failure attribution for each agent's solution failures.}
  \label{tab:failure-code-verification}
  \begingroup
  \definecolor{lbFailureHeat}{HTML}{729ECE}
  \newcommand{\lbModelLogo}[1]{%
    \raisebox{-0.28ex}{\includegraphics[height=1.22em,keepaspectratio]{figures/#1}}}
  \newcommand{\lbAgentLogo}[1]{%
    \raisebox{-0.32ex}{\includegraphics[height=1.38em,keepaspectratio]{figures/#1}}}
  \newcommand{\lbHarnessCell}[2]{%
    \makecell[c]{\lbAgentLogo{#1}\\[-0.15ex]\textsc{#2}}}
  \newcommand{\lbClaudeOpus}{\lbModelLogo{claude.png}\,Opus 5}
  \newcommand{\lbGPT}{\lbModelLogo{openai.png}\,GPT-5.6 Sol}
  \newcommand{\lbKimiModel}{\lbModelLogo{kimi.png}\,Kimi K3}
  \newcommand{\lbGLMModel}{\lbModelLogo{glm.png}\,GLM-5.2}
  \newcommand{\lbMiniMaxModel}{\lbModelLogo{minimax.png}\,MiniMax M3}
  \newcommand{\lbDeepSeekModel}{\lbModelLogo{deepseek.png}\,DeepSeek V4 Flash}
  \small
  \setlength{\tabcolsep}{3.0pt}
  \renewcommand{\arraystretch}{0.96}
  \scalebox{0.8}{%
  \begin{tabular}{@{}clr@{\hspace{7pt}}rrrr@{}}
    \toprule
    \multirow{2}{*}{\textbf{Scaffold}}
      & \multirow{2}{*}{\textbf{Model}}
      & \multicolumn{1}{c}{\textbf{Phase}}
      & \multicolumn{4}{c}{\textbf{Failure Modes}} \\
    \cmidrule(lr){3-3}\cmidrule(lr){4-7}
      &
      & \makecell{\textbf{Code}\\\textbf{Verification}}
      & \makecell{\textbf{No Validation}\\\textbf{Attempted}}
      & \makecell{\textbf{No Test After}\\\textbf{Final Edit}}
      & \makecell{\textbf{Authored Test}\\\textbf{Never Run}}
      & \makecell{\textbf{Test Result}\\\textbf{Misinterpreted}} \\
    \midrule
      \multirow[c]{5}{*}{\lbHarnessCell{claude.png}{Claude Code}} & \lbDeepSeekModel & \cellcolor{lbFailureHeat!14}13.8 & \cellcolor{lbFailureHeat!10}10.0 & \cellcolor{lbFailureHeat!2}2.3 & \cellcolor{lbFailureHeat!1}0.7 & \cellcolor{lbFailureHeat!1}0.8 \\
       & \lbClaudeOpus & \cellcolor{lbFailureHeat!17}13.8 & \cellcolor{lbFailureHeat!13}10.1 & \cellcolor{lbFailureHeat!3}2.3 & \cellcolor{lbFailureHeat!1}0.9 & \cellcolor{lbFailureHeat!1}0.5 \\
       & \lbKimiModel & \cellcolor{lbFailureHeat!11}9.8 & \cellcolor{lbFailureHeat!5}4.9 & \cellcolor{lbFailureHeat!4}3.3 & \cellcolor{lbFailureHeat!1}0.7 & \cellcolor{lbFailureHeat!1}0.9 \\
       & \lbGLMModel & \cellcolor{lbFailureHeat!13}12.9 & \cellcolor{lbFailureHeat!11}10.9 & \cellcolor{lbFailureHeat!1}1.1 & \cellcolor{lbFailureHeat!1}0.3 & \cellcolor{lbFailureHeat!1}0.6 \\
       & \lbMiniMaxModel & \cellcolor{lbFailureHeat!13}12.2 & \cellcolor{lbFailureHeat!10}9.9 & \cellcolor{lbFailureHeat!1}0.6 & \cellcolor{lbFailureHeat!1}1.3 & \cellcolor{lbFailureHeat!1}0.4 \\
    \midrule
      \multirow[c]{5}{*}{\lbHarnessCell{codex.png}{Codex}} & \lbDeepSeekModel & \cellcolor{lbFailureHeat!18}18.3 & \cellcolor{lbFailureHeat!14}13.5 & \cellcolor{lbFailureHeat!4}4.3 & \cellcolor{lbFailureHeat!1}0.3 & \cellcolor{lbFailureHeat!1}0.2 \\
       & \lbKimiModel & \cellcolor{lbFailureHeat!17}16.8 & \cellcolor{lbFailureHeat!10}10.1 & \cellcolor{lbFailureHeat!5}5.0 & \cellcolor{lbFailureHeat!2}1.7 & \cellcolor{lbFailureHeat!0}0.0 \\
       & \lbGPT & \cellcolor{lbFailureHeat!21}19.8 & \cellcolor{lbFailureHeat!17}16.0 & \cellcolor{lbFailureHeat!2}2.1 & \cellcolor{lbFailureHeat!1}1.2 & \cellcolor{lbFailureHeat!1}0.5 \\
       & \lbGLMModel & \cellcolor{lbFailureHeat!16}16.3 & \cellcolor{lbFailureHeat!14}13.9 & \cellcolor{lbFailureHeat!2}2.2 & \cellcolor{lbFailureHeat!0}0.0 & \cellcolor{lbFailureHeat!1}0.2 \\
       & \lbMiniMaxModel & \cellcolor{lbFailureHeat!16}15.8 & \cellcolor{lbFailureHeat!12}12.0 & \cellcolor{lbFailureHeat!3}2.7 & \cellcolor{lbFailureHeat!1}0.7 & \cellcolor{lbFailureHeat!1}0.4 \\
    \midrule
      \multirow[c]{6}{*}{\lbHarnessCell{opencode.png}{OpenCode}} & \lbDeepSeekModel & \cellcolor{lbFailureHeat!13}12.8 & \cellcolor{lbFailureHeat!11}10.3 & \cellcolor{lbFailureHeat!1}1.3 & \cellcolor{lbFailureHeat!1}0.4 & \cellcolor{lbFailureHeat!1}0.8 \\
       & \lbClaudeOpus & \cellcolor{lbFailureHeat!12}9.5 & \cellcolor{lbFailureHeat!8}6.2 & \cellcolor{lbFailureHeat!3}2.6 & \cellcolor{lbFailureHeat!1}0.7 & \cellcolor{lbFailureHeat!0}0.0 \\
       & \lbKimiModel & \cellcolor{lbFailureHeat!14}13.7 & \cellcolor{lbFailureHeat!10}9.9 & \cellcolor{lbFailureHeat!2}2.2 & \cellcolor{lbFailureHeat!1}1.0 & \cellcolor{lbFailureHeat!1}0.6 \\
       & \lbGPT & \cellcolor{lbFailureHeat!14}6.5 & \cellcolor{lbFailureHeat!9}4.1 & \cellcolor{lbFailureHeat!4}1.9 & \cellcolor{lbFailureHeat!1}0.1 & \cellcolor{lbFailureHeat!1}0.4 \\
       & \lbGLMModel & \cellcolor{lbFailureHeat!13}12.6 & \cellcolor{lbFailureHeat!11}10.7 & \cellcolor{lbFailureHeat!1}1.4 & \cellcolor{lbFailureHeat!1}0.5 & \cellcolor{lbFailureHeat!0}0.0 \\
       & \lbMiniMaxModel & \cellcolor{lbFailureHeat!11}11.2 & \cellcolor{lbFailureHeat!9}9.1 & \cellcolor{lbFailureHeat!2}1.5 & \cellcolor{lbFailureHeat!1}0.6 & \cellcolor{lbFailureHeat!0}0.0 \\
    \midrule
      \multirow[c]{6}{*}{\lbHarnessCell{pi.png}{Pi}} & \lbDeepSeekModel & \cellcolor{lbFailureHeat!11}10.5 & \cellcolor{lbFailureHeat!9}9.1 & \cellcolor{lbFailureHeat!1}0.9 & \cellcolor{lbFailureHeat!1}0.1 & \cellcolor{lbFailureHeat!1}0.4 \\
       & \lbClaudeOpus & \cellcolor{lbFailureHeat!11}8.5 & \cellcolor{lbFailureHeat!9}7.5 & \cellcolor{lbFailureHeat!1}0.7 & \cellcolor{lbFailureHeat!0}0.0 & \cellcolor{lbFailureHeat!1}0.3 \\
       & \lbKimiModel & \cellcolor{lbFailureHeat!11}10.7 & \cellcolor{lbFailureHeat!8}7.6 & \cellcolor{lbFailureHeat!2}1.7 & \cellcolor{lbFailureHeat!1}1.0 & \cellcolor{lbFailureHeat!1}0.4 \\
       & \lbGPT & \cellcolor{lbFailureHeat!12}11.7 & \cellcolor{lbFailureHeat!11}10.4 & \cellcolor{lbFailureHeat!1}1.0 & \cellcolor{lbFailureHeat!0}0.0 & \cellcolor{lbFailureHeat!1}0.3 \\
       & \lbGLMModel & \cellcolor{lbFailureHeat!11}10.7 & \cellcolor{lbFailureHeat!10}9.7 & \cellcolor{lbFailureHeat!1}0.4 & \cellcolor{lbFailureHeat!1}0.1 & \cellcolor{lbFailureHeat!1}0.5 \\
       & \lbMiniMaxModel & \cellcolor{lbFailureHeat!9}8.5 & \cellcolor{lbFailureHeat!8}7.6 & \cellcolor{lbFailureHeat!0}0.0 & \cellcolor{lbFailureHeat!1}0.5 & \cellcolor{lbFailureHeat!1}0.4 \\
    \midrule
      \multirow[c]{6}{*}{\lbHarnessCell{mini-swe-agent.png}{mini-SWE-agent}} & \lbDeepSeekModel & \cellcolor{lbFailureHeat!25}24.9 & \cellcolor{lbFailureHeat!18}17.5 & \cellcolor{lbFailureHeat!6}6.3 & \cellcolor{lbFailureHeat!1}0.8 & \cellcolor{lbFailureHeat!1}0.3 \\
       & \lbClaudeOpus & \cellcolor{lbFailureHeat!22}16.6 & \cellcolor{lbFailureHeat!16}11.5 & \cellcolor{lbFailureHeat!5}3.4 & \cellcolor{lbFailureHeat!2}1.5 & \cellcolor{lbFailureHeat!1}0.2 \\
       & \lbKimiModel & \cellcolor{lbFailureHeat!24}23.0 & \cellcolor{lbFailureHeat!16}15.5 & \cellcolor{lbFailureHeat!5}4.7 & \cellcolor{lbFailureHeat!2}1.8 & \cellcolor{lbFailureHeat!1}1.0 \\
       & \lbGPT & \cellcolor{lbFailureHeat!14}13.7 & \cellcolor{lbFailureHeat!9}8.5 & \cellcolor{lbFailureHeat!4}4.0 & \cellcolor{lbFailureHeat!1}0.4 & \cellcolor{lbFailureHeat!1}0.8 \\
       & \lbGLMModel & \cellcolor{lbFailureHeat!21}20.3 & \cellcolor{lbFailureHeat!16}16.3 & \cellcolor{lbFailureHeat!3}3.3 & \cellcolor{lbFailureHeat!1}0.7 & \cellcolor{lbFailureHeat!0}0.0 \\
       & \lbMiniMaxModel & \cellcolor{lbFailureHeat!17}17.0 & \cellcolor{lbFailureHeat!13}13.3 & \cellcolor{lbFailureHeat!2}2.2 & \cellcolor{lbFailureHeat!1}1.3 & \cellcolor{lbFailureHeat!1}0.2 \\
    \bottomrule
  \end{tabular}
  }
  \endgroup
\end{table*}

\begin{table*}[t]
  \centering
  \caption{Failure mode distribution for Self-Repair failures across the 28 agents. The phase column reproduces the corresponding failure attribution in Table~\ref{tab:failure-composition} and equals the sum of failure mode values in every row. Darker backgrounds indicate a larger share of total failure attribution for each agent's solution failures.}
  \label{tab:failure-self-repair-loop}
  \begingroup
  \definecolor{lbFailureHeat}{HTML}{729ECE}
  \newcommand{\lbModelLogo}[1]{%
    \raisebox{-0.28ex}{\includegraphics[height=1.22em,keepaspectratio]{figures/#1}}}
  \newcommand{\lbAgentLogo}[1]{%
    \raisebox{-0.32ex}{\includegraphics[height=1.38em,keepaspectratio]{figures/#1}}}
  \newcommand{\lbHarnessCell}[2]{%
    \makecell[c]{\lbAgentLogo{#1}\\[-0.15ex]\textsc{#2}}}
  \newcommand{\lbClaudeOpus}{\lbModelLogo{claude.png}\,Opus 5}
  \newcommand{\lbGPT}{\lbModelLogo{openai.png}\,GPT-5.6 Sol}
  \newcommand{\lbKimiModel}{\lbModelLogo{kimi.png}\,Kimi K3}
  \newcommand{\lbGLMModel}{\lbModelLogo{glm.png}\,GLM-5.2}
  \newcommand{\lbMiniMaxModel}{\lbModelLogo{minimax.png}\,MiniMax M3}
  \newcommand{\lbDeepSeekModel}{\lbModelLogo{deepseek.png}\,DeepSeek V4 Flash}
  \small
  \setlength{\tabcolsep}{3.0pt}
  \renewcommand{\arraystretch}{0.96}
  \scalebox{0.8}{%
  \begin{tabular}{@{}clr@{\hspace{7pt}}rrrr@{}}
    \toprule
    \multirow{2}{*}{\textbf{Scaffold}}
      & \multirow{2}{*}{\textbf{Model}}
      & \multicolumn{1}{c}{\textbf{Phase}}
      & \multicolumn{4}{c}{\textbf{Failure Modes}} \\
    \cmidrule(lr){3-3}\cmidrule(lr){4-7}
      &
      & \makecell{\textbf{Self-}\\\textbf{Repair}}
      & \makecell{\textbf{Repair Not}\\\textbf{Reverified}}
      & \makecell{\textbf{Repeated Ineffective}\\\textbf{Attempt}}
      & \makecell{\textbf{Reverted Own}\\\textbf{Change}}
      & \makecell{\textbf{Retry Without}\\\textbf{Change}} \\
    \midrule
      \multirow[c]{5}{*}{\lbHarnessCell{claude.png}{Claude Code}} & \lbDeepSeekModel & \cellcolor{lbFailureHeat!11}11.0 & \cellcolor{lbFailureHeat!6}5.7 & \cellcolor{lbFailureHeat!4}3.9 & \cellcolor{lbFailureHeat!1}1.4 & \cellcolor{lbFailureHeat!0}0.0 \\
       & \lbClaudeOpus & \cellcolor{lbFailureHeat!12}9.9 & \cellcolor{lbFailureHeat!8}6.7 & \cellcolor{lbFailureHeat!3}2.5 & \cellcolor{lbFailureHeat!1}0.5 & \cellcolor{lbFailureHeat!1}0.2 \\
       & \lbKimiModel & \cellcolor{lbFailureHeat!21}18.8 & \cellcolor{lbFailureHeat!10}9.1 & \cellcolor{lbFailureHeat!8}7.3 & \cellcolor{lbFailureHeat!3}2.4 & \cellcolor{lbFailureHeat!0}0.0 \\
       & \lbGLMModel & \cellcolor{lbFailureHeat!12}11.6 & \cellcolor{lbFailureHeat!5}5.2 & \cellcolor{lbFailureHeat!4}4.2 & \cellcolor{lbFailureHeat!2}2.2 & \cellcolor{lbFailureHeat!0}0.0 \\
       & \lbMiniMaxModel & \cellcolor{lbFailureHeat!12}11.5 & \cellcolor{lbFailureHeat!2}1.8 & \cellcolor{lbFailureHeat!6}5.3 & \cellcolor{lbFailureHeat!4}4.3 & \cellcolor{lbFailureHeat!1}0.1 \\
    \midrule
      \multirow[c]{5}{*}{\lbHarnessCell{codex.png}{Codex}} & \lbDeepSeekModel & \cellcolor{lbFailureHeat!9}8.5 & \cellcolor{lbFailureHeat!6}5.9 & \cellcolor{lbFailureHeat!2}2.5 & \cellcolor{lbFailureHeat!0}0.0 & \cellcolor{lbFailureHeat!1}0.1 \\
       & \lbKimiModel & \cellcolor{lbFailureHeat!12}11.4 & \cellcolor{lbFailureHeat!5}5.3 & \cellcolor{lbFailureHeat!6}5.7 & \cellcolor{lbFailureHeat!1}0.4 & \cellcolor{lbFailureHeat!0}0.0 \\
       & \lbGPT & \cellcolor{lbFailureHeat!19}18.0 & \cellcolor{lbFailureHeat!10}9.1 & \cellcolor{lbFailureHeat!7}7.0 & \cellcolor{lbFailureHeat!1}1.2 & \cellcolor{lbFailureHeat!1}0.7 \\
       & \lbGLMModel & \cellcolor{lbFailureHeat!9}9.2 & \cellcolor{lbFailureHeat!6}5.9 & \cellcolor{lbFailureHeat!3}2.8 & \cellcolor{lbFailureHeat!0}0.0 & \cellcolor{lbFailureHeat!1}0.5 \\
       & \lbMiniMaxModel & \cellcolor{lbFailureHeat!9}9.4 & \cellcolor{lbFailureHeat!3}2.8 & \cellcolor{lbFailureHeat!6}6.3 & \cellcolor{lbFailureHeat!0}0.0 & \cellcolor{lbFailureHeat!1}0.3 \\
    \midrule
      \multirow[c]{6}{*}{\lbHarnessCell{opencode.png}{OpenCode}} & \lbDeepSeekModel & \cellcolor{lbFailureHeat!12}11.8 & \cellcolor{lbFailureHeat!4}4.4 & \cellcolor{lbFailureHeat!5}4.7 & \cellcolor{lbFailureHeat!3}2.7 & \cellcolor{lbFailureHeat!0}0.0 \\
       & \lbClaudeOpus & \cellcolor{lbFailureHeat!19}14.7 & \cellcolor{lbFailureHeat!11}8.5 & \cellcolor{lbFailureHeat!8}6.0 & \cellcolor{lbFailureHeat!1}0.2 & \cellcolor{lbFailureHeat!0}0.0 \\
       & \lbKimiModel & \cellcolor{lbFailureHeat!19}18.1 & \cellcolor{lbFailureHeat!8}7.3 & \cellcolor{lbFailureHeat!8}7.8 & \cellcolor{lbFailureHeat!3}3.0 & \cellcolor{lbFailureHeat!0}0.0 \\
       & \lbGPT & \cellcolor{lbFailureHeat!14}6.5 & \cellcolor{lbFailureHeat!6}2.6 & \cellcolor{lbFailureHeat!7}3.2 & \cellcolor{lbFailureHeat!2}0.7 & \cellcolor{lbFailureHeat!0}0.0 \\
       & \lbGLMModel & \cellcolor{lbFailureHeat!12}11.4 & \cellcolor{lbFailureHeat!4}3.5 & \cellcolor{lbFailureHeat!6}5.5 & \cellcolor{lbFailureHeat!2}2.4 & \cellcolor{lbFailureHeat!0}0.0 \\
       & \lbMiniMaxModel & \cellcolor{lbFailureHeat!13}13.1 & \cellcolor{lbFailureHeat!4}3.7 & \cellcolor{lbFailureHeat!5}4.7 & \cellcolor{lbFailureHeat!5}4.6 & \cellcolor{lbFailureHeat!1}0.1 \\
    \midrule
      \multirow[c]{6}{*}{\lbHarnessCell{pi.png}{Pi}} & \lbDeepSeekModel & \cellcolor{lbFailureHeat!8}7.6 & \cellcolor{lbFailureHeat!1}1.1 & \cellcolor{lbFailureHeat!6}6.0 & \cellcolor{lbFailureHeat!0}0.0 & \cellcolor{lbFailureHeat!1}0.5 \\
       & \lbClaudeOpus & \cellcolor{lbFailureHeat!10}8.2 & \cellcolor{lbFailureHeat!1}0.5 & \cellcolor{lbFailureHeat!9}7.4 & \cellcolor{lbFailureHeat!0}0.0 & \cellcolor{lbFailureHeat!1}0.3 \\
       & \lbKimiModel & \cellcolor{lbFailureHeat!11}11.1 & \cellcolor{lbFailureHeat!1}1.3 & \cellcolor{lbFailureHeat!9}9.0 & \cellcolor{lbFailureHeat!0}0.0 & \cellcolor{lbFailureHeat!1}0.8 \\
       & \lbGPT & \cellcolor{lbFailureHeat!12}11.7 & \cellcolor{lbFailureHeat!5}5.0 & \cellcolor{lbFailureHeat!7}6.7 & \cellcolor{lbFailureHeat!0}0.0 & \cellcolor{lbFailureHeat!0}0.0 \\
       & \lbGLMModel & \cellcolor{lbFailureHeat!8}8.2 & \cellcolor{lbFailureHeat!1}1.3 & \cellcolor{lbFailureHeat!7}6.9 & \cellcolor{lbFailureHeat!0}0.0 & \cellcolor{lbFailureHeat!0}0.0 \\
       & \lbMiniMaxModel & \cellcolor{lbFailureHeat!10}10.4 & \cellcolor{lbFailureHeat!2}1.7 & \cellcolor{lbFailureHeat!8}7.5 & \cellcolor{lbFailureHeat!0}0.0 & \cellcolor{lbFailureHeat!1}1.2 \\
    \midrule
      \multirow[c]{6}{*}{\lbHarnessCell{mini-swe-agent.png}{mini-SWE-agent}} & \lbDeepSeekModel & \cellcolor{lbFailureHeat!5}5.1 & \cellcolor{lbFailureHeat!3}2.9 & \cellcolor{lbFailureHeat!2}2.0 & \cellcolor{lbFailureHeat!0}0.0 & \cellcolor{lbFailureHeat!1}0.2 \\
       & \lbClaudeOpus & \cellcolor{lbFailureHeat!15}10.8 & \cellcolor{lbFailureHeat!12}9.0 & \cellcolor{lbFailureHeat!2}1.8 & \cellcolor{lbFailureHeat!0}0.0 & \cellcolor{lbFailureHeat!0}0.0 \\
       & \lbKimiModel & \cellcolor{lbFailureHeat!12}11.7 & \cellcolor{lbFailureHeat!8}7.6 & \cellcolor{lbFailureHeat!4}4.1 & \cellcolor{lbFailureHeat!0}0.0 & \cellcolor{lbFailureHeat!0}0.0 \\
       & \lbGPT & \cellcolor{lbFailureHeat!19}18.2 & \cellcolor{lbFailureHeat!15}14.6 & \cellcolor{lbFailureHeat!3}2.8 & \cellcolor{lbFailureHeat!0}0.0 & \cellcolor{lbFailureHeat!1}0.8 \\
       & \lbGLMModel & \cellcolor{lbFailureHeat!9}8.8 & \cellcolor{lbFailureHeat!6}5.9 & \cellcolor{lbFailureHeat!3}2.7 & \cellcolor{lbFailureHeat!0}0.0 & \cellcolor{lbFailureHeat!1}0.2 \\
       & \lbMiniMaxModel & \cellcolor{lbFailureHeat!11}11.2 & \cellcolor{lbFailureHeat!5}5.1 & \cellcolor{lbFailureHeat!6}5.8 & \cellcolor{lbFailureHeat!0}0.0 & \cellcolor{lbFailureHeat!1}0.3 \\
    \bottomrule
  \end{tabular}
  }
  \endgroup
\end{table*}

\begin{table*}[t]
  \centering
  \caption{Failure mode distribution for Tool Use failures across the 28 agents. The phase column reproduces the corresponding failure attribution in Table~\ref{tab:failure-composition} and equals the sum of failure mode values in every row. Darker backgrounds indicate a larger share of total failure attribution for each agent's solution failures.}
  \label{tab:failure-tool-use}
  \begingroup
  \definecolor{lbFailureHeat}{HTML}{729ECE}
  \newcommand{\lbModelLogo}[1]{%
    \raisebox{-0.28ex}{\includegraphics[height=1.22em,keepaspectratio]{figures/#1}}}
  \newcommand{\lbAgentLogo}[1]{%
    \raisebox{-0.32ex}{\includegraphics[height=1.38em,keepaspectratio]{figures/#1}}}
  \newcommand{\lbHarnessCell}[2]{%
    \makecell[c]{\lbAgentLogo{#1}\\[-0.15ex]\textsc{#2}}}
  \newcommand{\lbClaudeOpus}{\lbModelLogo{claude.png}\,Opus 5}
  \newcommand{\lbGPT}{\lbModelLogo{openai.png}\,GPT-5.6 Sol}
  \newcommand{\lbKimiModel}{\lbModelLogo{kimi.png}\,Kimi K3}
  \newcommand{\lbGLMModel}{\lbModelLogo{glm.png}\,GLM-5.2}
  \newcommand{\lbMiniMaxModel}{\lbModelLogo{minimax.png}\,MiniMax M3}
  \newcommand{\lbDeepSeekModel}{\lbModelLogo{deepseek.png}\,DeepSeek V4 Flash}
  \small
  \setlength{\tabcolsep}{3.0pt}
  \renewcommand{\arraystretch}{0.96}
  \scalebox{0.8}{%
  \begin{tabular}{@{}clr@{\hspace{7pt}}rrr@{}}
    \toprule
    \multirow{2}{*}{\textbf{Scaffold}}
      & \multirow{2}{*}{\textbf{Model}}
      & \multicolumn{1}{c}{\textbf{Phase}}
      & \multicolumn{3}{c}{\textbf{Failure Modes}} \\
    \cmidrule(lr){3-3}\cmidrule(lr){4-6}
      &
      & \makecell{\textbf{Tool}\\\textbf{Use}}
      & \makecell{\textbf{Hallucinated Path}\\\textbf{Unrecovered}}
      & \makecell{\textbf{Tool Failure}\\\textbf{Not Retried}}
      & \makecell{\textbf{Repeated Patch}\\\textbf{Apply Failure}} \\
    \midrule
      \multirow[c]{5}{*}{\lbHarnessCell{claude.png}{Claude Code}} & \lbDeepSeekModel & \cellcolor{lbFailureHeat!1}0.1 & \cellcolor{lbFailureHeat!1}0.1 & \cellcolor{lbFailureHeat!0}0.0 & \cellcolor{lbFailureHeat!0}0.0 \\
       & \lbClaudeOpus & \cellcolor{lbFailureHeat!0}0.0 & \cellcolor{lbFailureHeat!0}0.0 & \cellcolor{lbFailureHeat!0}0.0 & \cellcolor{lbFailureHeat!0}0.0 \\
       & \lbKimiModel & \cellcolor{lbFailureHeat!1}0.6 & \cellcolor{lbFailureHeat!1}0.1 & \cellcolor{lbFailureHeat!1}0.3 & \cellcolor{lbFailureHeat!1}0.2 \\
       & \lbGLMModel & \cellcolor{lbFailureHeat!0}0.0 & \cellcolor{lbFailureHeat!0}0.0 & \cellcolor{lbFailureHeat!0}0.0 & \cellcolor{lbFailureHeat!0}0.0 \\
       & \lbMiniMaxModel & \cellcolor{lbFailureHeat!1}0.2 & \cellcolor{lbFailureHeat!1}0.1 & \cellcolor{lbFailureHeat!1}0.1 & \cellcolor{lbFailureHeat!0}0.0 \\
    \midrule
      \multirow[c]{5}{*}{\lbHarnessCell{codex.png}{Codex}} & \lbDeepSeekModel & \cellcolor{lbFailureHeat!1}0.3 & \cellcolor{lbFailureHeat!1}0.3 & \cellcolor{lbFailureHeat!0}0.0 & \cellcolor{lbFailureHeat!0}0.0 \\
       & \lbKimiModel & \cellcolor{lbFailureHeat!1}0.5 & \cellcolor{lbFailureHeat!1}0.5 & \cellcolor{lbFailureHeat!0}0.0 & \cellcolor{lbFailureHeat!0}0.0 \\
       & \lbGPT & \cellcolor{lbFailureHeat!0}0.0 & \cellcolor{lbFailureHeat!0}0.0 & \cellcolor{lbFailureHeat!0}0.0 & \cellcolor{lbFailureHeat!0}0.0 \\
       & \lbGLMModel & \cellcolor{lbFailureHeat!0}0.0 & \cellcolor{lbFailureHeat!0}0.0 & \cellcolor{lbFailureHeat!0}0.0 & \cellcolor{lbFailureHeat!0}0.0 \\
       & \lbMiniMaxModel & \cellcolor{lbFailureHeat!2}1.5 & \cellcolor{lbFailureHeat!1}0.6 & \cellcolor{lbFailureHeat!0}0.0 & \cellcolor{lbFailureHeat!1}0.9 \\
    \midrule
      \multirow[c]{6}{*}{\lbHarnessCell{opencode.png}{OpenCode}} & \lbDeepSeekModel & \cellcolor{lbFailureHeat!1}0.4 & \cellcolor{lbFailureHeat!1}0.4 & \cellcolor{lbFailureHeat!0}0.0 & \cellcolor{lbFailureHeat!0}0.0 \\
       & \lbClaudeOpus & \cellcolor{lbFailureHeat!0}0.0 & \cellcolor{lbFailureHeat!0}0.0 & \cellcolor{lbFailureHeat!0}0.0 & \cellcolor{lbFailureHeat!0}0.0 \\
       & \lbKimiModel & \cellcolor{lbFailureHeat!1}0.5 & \cellcolor{lbFailureHeat!1}0.4 & \cellcolor{lbFailureHeat!0}0.0 & \cellcolor{lbFailureHeat!1}0.1 \\
       & \lbGPT & \cellcolor{lbFailureHeat!0}0.0 & \cellcolor{lbFailureHeat!0}0.0 & \cellcolor{lbFailureHeat!0}0.0 & \cellcolor{lbFailureHeat!0}0.0 \\
       & \lbGLMModel & \cellcolor{lbFailureHeat!0}0.0 & \cellcolor{lbFailureHeat!0}0.0 & \cellcolor{lbFailureHeat!0}0.0 & \cellcolor{lbFailureHeat!0}0.0 \\
       & \lbMiniMaxModel & \cellcolor{lbFailureHeat!0}0.0 & \cellcolor{lbFailureHeat!0}0.0 & \cellcolor{lbFailureHeat!0}0.0 & \cellcolor{lbFailureHeat!0}0.0 \\
    \midrule
      \multirow[c]{6}{*}{\lbHarnessCell{pi.png}{Pi}} & \lbDeepSeekModel & \cellcolor{lbFailureHeat!1}0.6 & \cellcolor{lbFailureHeat!1}0.2 & \cellcolor{lbFailureHeat!1}0.4 & \cellcolor{lbFailureHeat!0}0.0 \\
       & \lbClaudeOpus & \cellcolor{lbFailureHeat!2}1.4 & \cellcolor{lbFailureHeat!1}0.1 & \cellcolor{lbFailureHeat!2}1.3 & \cellcolor{lbFailureHeat!0}0.0 \\
       & \lbKimiModel & \cellcolor{lbFailureHeat!1}0.6 & \cellcolor{lbFailureHeat!0}0.0 & \cellcolor{lbFailureHeat!1}0.6 & \cellcolor{lbFailureHeat!0}0.0 \\
       & \lbGPT & \cellcolor{lbFailureHeat!1}0.1 & \cellcolor{lbFailureHeat!0}0.0 & \cellcolor{lbFailureHeat!1}0.1 & \cellcolor{lbFailureHeat!0}0.0 \\
       & \lbGLMModel & \cellcolor{lbFailureHeat!1}0.4 & \cellcolor{lbFailureHeat!1}0.2 & \cellcolor{lbFailureHeat!1}0.2 & \cellcolor{lbFailureHeat!0}0.0 \\
       & \lbMiniMaxModel & \cellcolor{lbFailureHeat!1}1.2 & \cellcolor{lbFailureHeat!1}0.6 & \cellcolor{lbFailureHeat!1}0.6 & \cellcolor{lbFailureHeat!0}0.0 \\
    \midrule
      \multirow[c]{6}{*}{\lbHarnessCell{mini-swe-agent.png}{mini-SWE-agent}} & \lbDeepSeekModel & \cellcolor{lbFailureHeat!0}0.0 & \cellcolor{lbFailureHeat!0}0.0 & \cellcolor{lbFailureHeat!0}0.0 & \cellcolor{lbFailureHeat!0}0.0 \\
       & \lbClaudeOpus & \cellcolor{lbFailureHeat!1}0.6 & \cellcolor{lbFailureHeat!1}0.4 & \cellcolor{lbFailureHeat!0}0.0 & \cellcolor{lbFailureHeat!1}0.2 \\
       & \lbKimiModel & \cellcolor{lbFailureHeat!1}1.1 & \cellcolor{lbFailureHeat!1}1.1 & \cellcolor{lbFailureHeat!0}0.0 & \cellcolor{lbFailureHeat!0}0.0 \\
       & \lbGPT & \cellcolor{lbFailureHeat!1}0.2 & \cellcolor{lbFailureHeat!0}0.0 & \cellcolor{lbFailureHeat!0}0.0 & \cellcolor{lbFailureHeat!1}0.2 \\
       & \lbGLMModel & \cellcolor{lbFailureHeat!1}0.2 & \cellcolor{lbFailureHeat!1}0.2 & \cellcolor{lbFailureHeat!0}0.0 & \cellcolor{lbFailureHeat!0}0.0 \\
       & \lbMiniMaxModel & \cellcolor{lbFailureHeat!1}1.1 & \cellcolor{lbFailureHeat!1}0.4 & \cellcolor{lbFailureHeat!0}0.0 & \cellcolor{lbFailureHeat!1}0.7 \\
    \bottomrule
  \end{tabular}
  }
  \endgroup
\end{table*}

\section{Limitations}\label{app:limitation}

Our study has the following limitations.

\textbf{\bench evaluates only functional correctness.}
\bench evaluates implementations through F2P tests and P2P regression checks, but does not separately assess readability, maintainability, or consistency with project design conventions. Solutions with similar verifier scores may differ substantially in code duplication, abstraction quality, and ease of future modification. Successful task resolution therefore establishes that an implementation satisfies the evaluated behavioral requirements, while its
broader engineering quality remains unassessed. Complementary expert code review would help determine whether such implementations are suitable for long-term maintenance.

\textbf{Tasks in \bench are limited to some software systems.}
High-quality EPs for large software systems are primarily available in well-maintained open-source communities with established proposal and review processes. Relying on these communities limits the diversity of our data sources and the number of eligible proposals. Our tasks therefore cover a limited set of software systems, and the findings may not generalize to projects with different development practices or less formal requirements documentation.

\textbf{The evaluated models and scaffolds are limited.}
Computational cost restricts our evaluation to 28 agents built from six models and five scaffolds. This selection covers only part of the possible agents and execution settings. Alternative models and scaffolds may produce different results. We also use one attempt per task and agent, except when provider or infrastructure failures require a retry, limiting our characterization of run-to-run variability. The findings therefore describe the evaluated agents under the specified execution conditions and may not generalize to other agents or resource budgets.

\textbf{The instructions may omit important context.}
EPs may rely on assumptions or clarifications recorded in issue threads, design discussions, or review comments. Our tasks provide the curated EP and repository but cannot exhaustively reproduce this surrounding discussion history. Omitted details could introduce ambiguity about expected behavior, compatibility constraints, or design choices. Consequently, some task difficulty may arise from incomplete specification context in addition to the challenge of grounding a proposal in a large codebase.

\textbf{Failure analysis relies partly on LLM judges.}
LLM judges infer semantic failure causes from recorded trajectories, and their judgments may depend on the model, prompt, and interpretation of incomplete evidence. Although we combine predefined rules with outcome-blinded judgments, these controls do not completely eliminate potential bias or ambiguity. Different judges may assign the same failure to different phases or failure modes, particularly when several errors co-occur. The reported
attribution distributions should therefore be interpreted as diagnostic estimates. Replication across judges and systematic comparison with human annotations would further establish their robustness.

\section{A Task Example}\label{app:task}

We reproduce the original instructions for the \texttt{cpython\_1} task instance, PEP~617 (\emph{New PEG parser for CPython}), including the enhancement proposal as supplied to the agent. We remove a few sections in the EP, such as the test plan and validation sections, since they may disclose the hidden tests.

\begin{lstlisting}[
  basicstyle=\fontencoding{T1}\ttfamily\footnotesize,
  columns=fullflexible,
  keepspaces=true,
  breaklines=true,
  breakatwhitespace=false,
  showstringspaces=false,
  upquote=true,
  frame=single,
  rulecolor=\color{black!25},
  framesep=6pt,
  xleftmargin=6pt,
  xrightmargin=6pt,
  literate={ε}{{\ensuremath{\varepsilon}}}1
           {—}{{\textemdash}}1
           {‘}{{\textquoteleft}}1
           {’}{{\textquoteright}}1
]
> Implement the requirement described below in the project's source tree.
> Put implementation changes in `solution.patch`. If you add tests, put
> them in `test.patch`; tests are optional and must not be included in
> `solution.patch`.
>
> This environment has no outbound internet access — `curl`/`wget`, `git fetch`/`clone`, package installs, and web fetch/search will all fail. Implement the requirements using only the code already in the workspace and your own knowledge; do not attempt to fetch or search external resources.

---

# PEP 617: New PEG parser for CPython

## Overview
This PEP proposes replacing the current LL(1)-based parser of CPython
with a new PEG-based parser. This new parser would allow the elimination of multiple
"hacks" that exist in the current grammar to circumvent the LL(1)-limitation.
It would substantially reduce the maintenance costs in some areas related to the
compiling pipeline such as the grammar, the parser and the AST generation. The new PEG
parser will also lift the LL(1) restriction on the current Python grammar.

## Background on LL(1) parsers
The current Python grammar is an LL(1)-based grammar. A grammar can be said to be
LL(1) if it can be parsed by an LL(1) parser, which in turn is defined as a
top-down parser that parses the input from left to right, performing leftmost
derivation of the sentence, with just one token of lookahead.
The traditional approach to constructing or generating an LL(1) parser is to
produce a *parse table* which encodes the possible transitions between all possible
states of the parser. These tables are normally constructed from the *first sets*
and the *follow sets* of the grammar:

* Given a rule, the *first set* is the collection of all terminals that can occur
  first in a full derivation of that rule. Intuitively, this helps the parser decide
  among the alternatives in a rule. For
  instance, given the rule:

```
rule: A | B
```
  if only `A` can start with the terminal *a* and only `B` can start with the
  terminal *b* and the parser sees the token *b* when parsing this rule, it knows
  that it needs to follow the non-terminal `B`.

* An extension to this simple idea is needed when a rule may expand to the empty string.
  Given a rule, the *follow set* is the collection of terminals that can appear
  immediately to the right of that rule in a partial derivation. Intuitively, this
  solves the problem of the empty alternative. For instance,
  given this rule:

```
rule: A 'b'
```
  if the parser has the token *b* and the non-terminal `A` can only start
  with the token *a*, then the parser can tell that this is an invalid program.
  But if `A` could expand to the empty string (called an ε-production),
  then the parser would recognise a valid empty `A`,
  since the next token *b* is in the *follow set*  of `A`.

The current Python grammar does not contain ε-productions, so the *follow sets* are not
needed when creating the parse tables. Currently, in CPython, a parser generator
program reads the grammar and produces a parsing table representing a set of
deterministic finite automata (DFA) that can be included in a C program, the
parser. The parser is a pushdown automaton that uses this data to produce a Concrete
Syntax Tree (CST) sometimes known directly as a "parse tree". In this process, the
*first sets* are used indirectly when generating the DFAs.

LL(1) parsers and grammars are usually efficient and simple to implement
and generate. However, it is not possible, under the LL(1) restriction,
to express certain common constructs in a way natural to the language
designer and the reader. This includes some in the Python language.

As LL(1) parsers can only look one token ahead to distinguish
possibilities, some rules in the grammar may be ambiguous. For instance the rule:

```
rule: A | B
```
is ambiguous if the *first sets* of both `A` and `B` have some elements in
common. When the parser sees a token in the input
program that both *A* and *B* can start with, it is impossible for it to deduce
which option to expand, as no further token of the program can be examined to
disambiguate.
The rule may be transformed to equivalent LL(1) rules, but then it may
be harder for a human reader to grasp its meaning.
Examples later in this document show that the current LL(1)-based
grammar suffers a lot from this scenario.

Another broad class of rules precluded by LL(1) is left-recursive rules.
A rule is left-recursive if it can derive to a
sentential form with itself as the leftmost symbol. For instance this rule:

```
rule: rule 'a'
```
is left-recursive because the rule can be expanded to an expression that starts
with itself. As will be described later, left-recursion is the natural way to
express certain desired language properties directly in the grammar.

## Background on PEG parsers
A PEG (Parsing Expression Grammar) grammar differs from a context-free grammar
(like the current one) in the fact that the way it is written more closely
reflects how the parser will operate when parsing it. The fundamental technical
difference is that the choice operator is ordered. This means that when writing:

```
rule: A | B | C
```
a context-free-grammar parser (like an LL(1) parser) will generate constructions
that given an input string will *deduce* which alternative (`A`, `B` or `C`)
must be expanded, while a PEG parser will check if the first alternative succeeds
and only if it fails, will it continue with the second or the third one in the
order in which they are written. This makes the choice operator not commutative.

Unlike LL(1) parsers, PEG-based parsers cannot be ambiguous: if a string parses,
it has exactly one valid parse tree. This means that a PEG-based parser cannot
suffer from the ambiguity problems described in the previous section.

PEG parsers are usually constructed as a recursive descent parser in which every
rule in the grammar corresponds to a function in the program implementing the
parser and the parsing expression (the "expansion" or "definition" of the rule)
represents the "code" in said function. Each parsing function conceptually takes
an input string as its argument, and yields one of the following results:

* A "success" result. This result indicates that the expression can be parsed by
  that rule and the function may optionally move forward or consume one or more
  characters of the input string supplied to it.
* A "failure" result, in which case no input is consumed.

Notice that "failure" results do not imply that the program is incorrect or a
parsing failure because as the choice operator is ordered, a "failure" result
merely indicates "try the following option". A direct implementation of a PEG
parser as a recursive descent parser will present exponential time performance in
the worst case as compared with LL(1) parsers, because PEG parsers have infinite lookahead
(this means that they can consider an arbitrary number of tokens before deciding
for a rule). Usually, PEG parsers avoid this exponential time complexity with a
technique called "packrat parsing" [1]_ which not only loads the entire
program in memory before parsing it but also allows the parser to backtrack
arbitrarily. This is made efficient by memoizing the rules already matched for
each position. The cost of the memoization cache is that the parser will naturally
use more memory than a simple LL(1) parser, which normally are table-based. We
will explain later in this document why we consider this cost acceptable.

## Rationale
In this section, we describe a list of problems that are present in the current parser
machinery in CPython that motivates the need for a new parser.

### Some rules are not actually LL(1)
Although the Python grammar is technically an LL(1) grammar (because it is parsed by
an LL(1) parser) several rules are not LL(1) and several workarounds are
implemented in the grammar and in other parts of CPython to deal with this. For
example, consider the rule for assignment expressions:

```
namedexpr_test: [NAME ':='] test
```
This simple rule is not compatible with the Python grammar as *NAME* is among the
elements of the *first set* of the rule *test*. To work around this limitation the
actual rule that appears in the current grammar is:

```
namedexpr_test: test [':=' test]
```
Which is a much broader rule than the previous one allowing constructs like ``[x
for x in y] := [1,2,3]``. The way the rule is limited to its desired form is by
disallowing these unwanted constructions when transforming the parse tree to the
abstract syntax tree. This is not only inelegant but a considerable maintenance
burden as it forces the AST creation routines and the compiler into a situation in
which they need to know how to separate valid programs from invalid programs,
which should be a responsibility solely of the parser. This also leads to the
actual grammar file not reflecting correctly what the *actual* grammar is (that
is, the collection of all valid Python programs).

Similar workarounds appear in multiple other rules of the current grammar.
Sometimes this problem is unsolvable. For instance, [bpo-12782: Multiple context expressions do not support parentheses for continuation across lines](https://github.com/python/cpython/issues/56991) shows how making an LL(1) rule that supports
writing:

```
with (
    open("a_really_long_foo") as foo,
    open("a_really_long_baz") as baz,
    open("a_really_long_bar") as bar
):
  ...
```
is not possible since the first sets of the grammar items that can
appear as context managers include the open parenthesis, making the rule
ambiguous. This rule is not only consistent with other parts of the language (like
the rule for multiple imports), but is also very useful to auto-formatting tools,
as parenthesized groups are normally used to group elements to be
formatted together (in the same way the tools operate on the contents of lists,
sets...).

### Complicated AST parsing
Another problem of the current parser is that there is a huge coupling between the
AST generation routines and the particular shape of the produced parse trees. This
makes the code for generating the AST especially complicated as many actions and
choices are implicit. For instance, the AST generation code knows what
alternatives of a certain rule are produced based on the number of child nodes
present in a given parse node. This makes the code difficult to follow as this
property is not directly related to the grammar file and is influenced by
implementation details. As a result of this, a considerable amount of the AST
generation code needs to deal with inspecting and reasoning about the particular
shape of the parse trees that it receives.

### Lack of left recursion
As described previously, a limitation of LL(1) grammars is that they cannot allow
left-recursion. This makes writing some rules very unnatural and far from how
programmers normally think about the program. For instance this construct (a simpler
variation of several rules present in the current grammar):

```
expr: expr '+' term | term
```
cannot be parsed by an LL(1) parser. The traditional remedy is to rewrite the
grammar to circumvent the problem:

```
expr: term ('+' term)*
```
The problem that appears with this form is that the parse tree is forced to have a
very unnatural shape. This is because with this rule, for the input program ``a +
b + c` the parse tree will be flattened (`['a', '+', 'b', '+', 'c']``) and must
be post-processed to construct a left-recursive parse tree (``[['a', '+', 'b'],
'+', 'c']``). Being forced to write the second rule not only leads to the parse
tree not correctly reflecting the desired associativity, but also imposes further
pressure on later compilation stages to detect and post-process these cases.

### Intermediate parse tree
The last problem present in the current parser is the intermediate creation of a
parse tree or Concrete Syntax Tree that is later transformed to an Abstract Syntax
Tree. Although the construction of a CST is very common in parser and compiler
pipelines, in CPython this intermediate CST is not used by anything else (it is
only indirectly exposed by the *parser* module and a surprisingly small part of
the code in the CST production is reused in the module). Which is worse: the whole
tree is kept in memory, keeping many branches that consist of chains of nodes with
a single child. This has been shown to consume a considerable amount of memory (for
instance in [bpo-26415: Excessive peak memory consumption by the Python parser](https://github.com/python/cpython/issues/70603)).

Having to produce an intermediate result between the grammar and the AST is not only
undesirable but also makes the AST generation step much more complicated, raising
considerably the maintenance burden.

## The new proposed PEG parser
The new proposed PEG parser contains the following pieces:

* A parser generator that can read a grammar file and produce a PEG parser
  written in Python or C that can parse said grammar.

* A PEG meta-grammar that automatically generates a Python parser that is used
  for the parser generator itself (this means that there are no manually-written
  parsers).

* A generated parser (using the parser generator) that can directly produce C and
  Python AST objects.

On the implementation side, the Python parser generator exposes these steps as `parse_string`, which runs a parser class over grammar source text; `generate_parser`, which turns a parsed `Grammar` into a parser class; and `make_parser`, which composes the two to build a parser directly from grammar source.
Expose the parser-generator package under `Tools/peg_generator/pegen/`; in
particular, provide `pegen.grammar_parser.GeneratedParser` (commonly used as
`GrammarParser`), `pegen.testutil.parse_string`,
`pegen.testutil.generate_parser`, `pegen.testutil.make_parser`,
`pegen.testutil.generate_parser_c_extension`,
`pegen.testutil.generate_c_parser_source`, `pegen.grammar.Grammar`,
`pegen.grammar.GrammarError`, `pegen.grammar.GrammarVisitor`, and
`pegen.first_sets.FirstSetCalculator`.

### Left recursion
PEG parsers normally do not support left recursion but we have implemented a
technique similar to the one described in Medeiros et al. [2]_ but using the
memoization cache instead of static variables. This approach is closer to the one
described in Warth et al. [3]_. This allows us to write not only simple left-recursive
rules but also more complicated rules that involve indirect left-recursion like:

```
rule1: rule2 | 'a'
rule2: rule3 | 'b'
rule3: rule1 | 'c'
```
and "hidden left-recursion" like:

```
rule: 'optional'? rule '@' some_other_rule
```
### Syntax
The grammar consists of a sequence of rules of the form:

```
rule_name: expression
```
Optionally, a type can be included right after the rule name, which
specifies the return type of the C or Python function corresponding to
the rule:

```
rule_name[return_type]: expression
```
If the return type is omitted, then a `void *` is returned in C and an
`Any` in Python.

#### Grammar Expressions
`# comment`
'''''''''''''

Python-style comments.

`e1 e2`
'''''''''

Match e1, then match e2.

```PEG
rule_name: first_rule second_rule
```
`e1 | e2`
'''''''''''

Match e1 or e2.

The first alternative can also appear on the line after the rule name
for formatting purposes. In that case, a \| must be used before the
first alternative, like so:

```PEG
rule_name[return_type]:
    | first_alt
    | second_alt
```
`( e )`
'''''''''

Match e.

```PEG
rule_name: (e)
```
A slightly more complex and useful example includes using the grouping
operator together with the repeat operators:

```PEG
rule_name: (e1 e2)*
```
`[ e ] or e?`
'''''''''''''''

Optionally match e.

```PEG
rule_name: [e]
```
A more useful example includes defining that a trailing comma is
optional:

```PEG
rule_name: e (',' e)* [',']
```
`e*`
''''''

Match zero or more occurrences of e.

```PEG
rule_name: (e1 e2)*
```
`e+`
''''''

Match one or more occurrences of e.

```PEG
rule_name: (e1 e2)+
```
`s.e+`
''''''''

Match one or more occurrences of e, separated by s. The generated parse
tree does not include the separator. This is otherwise identical to
`(e (s e)*)`.

```PEG
rule_name: ','.e+
```
`&e`
''''''

Succeed if e can be parsed, without consuming any input.

`!e`
''''''

Fail if e can be parsed, without consuming any input.

An example taken from the proposed Python grammar specifies that a primary
consists of an atom, which is not followed by a `.` or a `(` or a
`[`:

```PEG
primary: atom !'.' !'(' !'['
```
`~`
''''''

Commit to the current alternative, even if it fails to parse.

```PEG
rule_name: '(' ~ some_rule ')' | some_alt
```
In this example, if a left parenthesis is parsed, then the other
alternative won’t be considered, even if some_rule or ‘)’ fail to be
parsed.

#### Variables in the Grammar
A subexpression can be named by preceding it with an identifier and an
`=` sign. The name can then be used in the action (see below), like this:

```
rule_name[return_type]: '(' a=some_other_rule ')' { a }
```
### Grammar actions
To avoid the intermediate steps that obscure the relationship between the
grammar and the AST generation the proposed PEG parser allows directly
generating AST nodes for a rule via grammar actions. Grammar actions are
language-specific expressions that are evaluated when a grammar rule is
successfully parsed. These expressions can be written in Python or C
depending on the desired output of the parser generator. This means that if
one would want to generate a parser in Python and another in C, two grammar
files should be written, each one with a different set of actions, keeping
everything else apart from said actions identical in both files. As an
example of a grammar with Python actions, the piece of the parser generator
that parses grammar files is bootstrapped from a meta-grammar file with
Python actions that generate the grammar tree as a result of the parsing.

In the specific case of the new proposed PEG grammar for Python, having
actions allows directly describing how the AST is composed in the grammar
itself, making it more clear and maintainable. This AST generation process is
supported by the use of some helper functions that factor out common AST
object manipulations and some other required operations that are not directly
related to the grammar.

To indicate these actions each alternative can be followed by the action code
inside curly-braces, which specifies the return value of the alternative:

```
rule_name[return_type]:
    | first_alt1 first_alt2 { first_alt1 }
    | second_alt1 second_alt2 { second_alt1 }
```
If the action is omitted and C code is being generated, then there are two
different possibilities:

1. If there’s a single name in the alternative, this gets returned.
2. If not, a dummy name object gets returned (this case should be avoided).

If the action is omitted and Python code is being generated, then a list
with all the parsed expressions gets returned (this is meant for debugging).

The full meta-grammar for the grammars supported by the PEG generator is:

```PEG
start[Grammar]: grammar ENDMARKER { grammar }

grammar[Grammar]:
    | metas rules { Grammar(rules, metas) }
    | rules { Grammar(rules, []) }

metas[MetaList]:
    | meta metas { [meta] + metas }
    | meta { [meta] }

meta[MetaTuple]:
    | "@" NAME NEWLINE { (name.string, None) }
    | "@" a=NAME b=NAME NEWLINE { (a.string, b.string) }
    | "@" NAME STRING NEWLINE { (name.string, literal_eval(string.string)) }

rules[RuleList]:
    | rule rules { [rule] + rules }
    | rule { [rule] }

rule[Rule]:
    | rulename ":" alts NEWLINE INDENT more_alts DEDENT {
          Rule(rulename[0], rulename[1], Rhs(alts.alts + more_alts.alts)) }
    | rulename ":" NEWLINE INDENT more_alts DEDENT { Rule(rulename[0], rulename[1], more_alts) }
    | rulename ":" alts NEWLINE { Rule(rulename[0], rulename[1], alts) }

rulename[RuleName]:
    | NAME '[' type=NAME '*' ']' {(name.string, type.string+"*")}
    | NAME '[' type=NAME ']' {(name.string, type.string)}
    | NAME {(name.string, None)}

alts[Rhs]:
    | alt "|" alts { Rhs([alt] + alts.alts)}
    | alt { Rhs([alt]) }

more_alts[Rhs]:
    | "|" alts NEWLINE more_alts { Rhs(alts.alts + more_alts.alts) }
    | "|" alts NEWLINE { Rhs(alts.alts) }

alt[Alt]:
    | items '$' action { Alt(items + [NamedItem(None, NameLeaf('ENDMARKER'))], action=action) }
    | items '$' { Alt(items + [NamedItem(None, NameLeaf('ENDMARKER'))], action=None) }
    | items action { Alt(items, action=action) }
    | items { Alt(items, action=None) }

items[NamedItemList]:
    | named_item items { [named_item] + items }
    | named_item { [named_item] }

named_item[NamedItem]:
    | NAME '=' ~ item {NamedItem(name.string, item)}
    | item {NamedItem(None, item)}
    | it=lookahead {NamedItem(None, it)}

lookahead[LookaheadOrCut]:
    | '&' ~ atom {PositiveLookahead(atom)}
    | '!' ~ atom {NegativeLookahead(atom)}
    | '~' {Cut()}

item[Item]:
    | '[' ~ alts ']' {Opt(alts)}
    |  atom '?' {Opt(atom)}
    |  atom '*' {Repeat0(atom)}
    |  atom '+' {Repeat1(atom)}
    |  sep=atom '.' node=atom '+' {Gather(sep, node)}
    |  atom {atom}

atom[Plain]:
    | '(' ~ alts ')' {Group(alts)}
    | NAME {NameLeaf(name.string) }
    | STRING {StringLeaf(string.string)}

## Mini-grammar for the actions

action[str]: "{" ~ target_atoms "}" { target_atoms }

target_atoms[str]:
    | target_atom target_atoms { target_atom + " " + target_atoms }
    | target_atom { target_atom }

target_atom[str]:
    | "{" ~ target_atoms "}" { "{" + target_atoms + "}" }
    | NAME { name.string }
    | NUMBER { number.string }
    | STRING { string.string }
    | "?" { "?" }
    | ":" { ":" }
```
As an illustrative example this simple grammar file allows directly
generating a full parser that can parse simple arithmetic expressions and that
returns a valid C-based Python AST:

```PEG
start[mod_ty]: a=expr_stmt* $ { Module(a, NULL, p->arena) }
expr_stmt[stmt_ty]: a=expr NEWLINE { _Py_Expr(a, EXTRA) }
expr[expr_ty]:
    | l=expr '+' r=term { _Py_BinOp(l, Add, r, EXTRA) }
    | l=expr '-' r=term { _Py_BinOp(l, Sub, r, EXTRA) }
    | t=term { t }

term[expr_ty]:
    | l=term '*' r=factor { _Py_BinOp(l, Mult, r, EXTRA) }
    | l=term '/' r=factor { _Py_BinOp(l, Div, r, EXTRA) }
    | f=factor { f }

factor[expr_ty]:
    | '(' e=expr ')' { e }
    | a=atom { a }

atom[expr_ty]:
    | n=NAME { n }
    | n=NUMBER { n }
    | s=STRING { s }
```
Here `EXTRA` is a macro that expands to ``start_lineno, start_col_offset,
end_lineno, end_col_offset, p->arena``, those being variables automatically
injected by the parser; `p` points to an object that holds on to all state
for the parser.

A similar grammar written to target Python AST objects:

```PEG
start: expr NEWLINE? ENDMARKER { ast.Expression(expr) }
expr:
    | expr '+' term { ast.BinOp(expr, ast.Add(), term) }
    | expr '-' term { ast.BinOp(expr, ast.Sub(), term) }
    | term { term }

term:
    | l=term '*' r=factor { ast.BinOp(l, ast.Mult(), r) }
    | term '/' factor { ast.BinOp(term, ast.Div(), factor) }
    | factor { factor }

factor:
    | '(' expr ')' { expr }
    | atom { atom }

atom:
    | NAME { ast.Name(id=name.string, ctx=ast.Load()) }
    | NUMBER { ast.Constant(value=ast.literal_eval(number.string)) }
```
## Migration plan
This section describes the migration plan when porting to the new PEG-based parser
if this PEP is accepted. The migration will be executed in a series of steps that allow
initially to fallback to the previous parser if needed:

1.  Starting with Python 3.9 alpha 6, include the new PEG-based parser machinery in CPython
    with a command-line flag and environment variable that allows switching between
    the new and the old parsers together with explicit APIs that allow invoking the
    new and the old parsers independently. At this step, all Python APIs like `ast.parse`
    and `compile` will use the parser set by the flags or the environment variable and
    the default parser will be the new PEG-based parser. Which parser is active is
    observable at runtime through `sys.flags.use_peg`, a new member of the `sys.flags`
    struct sequence that is nonzero while the PEG parser is in effect.

2.  Between Python 3.9 and Python 3.10, the old parser and related code (like the
    "parser" module) will be kept until a new Python release happens (Python 3.10). In
    the meanwhile and until the old parser is removed, **no new Python Grammar
    addition will be added that requires the PEG parser**. This means that the grammar
    will be kept LL(1) until the old parser is removed.

3.  In Python 3.10, remove the old parser, the command-line flag, the environment
    variable and the "parser" module and related code.

## Performance and validation
We have done extensive timing and validation of the new parser, and
this gives us confidence that the new parser is of high enough quality
to replace the current parser.

### Performance
We have tuned the performance of the new parser to come within 10% of
the current parser both in speed and memory consumption. While the
PEG/packrat parsing algorithm inherently consumes more memory than the
current LL(1) parser, we have an advantage because we don't construct
an intermediate CST.

Below are some benchmarks. These are focused on compiling source code
to bytecode, because this is the most realistic situation. Returning
an AST to Python code is not as representative, because the process to
convert the *internal* AST (only accessible to C code) to an
*external* AST (an instance of `ast.AST`) takes more time than the
parser itself.

All measurements reported here are done on a recent MacBook Pro,
taking the median of three runs. No particular care was taken to stop
other applications running on the same machine.

The first timings are for our canonical test file, which has 100,000
lines endlessly repeating the following three lines:

```python
1 + 2 + 4 + 5 + 6 + 7 + 8 + 9 + 10 + ((((((11 * 12 * 13 * 14 * 15 + 16 * 17 + 18 * 19 * 20))))))
2*3 + 4*5*6
12 + (2 * 3 * 4 * 5 + 6 + 7 * 8)
```
- Just parsing and throwing away the internal AST takes 1.16 seconds
  with a max RSS of 681 MiB.

- Parsing and converting to `ast.AST` takes 6.34 seconds, max RSS
  1029 MiB.

- Parsing and compiling to bytecode takes 1.28 seconds, max RSS 681
  MiB.

- With the current parser, parsing and compiling takes 1.44 seconds,
  max RSS 836 MiB.

For this particular test file, the new parser is faster and uses less
memory than the current parser (compare the last two bullets).

We also did timings with a more realistic payload, the entire Python
3.8 stdlib. This payload consists of 1,641 files, 749,570 lines,
27,622,497 bytes. (Though 11 files can't be compiled by any Python 3
parser due to encoding issues, sometimes intentional.)

- Compiling and throwing away the internal AST took 2.141 seconds.
  That's 350,040 lines/sec, or 12,899,367 bytes/sec. The max RSS was
  74 MiB (the largest file in the stdlib is much smaller than our
  canonical test file).

- Compiling to bytecode took 3.290 seconds. That's 227,861 lines/sec,
  or 8,396,942 bytes/sec. Max RSS 77 MiB.

- Compiling to bytecode using the current parser took 3.367 seconds.
  That's 222,620 lines/sec, or 8,203,780 bytes/sec. Max RSS 70 MiB.

Comparing the last two bullets we find that the new parser is slightly
faster but uses slightly (about 10%) more memory. We believe this is
acceptable. (Also, there are probably some more tweaks we can make to
reduce memory usage.)

## Rejected Alternatives
We did not seriously consider alternative ways to implement the new
parser, but here's a brief discussion of LALR(1).

Thirty years ago the first author decided to go his own way with
Python's parser rather than using LALR(1), which was the industry
standard at the time (e.g. Bison and Yacc).  The reasons were
primarily emotional (gut feelings, intuition), based on past experience
using Yacc in other projects, where grammar development took more
effort than anticipated (in part due to shift-reduce conflicts).  A
specific criticism of Bison and Yacc that still holds is that their
meta-grammar (the notation used to feed the grammar into the parser
generator) does not support EBNF conveniences like
`[optional_clause]` or `(repeated_clause)*`.  Using a custom
parser generator, a syntax tree matching the structure of the grammar
could be generated automatically, and with EBNF that tree could match
the "human-friendly" structure of the grammar.

Other variants of LR were not considered, nor was LL (e.g. ANTLR).
PEG was selected because it was easy to understand given a basic
understanding of recursive-descent parsing.
\end{lstlisting}

\end{document}